\documentclass[a4paper,11pt]{article}
\pdfoutput=1 % if your are submitting a pdflatex (i.e. if you have
\usepackage{amsmath} 
\allowdisplaybreaks
\usepackage{jheppub} % for details on the use of the package, please
\usepackage{empheq}
\usepackage{bm}
\usepackage{slashed}
\usepackage[T1]{fontenc} % if needed
\usepackage[all]{xy}
\usepackage{rotating}
\usepackage{mathtools}
\usepackage{bbm}% mathbb
\usepackage{dcolumn}
\usepackage{multirow}
\usepackage{mathrsfs}% mathscr
\usepackage{arydshln}% dashed line
\usepackage{verbatim}% comment
\usepackage{graphics}
\usepackage{longtable}
\usepackage{tcolorbox}

\def\diag{\mathop{\rm diag}\nolimits}

\DeclareMathOperator{\Tr}{Tr}

\def\Re{\mathop{\mathrm{Re}}}

\usepackage{slashed}

\def\a{\alpha}

\def\CA{{\cal A}}
\def\CB{{\cal B}}
\def\CC{{\cal C}}
\def\CD{{\cal D}}

\def\CH{{\cal H}}
\def\CI{{\cal I}}

\def\CL{{\cal L}}
\def\CM{{\cal M}}
\def\CN{{\cal N}}
\def\CO{{\cal O}}

\def\CR{{\cal R}}
\def\CS{{\cal S}}
\def\CT{{\cal T}}

\def\CW{{\cal W}}

\def\CZ{{\cal Z}}

\def\beq#1\eeq{\begin{align}#1\end{align}}

  \let\n=\nu

\newcommand{\be}{\begin{equation}}
	\newcommand{\ee}{\end{equation}}
\newcommand{\ba}{\begin{align}}
	\newcommand{\ea}{\end{align}}%Verbatim cannot recognize this command. 

\newcommand{\bi}{\begin{itemize}}
	\newcommand{\ei}{\end{itemize}}

\makeatletter
\newcommand*{\rom}[1]{\expandafter\romannumeral #1}
\makeatother%For Roman Numeral

{\preprint{KIAS-P23052}}

\title{\boldmath  A non-unitary bulk-boundary correspondence
\\  :Non-unitary Haagerup RCFTs from S-fold SCFTs}

	\abstract{We introduce a novel class of two-dimensional non-unitary  rational conformal field theories (RCFTs) whose modular data are identical 
    to the generalized Haagerup-Izumi modular data. Via the bulk-boundary correspondence, 
    they are related to the three-dimensional non-unitary Haagerup topological field theories, 
    recently constructed by a topological twisting of three-dimensional ${\cal N}=4$ rank-zero superconformal field theories (SCFTs), called S-fold 
    SCFTs. We propose that, up to the overall factors, the half-indices of the rank-zero SCFTs give the explicit Nahm representation of  four conformal characters of 
    the RCFTs including the vacuum character. Using the theory of Bantay-Gannon, we can successfully complete them into 
    the full admissible conformal characters of the RCFTs.  
 }
	
	\author[a]{Dongmin Gang,}
	\author[a,b]{Dongyeob Kim,}
	\author[c]{Sungjay Lee}

	\affiliation[a]{
		Department of Physics and Astronomy $\&$ Center for Theoretical Physics,
		\\
		Seoul National University, 1 Gwanak-ro, Seoul 08826, Korea}
	\affiliation[b]{Department of Physics, Princeton University, Princeton, NJ 08544, USA}
	\affiliation[c]{Korea Institute for Advanced Study, 85 Hoegiro, Dongdaemun-Gu, Seoul 02455, Korea}
	
	\emailAdd{arima275@snu.ac.kr}
 	\emailAdd{dk2846@princeton.edu}
	\emailAdd{sjlee@kias.re.kr}
	
\begin{document} 
		\maketitle
		\flushbottom	
		
	\section{Introduction}

A characteristic property of (2+1)-dimensional topological matters is the presence of chiral gapless states 
on their boundary. It is well-studied that, when a topological phase in the bulk can be described by 
a unitary topological quantum field theory (TQFT), the universal behaviors of gapless boundary states 
can be captured by a two-dimensional unitary rational conformal field theory (RCFT). This connection 
is often referred to as the bulk-boundary correspondenc  \cite{zbMATH04092352,moore1989lectures}.  A prominent example is the correspondence 
between the three-dimensional Chern-Simons theories and the two-dimensional Wess-Zumino-Witten (WZW) models. 
The Chern-Simons theories are low-energy effective theories for certain quantum Hall 
systems, typical many-body systems with a mass gap. 

The bulk-boundary correspondence however becomes less clear in non-unitary cases, primarily due to the limited understanding of the physical nature of the non-unitary TQFTs. Mathematically, the non-unitary TQFTs 
can be defined in a manner nearly identical to the unitary ones with certain unitarity conditions being relaxed.  
In particular, the bulk observables in both theories are encoded in rigid mathematical structure known as the modular tensor category (MTC) \cite{moore1989classical,turaev1992modular}. 
However, the manifestation of a non-unitary TQFT equipped with MTC in a physical system has been a formidable challenge
until the recent proposals \cite{Gang:2018huc,Gang:2021hrd}. 

Specifically, it was argued that a family of non-unitary TQFTs can be realized as a topological twisting of an exotic class of three-dimensional ${\cal N}=4$ superconformal field theories (SCFTs), called the rank-zero theories. Those rank-zero theories can be characterized by the absence of both Coulomb and Higgs branches, and have an isolated vacuum. 
Despite the extensive studies of the landscape of the three-dimensional ${\cal N}=4$ SCFTs, 
this class of exotic SCFTs has largely remained unexplored. This is essentially because 
most of them are not connected to weakly-coupled UV theories with manifest 
${\cal N}=4$ supersymmetry by RG flows. Rather, their ${\cal N}=4$ supersymmetry is accidental in the infrared limit, namely their SUSY get enhanced dynamically.  

In three dimensions, the ${\cal N}=4$ SUSY is the minimal SUSY that admits a topological twisting. 
When a given theory is topologically twisted, the spin-statistics theorem is no longer valid and the 
unitarity becomes violated. For the theories of rank zero, the topological twisting results in 
genuine (semi-simple) TQFTs associated with MTC. Thus,  one can expect that
the topologically twisted ${\cal N}=4$ rank-zero theories physically realizes the non-unitary TQFTs. We summarize the chains of correspondences in the following diagram:
\begin{align}
	\fbox{Rank-zero SCFT} \xrightarrow{  \rm top'l \; twisting  } \fbox{Non-unitary TQFT}  \xrightarrow{  \textrm{bulk-boundary}} \fbox{Non-unitary RCFT}
	\nonumber % \label{non-unitary bulk-boundary}
\end{align}

In the present work, we examine the non-unitary version of bulk-boundary correspondence for the S-fold SCFTs, defined in \eqref{S-fold from T[SU(2)]}, as concrete examples. As demonstrated in \cite{Gang:2022kpe}, the topological twisting 
of such S-fold SCFTs gives rise to the TQFTs with modular data generalizing the non-unitary Haagerup-Izumi modular data.  For this reason, we refer to them as the generalized Haagerup TQFTs. 
Is there a boundary RCFT corresponding to a generalized Haagerup TQFT?  
To see this, we utilize the so-called Riemann-Hilbert method \cite{Bantay:2005vk,Ballin:2022rto} to 
explicitly construct the conformal characters compatible with 
a given generalized Haagerup-Izumi modular data. Herein, the half-index computation of the aforementioned rank-zero 
theories play a key role in making this approach effective. Note that the central charge read off from the vacuum conformal character is always negative, which is consistent with the fact that
the modular data of our interest give negative quantum dimensions. 
It strongly supports that there exist non-unitary RCFTs associated with the Haagerup TQFTs, 
and suggests that the non-unitary version of the bulk-boundary correspondence actually does work.

The rest of the paper is organized as follows.   In section \ref{sec 2}, we introduce the S-fold SCFTs labeled by an integer $k\geq 3$ and its topologically twisted theories. We also propose a dual field theory description of the SCFTs  in terms of  abelian gauge fields. We review the computation of modular data of the topologically twisted theories and explain why we call them  generalized non-unitary Haagerup TQFTs.  In section \ref{sec 3}, we introduce generalized non-unitary Haagerup RCFTs, which live on the boundary of the TQFTs.  From the computation of half-indices using the abelian dual description, we propose fermionic sum representations of some characters of the non-unitary RCFTs. We then verify  that the characters  can be completed into full admissible RCFT characters  by using either modular linear differential equation  (MLDE) or  Riemann-Hilbert method. We present the full characters  up to $k=11$ but the analysis can be in principle  extended to arbitrarily higher $k$.  In  appendix, we give a derivation of the dual abelian description  using  3D-3D correspondence.

	\section{3D S-fold SCFTs and Generalized Non-unitary Haagerup TQFTs} \label{sec 2}

	\subsection{S-fold SCFT $\mathcal{S}_k$ }
	
	Here we introduce two UV descriptions of the  S-fold SCFTs. 
	One is using $T[SU(2)]$ theory   which is the conventional one. The other is based on $\CN=2$ abelian gauge fields coupled to chiral multiplets. The abelian description will be used in defining the generalized Haagerup  RCFTs, the main hero of this paper, and computing the half-indices, which give  characters of the RCFTs .

	\subsubsection{From $T[SU(2)]$ theory}
	We define the S-fold superconformal field theory (SCFT) $\CS_{k}$ by 
	\begin{align}
		\begin{split}
			&\CS_k := \frac{T[SU(2)]}{SU(2)_k^{\rm diag}}\;
			\\&:= (\textrm{Gauging diagonal $SU(2)$ of $T[SU(2)]$ theory}
			\\ & \quad \quad  \textrm{with $\CN=3$ Chern-Simons term of level $k$})\;. \label{S-fold from T[SU(2)]}
		\end{split}
	\end{align}
	The $T[SU(2)]$ theory is a 3D $\CN=4$ SQED with two hypermultiplets \cite{Gaiotto:2008ak}. The theory has manifest $U(1)\times SU(2)$ flavor symmetry which is enhanced to $SU(2)\times SU(2)$ in the infra-red (IR).  By gauging the diagonal $SU(2)$ with non-zero Chern-Simons level $k$, we have the S-fold theory.  The gauging breaks the $\CN=4$ SUSY to $\CN=3$  but the $\CN=4$ SUSY is restored in the IR and the theory flows to an non-trivial $\CN=4$ SCFT (resp. direct product of  $\CN=4$ SCFTs) in the IR when $|k|\geq 4$ (resp. $|k|=3$).  Both Coulomb and Higgs branches of the $T[SU(2)]$ theory are lifted during the gauging and the S-fold SCFT has trivial Higgs and Coulomb branches and such a  $\CN=4$ theory is called 3D $\CN=4$ rank-zero theory \cite{Gang:2021hrd}.  Refer to  \cite{Assel:2018vtq,Garozzo:2018kra,Garozzo:2019hbf,Garozzo:2019ejm,Beratto:2020qyk,Arav:2021gra,Bobev:2021yya} for recent studies on the theory. 
 	
	\subsubsection{Dual Abelian description}
	The S-fold SCFT $\CS_k$ can be obtained from a twisted compactification of 6D  (2,0) theory of $A_1$ type on a 3-manifold, once-punctured torus bundle with a monodromy element $\varphi = ST^k \in SL(2,\mathbb{Z})$ \cite{Terashima:2011qi}. 
	Using the 3D-3D correspondence,  we propose following dual description for the S-fold SCFT \cite{Dimofte:2011ju,Gang:2013sqa},
\begin{align}
	\CS_k =\begin{cases} \textrm{3D $\CN=2$ $U(1)_K^2$ gauge theory coupled to two chirals $\Phi_{a=1,2}$}   \;, \quad k=3
		\\
		\left( U(1)_K^{r=k-1} \textrm{ coupled to $r$ chirals $\Phi_{a=1,\ldots, r}$} \textrm{ with } \CW_{\rm sup} = \sum_{m=1}^{r-1} \CO_{m}   \right) \;, \quad k\geq 4 \label{S-fold from abelian}
	\end{cases}
\end{align}
See Appendix \ref{App : dual of Sk} for a detailed  derivation. 
Here $\Phi_{1\leq a \leq  r}$ is a chiral multiplet whose  charge under the $b$-th  $U(1)$  gauge symmetry  is $\delta_{ab} Q_a$ with 
\begin{align}\label{gauge charge}
Q_a = \begin{cases}
	2, \quad a=1
	\\
	1, \quad 2\leq a \leq r
	\end{cases}
\end{align}
The mixed Chern-Simons level $K$ for the $\CN=2$ $U(1)^r$ gauge theory is given by\footnote{Here we use "$U(1)_{-1/2}$ quantization" of the chiral multiplet, i.e. we turn on background CS level $-1/2$ for the $U(1)$ symmetry of $\Phi$. In the usual convention, the (UV effective) mixed CS level is $K_{ab} - \frac{1}2 \delta_{ab}Q_a^2$. } 
\begin{align}
	\begin{split}
		&K= \begin{pmatrix}
			2(k-2) & 0 & 2 & 4 & 6 & \ldots  & 2(k-3)\\
			0 & 2 & 2 & 2 & 2& \ldots & 2\\
			2 & 2 & 4 & 4 & 4 & \ldots & 4\\
			4 & 2 & 4 & 6 & 6 & \ldots & 6\\
			6 & 2 & 4 & 6 & 8 & \ldots  & 8 \\
			\ldots & \ldots & \ldots & \ldots & \ldots & \ldots \\
			2(k-3) & 2 & 4 & 6 & 8 & \ldots  & 2(k-2)
		\end{pmatrix}
	\end{split}\;.\label{K-matrix}
\end{align} 
$\CO_m$'s in the superpotential $\CW_{\rm sup}$ are gauge-invariant $1/2$ BPS monopole operators (more precisely, chiral primary multiplets containing the monopole operators):
\begin{align}
	\begin{split}
		&\CO_{m=1} = \begin{cases} 
						V_{(-1,-2,2)} \;, \quad k=4
						\\
		 	V_{(-1,-1, \mathbf{0}_{k-5} , -1, 2)} \;, \quad k \geq 5
		 	\end{cases}\;,
		\\
		&\CO_{m=2} = V_{(0, 2,-1, \mathbf{0}_{k-4})} \Phi_1\;, 
		\\
		& \CO_{m=3} = V_{(0,-1,2,-1,\mathbf{0}_{k-5})}\;,
		\\
		&\CO_{m=4} = V_{(0,0,-1,2,-1,\mathbf{0}_{k-6})}\;,
		\\
		& \ldots
		\\
		&\CO_{m=r-1} = V_{(0,0,0,\ldots, -1,2,-1)}\;.
	\end{split} \;.  \label{monopole operators}
\end{align}
Here $\mathbf{0}_n= \overbrace{0,0,\ldots,0}^{n}$ and $V_{\mathfrak{m}=(m_1, \ldots, m_r)}$ denotes the 1/2 BPS bare monopole operator with flux $\mathfrak{m}$. The charge $q_a$ under the $a$-th $U(1)$ gauge symmetry of the monopole operator is
\begin{align}
q_a (V_{\mathfrak{m}}) =\sum_{b=1}^r (K_{ab}-\frac{1}2 \delta_{ab} Q_a^2) m_b - \frac{Q_a^2}2 |  m_a|\;,
\end{align}
and one can confirm the gauge invariance of the superpotential $\CW_{\rm sup}$. The 2nd monopole operator $\CO_{m=2}$ is also $1/2$ BPS since it is purely electric or magnetic for each $U(1)$ factor in the gauge group. 

The dual description has only manifest $\CN=2$ supersymmetry which is expected to be enhanced to $\CN=4$ in the   IR. 
For $k\geq 4$,   the  superpotential deformation breaks the $U(1)^r$ topological symmetries (of  $U(1)^r$ gauge symmetry) to $U(1)_A$ symmetry whose charge  $A$ is 
\begin{align}
	A =(r-1) M_1+ M_2 + 2 M_3 +3 M_4 +\ldots + (r-1) M_r\;. \label{Axial symmetry}
\end{align}
Here $M_a$ is the charge of $U(1)$ topological symmetry for  $a$-th $U(1)$ gauge symmetry.  The theory also has $U(1)_R$ R-symmetry which can be mixed with the $U(1)_A$ flavor symmetry. Let the R-charge at the general mixing be
\begin{align}
R_{\nu}  = R_{\nu=0}+\nu A\;,
\end{align}
with a mixing parameter $\nu\in \mathbb{R}$. We choose  a reference R-charge $R_{\nu=0}$ to be the superconformal R-charge of the IR SCFT.  Under the SUSY enhancement, the $U(1)_R \times U(1)_A$ is expected to be enhanced to $SO(4)_R  \simeq  SU(2)_C \times SU(2)_H$ R-symmetry with the following embedding
\begin{align} 
R_\nu = \left(J_3^C+J_3^H  \right)+\nu \left( J_3^C- J_3^H \right)\;. \label{R-symmetry mixing}
\end{align}
Here $J_3^{C/H}$ denotes the Cartan generator of $SU(2)_{C/H}$ normalized as $J_3 \in  \mathbb{Z}/2$.

\paragraph{SUSY partition functions} The squashed 3-sphere partition function \cite{Hama:2010av,Hama:2011ea} of the $\CS_k$ in \eqref{S-fold from abelian} is (following the conventions in \cite{Gang:2019jut})
\begin{align}
	\begin{split}
	&\mathcal{Z}_{S^3_b} \left(m, \nu \right) = \int \frac{d^r \mathbf{Z}}{(2\pi \hbar)^{\frac{r}2}} \CI_\hbar(\mathbf{Z},m, \nu)\;, \textrm{ where } \mathbf{Z} = (Z_1, Z_2,\ldots, Z_r)  \textrm{ and}
	\\
	& \CI_\hbar (\mathbf{Z}, m, \nu ) =  \exp \left( \frac{\mathbf{Z}^T K \mathbf{Z}+2 W \left((r-1)Z_1 +\sum_{a=1}^{r-1} a Z_{a+1} \right)}{2\hbar }  \right) \prod_{a=1}^r \psi_\hbar (Q_a Z_a)
	\\
	& \textrm{ with }W = m + (i \pi +\frac{\hbar}2 ) (\nu-1)\;. \label{3-sphere of Sk}
	\end{split}
\end{align}
Here $\hbar = 2\pi i b^2$ with the squashing parameter $b$.  $m$ is a rescaled real mass parameter, $b \times (\textrm{real mass})$,  of the $U(1)_A$ symmetry and $\nu$ is the R-symmetry mixing parameter in \eqref{R-symmetry mixing}.\footnote{The overall additive constant of $\nu$ is fixed by requiring that the round 3-sphere partition function $|\CZ_{S^3_{b=1}} (m=0, \nu)|$ is minimized at $\nu=0$ \cite{Jafferis:2010un}. Actually,  the $|\CZ_{S^3_{b=1}} (m=0, \nu)|$ is invariant under a sign change of $\nu$ which is related to the self-mirror property of the $T[SU(2)]$  and is related to the $\mathbb{Z}_2$ Weyl-symmetry,  $M_A \leftrightarrow -M_A$,  of the $SL(2,\mathbb{C})$ Chern-Simons theory in \eqref{SL(2,C) CS ptn} through a 3D-3D relation \eqref{3D-3D correpondence-2}. }  
Using the following asymptotic expansion of the quantum dilogarithm function $\psi_\hbar (Z)$ \cite{Faddeev:1993rs},
\begin{align}
\log \psi_{\hbar} (Z) \xrightarrow {\quad \hbar \rightarrow 0 \quad } \frac{1}{\hbar } {\rm Li}_2 (e^{-Z}) - \frac{1}2 \log (1-e^{-Z})+ O (\hbar),
\end{align}
the integrand can be expanded  perturbatively in $\hbar$ as follows
\begin{align}
\begin{split}
	&\log \CI (\mathbf{Z}, m, \nu ) \xrightarrow{\quad \hbar \rightarrow 0 \quad }  \frac{1}{\hbar} \CW_0 (\mathbf{Z}, m, \nu) + \CW_1 (\mathbf{Z}, m, \nu)+ O (\hbar)\;,  \;\; \textrm{where}
	\\
	&\CW_0 = \sum_{a=1}^r {\rm Li}_2 (e^{-Q_a Z_a})+ \frac{1}2 \mathbf{Z}^T K \mathbf{Z}+ \left(m+ i \pi (\nu-1)\right) \left((r-1)Z_1 +\sum_{a=1}^{r-1} a Z_{a+1}\right)\;,
	\\
	&\CW_1 =  -\frac{1}2 \sum_{a=1}^r \log (1-e^{- Q_a Z_a}) +\frac{1}2 (\nu-1) \left((r-1)Z_1 +\sum_{a=1}^{r-1} a Z_{a+1}\right) \;. \label{twisted ptns 1}
\end{split}
\end{align}
Using the first two coefficients $\CW_0$ and $\CW_1$, the supersymmetric partition function $\mathcal{Z}_{(g,p)}$ on $\CM_{g,p}$, degree $p$ bundle over genus $g$ Riemann surface $\Sigma_g$, with $p\in 2\mathbb{Z}$ can be computed in the following way \cite{Closset:2017zgf,Closset:2018ghr,Gang:2019jut}
\begin{align}
\begin{split}
&\CZ_{(g,p)} (m, \nu) = \sum_{\mathbf{z}^\alpha \in \CS_{\rm B.E}(m,\nu)} \left( H_\alpha (m, \nu)\right)^{g-1} \left(F_\alpha (m, \nu)\right)^{p}\;,  \;\; \textrm{where}
\\
&\textrm{Bethe-vacua : }\CS_{\rm B.E}(m,\nu)  = \left\{\mathbf{z} \;:\; \exp(\partial_{Z_b} \CW_0)|_{\mathbf{Z} \rightarrow \log \mathbf{z}} =1, \; b=1,\ldots, r\right\} 
\\
&\qquad \qquad \qquad \qquad \quad \quad \quad = \{ \mathbf{z}^\alpha\}_{\alpha=0}^{|\CS_{\rm B.E}|-1}\;,
\\
&\textrm{Handle gluing operator : } H_\alpha (m, \nu) = \det_{a,b}  \left(\partial_{Z_a} \partial_{Z_b}\CW_0  \right) e^{-2\CW_1 } \big{|}_{\mathbf{Z} \rightarrow \log \mathbf{z}^{\alpha}}\;,
\\
& \textrm{Fibering operator : } F_{\alpha} (m, \nu) = \exp \left(-\frac{\CW_0 -\sum_a Z_a \partial_{Z_a}\CW_0- m \partial_m \CW_0}{2\pi i }\right)\bigg{|}_{\mathbf{Z}\rightarrow \log \mathbf{z}^{\alpha}}\;. \label{twisted ptns 2}
\end{split}
\end{align} 
 For $p \in 2\mathbb{Z}$, there are two distinct supersymmetric backgrounds, $\nu_R =0$ or $\nu_R = \frac{1}2$ in \cite{Closset:2018ghr},  depending on the spin-strucuture along the fiber $[S^1]$ direction.  The above twisted partition function corresponds to the spin-structure with  anti-periodic boundary condition ($\nu_R = \frac{1}2$). Especially when $p=0$, the partition function can be identified with the twisted index $\CI_{g}$ on a Riemann surface $\Sigma_g$ \cite{Gukov:2015sna,Benini:2015noa,Benini:2016hjo,Closset:2016arn},
\begin{align}
\CI_{g} (\eta, \nu) :=\textrm{Tr}_{\CH(\Sigma_g;\nu)} (-1)^{R_\nu}  \eta^A = \CZ_{(g,p=0)} (m=\log \eta, \nu)\;.
\end{align} 
We use the $U(1)_{R_\nu}$ R-symmetry for the topological twisting  on $\Sigma_g$. Thus, the twisted index is well-defined only when the mixing parameter $\nu$ satisfies the following Dirac quantization condition
\begin{align}
R_\nu (2-2g) \in \mathbb{Z}\;.  \label{Dirac quantization}
\end{align}

The twisted partition functions $\CZ_{(g,p)}$ for $\CS_k$ theory were computed in  \cite{Gang:2021hrd,Gang:2022kpe} using the  field theory description given in \eqref{S-fold from T[SU(2)]}.  One can check that they agree with the twisted partition functions given in  \eqref{twisted ptns 1} and \eqref{twisted ptns 2}.  The  non-trivial matches of various SUSY partition functions support the proposed IR duality between \eqref{S-fold from T[SU(2)]}  and \eqref{S-fold from abelian}.

	\subsection{Generalized Haagerup TQFTs }
	In \cite{Gang:2021hrd}, they study the non-unitary semi-simple topological field theories (TQFTs), ${\rm TFT}_- [\mathcal{T}]$  and ${\rm TFT}_+ [\mathcal{T}]$ , associated to  a 3D $\mathcal{N}=4$ rank-0 SCFT $\CT$.  The two TQFTs are believed to be identical with the topologically twisted theories using $SU(2)_H$ and $SU(2)_C$ R-symmetry respectively. In 3D, $\CN=4$ (8 supercharges) is the minimal number of SUSY for a topological twisting.  Being of rank-0 is important for the topologically twisted theory to be a genuine semi-simple TQFT which supports a  rational  chiral algebra on the boundary. For $\CN=4$ SCFTs of non-zero rank, the topologically twisted theories are non-semisimple and the corresponding 2D chiral algebras are generically  logarithmic instead of rational \cite{Gukov:2020lqm,Creutzig:2021ext}.  Unitarity is broken in the topological twisting procedure. From all these considerations, topologically twisted $\CN=4$ rank-0 SCFTs seem to provide a natural physical realization of non-unitary semi-simple TQFTs.	
	
	There have been several previous works realizing non-unitary TQFTs in physical (2+1)D systems. In \cite{Cho:2020ljj}, they obtained non-unitary TQFTs from M5 branes wrapped on some non-hyperbolic 3-manifolds. Later it is found  that all such  systems  flow to 3D rank-zero SCFTs in the IR \cite{Choi:2022dju}.  In \cite{Gang:to-appear}, they consider 3D  $\CN=2$ gauge theories whose half-indices give the characters of  non-unitary minimal models. The gauge theories also turned out to experience non-trivial SUSY enhancement and flow to 3D rank-0 SCFTs in the IR. So these realizations are actually equivalent to ours.  In \cite{Dedushenko:2018bpp}, they realize non-unitary TQFTs from a non-trivial $S^1$-reduction of 4D Argyres-Douglas theories. There is also an attempt to understand the bulk dual of a non-unitary minimal model $M(3,5)$  using a  candidate  bulk ground state wave function called Gaffnian state \cite{Jolicoeur:2014isa}. We need further study to check whether these realizations are eventually equivalent to ours  or not.    
	
	In the (rank-0 SCFT)/(non-unitary TQFTs) correspondence, it is claimed that\footnote{The two partition functions have subtle overall phase factors depending on  3-manifold framing, background Chern-Simons level of $U(1)_R$ symmetry and etc. We will not keep track of all these subtle choices and the equality should be understood as an equality up to an overall phase factor.  }
	\begin{align}
	\left( \mathcal{\CZ}_{(g,p) } (m=0, \nu=\pm 1) \textrm{ of $\CT$} \right) = \mathcal{Z}[\textrm{TFT}_{\pm}[\CT] \textrm{ on }\mathcal{M}_{g,p}] \;.
	\end{align}
	According to the claim, taking the degenerate limit, $(m,\nu) \rightarrow (0,+1)$ or $(m,\nu) \rightarrow (0,-1)$, on the BPS partition functions is somehow equivalent to  the procedure of topological twisting, using $SU(2)_C$ or $SU(2)_H$ R-symmetry.  The equivalence is manifest for the SUSY backgrounds with $p=0$ but is not quite obvious for general $\CZ_{(g,p)}$. Notice that the Dirac quantization condition \eqref{Dirac quantization} is automatically   met in the degenerate limits  since 
	\begin{align}
	R_{\nu=1} = 2J^C_3  \in \mathbb{Z} \textrm{ and } R_{\nu=-1} =2 J^H_3 \in \mathbb{Z}\;.
    \end{align}%
	Combining the localization result for $\CZ_{(g,p)}$ in \eqref{twisted ptns 2} with the following general formula for (2+1)D bosonic TQFT
	\begin{align}
	\begin{split}
	 &\mathcal{Z}[\textrm{TQFT} \textrm{ on }\mathcal{M}_{g,p}]  = \sum_{\alpha \;:\; \textrm{simple objects}} (S_{0\alpha})^{2-2g} (T_{\alpha \alpha})^{p}\;, \nonumber
	 \end{split}
	\end{align}
	we have the basic dictionaries in Table \ref{table : dictionaries}. 
	\begin{table}[h]
	\begin{center}
		\begin{tabular}{ |c|c|} 
			\hline
			 TFT$_{\pm} [\CT]$&  Rank-0 SCFT $\CT$
			  \\
			\hline
		    \quad $\alpha$ : simple object  \quad & Bethe-vacuum $\mathbf{z}^\alpha \in \CS_{\rm B.E} (m=0, \nu = \pm 1)$ \\ 
		     \hline
			 $S_{0 \alpha}^2 {\rm \; of \; TFT}_\pm $ & $H_\alpha^{-1} (m=0, \nu=\pm 1)$  \\ 
			 \hline
			 $\frac{T_{\alpha \alpha}}{T_{00}} {\rm \; of \; TFT}_\pm $ & $\frac{ F_\alpha (m=0, \nu=\pm 1)}{F_{\alpha=0} (m=0, \nu=\pm 1)}$  \\ 
			\hline 
		\end{tabular}
	\caption{\label{demo-table}Basic dictionaries of (rank-0 SCFT)/(non-unitary TQFTs) correspondence. See more dictionaries in \cite{Gang:2021hrd}. }  \label{table : dictionaries}
	\end{center}
\end{table}
In the table,    $(S_{\alpha \beta }, T_{\alpha \beta })$   are modular  matrices and the simple object $\alpha =0$ corresponds to the trivial one. Among  Bethe-vacua in $\CS_{\rm B.E}$, the  one $\mathbf{z}^{\alpha=0}$ corresponding to the trivial simple object  is chosen to satisfy the following consistency condition 
	\begin{align}
\big{|}\sum_\alpha  H_\alpha^{-1} F_{\a } \big{|}   = 	\big{|}\sum_\alpha  S_{0\alpha}^2 T_{\a \a} \big{|}  = |(STS)_{00}|= |(T^{-1} S T^{-1})_{00}|  = |S_{00}| = |H_{\alpha=0}|^{-1/2}\;. \label{modular data 2}
	\end{align}
	Here we use the $SL(2,\mathbb{Z})$ relations $S^2 = (ST)^3 = \CC$ where the charge conjugation $\CC$ is an identity matrix in our case. Using the relations in the above table and \eqref{modular data 2}, one can obtain partial information on the modular data of the non-unitary TQFTs, $\textrm{TFT}_{\pm}$. Imposing general consistency conditions on the top of the partial information, we obtain following consistent modular data ($k \geq 3$) \cite{Gang:2022kpe}\footnote{The modular data presented here is slightly different from the one in \cite{Gang:2022kpe}. We exchanged $\alpha =0 \leftrightarrow \alpha=1$ and permutated other $\alpha$'s from the previous one. Generally, an exchange of the vacuum $\alpha=0$ and $\alpha \neq 0$ could give an inconsistent modular data with some negative fusion coefficients. In our case, however, both are consistent modular data. The  two consistent choices of modular data are related to the ambiguity of choosing the $\mathbf{z}^{\alpha=0}$ from the relation in \eqref{modular data 2}   due to the fact that $|H_{\alpha=0}| = |H_{\alpha=1}|$. At the level of RCFT characters,  two choices are simply related to each other by exchanging $\chi_{\alpha=0} \leftrightarrow \chi_{\alpha=1}$  and permutating other $\chi_\alpha$'s accordingly.  } 
	\begin{align}
		\begin{split}
			&(S_{\alpha \beta}  \textrm { of  }{\rm TFT}_-[\CS_{k}])
			\\
			&= \setlength\arraycolsep{1pt} \left(\begin{array}{c|c|c|c}
				\begin{matrix}
					 (-1)^k a_0 & a_0 \\ a_0 &  a_0
				\end{matrix} & \begin{matrix}
					 {a_3} &   (-1)^k {a_3}  \\ {a_3} & {a_3} 
				\end{matrix} & \begin{matrix}
				  -a_1&  a_1 & -a_1 &\cdots  &  {(-1)^{k-3} a_1} \\	{a_1} & a_1 & a_1 &\cdots &a_1  \end{matrix} & \begin{matrix}
				 	{a_2} & -a_2 & a_2   \cdots &  {(-1)^{k+2} a_2}  \\ -{a_2}  & -a_2 & -a_2 \cdots & -{a_2}  
				\end{matrix}  \\
				\hline
				%-----------------
				%----------------------------
				\begin{matrix}   a_3  &    a_3 \\ (-1)^k a_3 &   a_3 \end{matrix}  & \begin{matrix}    { a_0} & {a_0} \\ {a_0} & (-1)^k {a_0}\end{matrix} & \begin{matrix}{a_1} & a_1 & a_1 & \cdots & {a_1}  \\  {-a_1} & a_1 & - a_1 & \cdots & {(-1)^{k-3}a_1}  \end{matrix} & \begin{matrix}   {a_2} & a_2 & a_2 &\cdots & {a_2}  \\ -{a_2} & a_2 & -a_2  & \cdots & {(-1)^{k+1}a_2}   \end{matrix} \\ 
				\hline
				\begin{matrix}
					-a_1 & a_1 \\ a_1 & a_1 \\  -a_1 & a_1 \\   \vdots & \vdots \\   (-1)^{k-3} a_1 &  a_1 
				\end{matrix} & 
				\begin{matrix}
					a_1 & -a_1  \\  a_1 & a_1  \\ a_1 & -a_1  \\ \vdots & \vdots \\  a_1 &  (-1)^{k-3} a_1 
				\end{matrix} &
				\mbox{\Large $2 a_1 \cos{\frac{i j \pi}{k-2} } $}\bigg{|}_{1\leq i, j\leq k-3} & \mbox{\Large 0}   \\
				\hline
				%-------------------------- 
				\begin{matrix}
					a_2 & -a_2 \\  -a_2 & -a_2 \\  a_2 & -a_2 \\  \vdots & \vdots  \\  (-1)^{k+2}a_2 &  -a_2
				\end{matrix} & \begin{matrix}
					a_2 & -a_2 \\  a_2 & a_2 \\  a_2 & -a_2 \\ \vdots & \vdots\\    a_2 & (-1)^{k+1} a_2
				\end{matrix}  & 
				\mbox{\Large 0} & \mbox{\Large $ 2 a_2 \cos{\frac{i j \pi}{k+2} } $} \bigg{|}_{1\leq i, j\leq k+1} 
			\end{array}\right) , 
			\\
			& (T_{\alpha \beta}  \textrm{ of  }{\rm TFT}_-[\CS_k]) = \delta_{\alpha \beta} \exp \left(2\pi i (h_\alpha - \frac{c}{24})\right) \textrm{ with }
			\\
			& \vec{h} = \bigg{ \{} \frac{k+2}{4} ,0, 0,\frac{k+2}{4},\frac{A^2}{4(k-2)}\bigg{|}_{A=1,\ldots, k-3}  , \frac{B^2}{4(k+2)} \bigg{|}_{B=1,\ldots, k+1}   \bigg{\}} -\frac{k+2}{4} \;  (\textrm{mod }1 )
			\\
			&\textrm{ and }  c =-(6k+11) \;  (\textrm{mod }24 )\;. \label{Modular S/T matrices}
		\end{split}
	\end{align} 
	They are square matrices of dimension $(2k+2)$. Here we introduce
	\begin{align}
	\begin{split}
	&(a_0, a_1, a_2, a_3)
	\\
	&= \left( \frac{1}{\sqrt{8(k-2)}}+\frac{1}{\sqrt{8(k+2)}}, \frac{1}{\sqrt{2 (k-2)}}, \frac{1}{\sqrt{2 (k+2)}} ,  \frac{1}{\sqrt{8(k-2)}}-\frac{1}{\sqrt{8(k+2)}} \right)\;.
	\end{split}
	\end{align}
	\paragraph{Bethe-vacuum to loop operator map} Generally, the Hilbert-space of a bosonic 3D TQFT on a two-torus $\mathbb{T}^2$ can be  given as follows
	 \begin{align}
	 \begin{split}
			&\mathcal{H}(\mathbb{T}^2)={\rm Span}\{ |\alpha \rangle \;:\; \alpha = 0,1, \ldots, n-1 \}\;,
			\\
			& \textrm{with } |\alpha \rangle = \hat{\CL}^{\alpha}_{(0,1)}|0\rangle\;.
	\end{split}
   \end{align} 
	For our case, i.e.  when the TQFT is ${\rm TFT}_- [\CS_k]$, the state $|\alpha \rangle$ can be identified with the Bethe-vacuum $\mathbf{z}^\alpha \in \CS_{\rm B.E}(m=0, \nu=-1)$ in  \eqref{twisted ptns 2}. The trivial vacuum $|\alpha =0 \rangle $ is identified with the Bethe-vacuum $\mathbf{z}^{\alpha=0}$ satisfying the relation in  \eqref{modular data 2}.  Here $\hat{\CL}^{\alpha}_{(p,q)}$ is the quantum loop operator of anyon type $\alpha$ supported on a $(p,q)$-cycle on the two-torus. From the analysis, we expect that there is a natural one-to-one map between the Bethe-vacua and the  types of  loop operators.  We claim that  the first four simple objects (or  Bethe-vacua) in ${\rm TFT}_- [\CS_k]$  correspond to following  loop operators
	\begin{align}
	\begin{split}
	&\alpha =0 \;:\; \textrm{Trivial loop operator $\mathbf{1}$}
	\\
	&\alpha =1\;:\; \textrm{Topological defect $\mathcal{L}^{\mathbb{Z}_2}$ associated to the $\mathbb{Z}_2$ 1-form symmetry}
	\\
		&\alpha =2\;:\; \textrm{SUSY Wilson loop $\mathcal{L}^{W}$ with gauge charge $\mathbf{Q}_W = (r-1, 1,2,\ldots , r-1)$} 
		\\
	&\alpha=3 \;:\; \mathcal{L}^{\mathbb{Z}_2}* \mathcal{L}^W   \label{SUSY loops}
	\end{split}
	\end{align}
   The map for $\alpha=0$ is obvious. The modular data  has $\mathbb{Z}_2$ 1-form symmetry generated by an anyon  $\alpha=1$. The symmetry is originated from the $\mathbb{Z}_2$ center subgroup of $SU(2)$ gauge theory in \eqref{S-fold from T[SU(2)]} or from the $\mathbb{Z}_2 = \{\pm 1\} $ subgroup of the first $U(1)$ gauge group in the dual abelian theory \eqref{S-fold from abelian}. The 1-form symmetry defines a codimension 2 topological defect, which is the $\CL^{\mathbb{Z}_2}$. 
   To understand  the  map for $\alpha=2$, consider the action of loop operators $\hat{\CL}^{\alpha}_{(1,0)}$  on the basis 
	\begin{align}
		\hat{\CL}^{\beta}_{(1,0)} |\alpha \rangle = \frac{S_{\beta \alpha}}{S_{0 \alpha }} |\alpha \rangle\;. \label{Loop on Bethe}
	\end{align}
   The supersymmetric Wilson loop $\hat{\CL}^W_{(1,0)}$ with gauge charge $\mathbf{Q}= (Q_1, \ldots, Q_r)$ acts on the Bethe-vacua as follows \cite{Closset:2016arn}
   \begin{align}
   \hat{\CL}^{W}_{(1,0)}  |\alpha \rangle =\left(  \prod_{a=1}^r z_a^{Q_a}|_{\mathbf{z}\rightarrow \mathbf{z}^\alpha}  \right) |\alpha \rangle.  \label{Loop on Bethe-2}
   \end{align}
   For the case when $\mathbf{Q}= \mathbf{Q}_W = (r-1, 1,2,\ldots, r-1)$, one can check that
   \begin{align}
   	\left(  \prod_{a=1}^r z_a^{Q_a}|_{\mathbf{z}\rightarrow \mathbf{z}^\alpha}  \right)   =\left( \frac{S_{\beta \alpha}}{S_{0 \alpha}} \textrm{ with }\beta =2\right)\;. \label{Loop on Bethe-3}
   \end{align}
From \eqref{Loop on Bethe}, \eqref{Loop on Bethe-2} and \eqref{Loop on Bethe-3}, we have the map for $\alpha=2$. Finally, the  map for $\alpha=3$ follows from the fusion rule $\CL^{\alpha=1} *\CL^{\alpha=2} = \CL^{\alpha=3}$ which can be read off from the S-matrix using Verlinde's formula.  We also tried to find the map for other Bethe-vacua but  were not successful in several attempts.

	 \paragraph{Gauging the $\mathbb{Z}_2$ 1-form symmetry} The $\mathbb{Z}_2$ 1-form symmetry  has non-trivial 't Hooft  anomaly  for odd $k$. It  follows from the fact that the topological spin $h_{\alpha} $ of the symmetry generating anyon $\alpha=1$  is $\pm \frac{1}4$ (mod 1) instead of $0$ or $\frac{1}2$ \cite{Hsin:2018vcg}. The 1-form symmetry is non-anomalous for $k\in 2\mathbb{Z}$. 
	 In the abelian UV description in \eqref{S-fold from abelian}, gauging the 1-form symmetry is equivalent to rescaling the first $U(1)$ gauge field $A$ to $\frac{A}2$. Then, the mixed CS level becomes $\frac{K_{ab}}{Q_b Q_b}$  which is not properly quantized for odd $k$. It means that  there is an obstruction of the gauging, i.e. a 't Hooft anomaly, for odd $k$. The `t Hooft anomay in the UV description \eqref{S-fold from T[SU(2)]} using $T[SU(2)]$ theory was found in \cite{Gang:2018wek}.
	 Using the $\mathbb{Z}_2$ 1-form symmetry, we define the $\mathbb{Z}_2$ gauged S-fold theory $\widetilde{\CS}_k$ as
	 \begin{align}
	 	\widetilde{\CS}_{k} :=  \begin{cases} 
	 		\frac{\CS_k}{\mathbb{Z}_2}\;, \quad k \in 2\mathbb{Z}
	 		\\
	 		\frac{\CS_k \otimes \CA^{\pm }_{\mathbb{Z}_2}}{\mathbb{Z}^{\rm diag}_2}\;, \quad k \in 4\mathbb{Z} \pm 1
	 		\end{cases} \label{Sk/Z2}
	 \end{align} 
Here $/\mathbb{Z}_2$ denotes  gauging the $\mathbb{Z}_2$ 1-form symmetry. Here $\CA^{\pm }_{\mathbb{Z}_2}$ is a topological field theory with  anomalous $\mathbb{Z}_2$ 1-form symmetry generated by an anyon with topological spin $\pm  \frac{1}4$.  Then, the diagonal $\mathbb{Z}_2$ 1-form symmetry is non-anomalous and thus can be gauged. The $\CA^{\pm }_{\mathbb{Z}_2}$ is expected to have two simple objects,  $\alpha=0$ and $\alpha=1$, related to each other by the $\mathbb{Z}_2$ 1-form symmetry   and has following S-matrix
\begin{align}
\left( S\textrm{-matrix of } \CA^\pm_{\mathbb{Z}_2}  \right)=  \frac{1}{\sqrt{2}} \begin{pmatrix} 1 & 1  \\ 1 & -1\end{pmatrix}\;.
\end{align}
Let $\textrm{TFT}_- [\widetilde{\CS}_{k}]$ be the non-unitary TQFT associated to the $\widetilde{\CS}_{k}$:
\begin{align}\label{defTQFT02}
	\textrm{TFT}_-[\widetilde{S}_k]:= \begin{cases}
		\frac{\textrm{TFT}_- [\CS_k]}{\mathbb{Z}_2}, \quad k \in 2\mathbb{Z}
			\\
		\frac{\textrm{TFT}_- [\CS_k]\otimes  \CA^{\pm }_{\mathbb{Z}_2}}{\mathbb{Z}^{\rm diag}_2}, \quad k \in 4\mathbb{Z} \pm 1
		\end{cases}
\end{align} 
 The gauged $\mathbb{Z}_2$ 1-form symmetry is fermionic for $k \in 4\mathbb{Z}$ and bosonic  otherwise.  Thus,  the topological theory $\textrm{TFT}_-[\widetilde{\CS}_k]$ after the gauging is fermionic (i.e. spin TQFT) when $k \in 4\mathbb{Z}$ and bosonic (i.e. non-spin TQFT) otherwise.  The modular data, $\widetilde{S}$ and $\widetilde{T}$, for the bosonic TQFT $\textrm{TFT}_-[\widetilde{\CS}_{k}]$ is given as follows
 \cite{Gang:2022kpe} 
 \begin{align}
 	\begin{split}
 		& \textrm{For $k \in 4\mathbb{Z}_{\geq 1}\pm 1$},
 		\\
 		&\widetilde{S}= \sqrt{2}\left(\begin{array}{c|c|c}
 			\begin{matrix}
 				a_0 & a_3 \\
 				a_3 & a_0
 			\end{matrix} &  \begin{matrix}
 				a_1 & \cdots & a_1 \\
 				a_1 & \cdots & a_1
 			\end{matrix}   &  \begin{matrix}
 				-a_2 & \cdots & -a_2 \\
 				a_2 & \cdots & a_2
 			\end{matrix}\\
 			\hline
 			\begin{matrix}
 				a_1 & a_1 \\ \vdots & \vdots \\ a_1 & a_1
 			\end{matrix} & \mbox{\large $ 2 a_1 \cos{\frac{4 i j \pi}{k-2} } $}\bigg{|}_{1\leq i,j \leq \frac{k-3}2}  &   \mbox{\large $ 0 $} \\
 			\hline 
 			\begin{matrix}
 				-a_2 & a_2 \\ \vdots & \vdots \\ -a_2 & a_2
 			\end{matrix} & \mbox{\large $ 0 $ } & 
 			\mbox{\large $ 2 a_2 \cos{\frac{4 i j \pi}{k+2} } $}\bigg{|}_{1\leq i,j \leq \frac{k+1}2}  
 		\end{array}\right),
 		\\
 		&\widetilde{T} = e^{-\frac{2\pi i c}{24} }\mathrm{diag}\left[\exp\left(2 \pi i \left\{0,0,\frac{A^2}{(k-2)}\bigg{|}_{A=1,\cdots,\frac{k-3}{2}}, \frac{B^2}{(k+2)}\bigg{|}_{B=1,\cdots,\frac{k+1}{2}}\right\}\right)\right]\;,
 		\\
 		& c= (1 \pm 1) \left( \textrm{mod 8}\right)\;.
 	\end{split} \label{Haagerup modular data}
 \end{align}
They are square matrices of dimension $(k+1)$.
 \begin{align}
 	\begin{split}
 		& \textrm{For $k \in 4\mathbb{Z}_{\geq 1}+2$},
 		\\
 		&\widetilde{S}= \\
 		&\setlength\arraycolsep{1pt}\left(\begin{array}{c|c|c|c|c}
 			\begin{matrix}
 				2 a_0 & 2 a_3 \\ 2 a_3 & 2 a_0
 			\end{matrix}   & \begin{matrix}
 				\phantom{11}2 a_1\phantom{11}& \cdots & \phantom{11}2 a_1\phantom{11} \\ 2 a_1 & \cdots & 2 a_1
 			\end{matrix} & \begin{matrix}
 				\phantom{1}a_1\phantom{1} & \phantom{1}a_1\phantom{1} \\ a_1 & a_1
 			\end{matrix} & \begin{matrix}
 				\phantom{1}- 2 a_2\phantom{1}& \cdots & \phantom{1}- 2 a_2\phantom{1} \\ 2 a_2 & \cdots & 2 a_2
 			\end{matrix} & \begin{matrix}
 				- a_2 & -a_2 \\ 
 				a_2 & a_2
 			\end{matrix} \\ 
 			\hline
 			%%%%%%%%%%%%%%%%%%%%%%%%%%%%%%%%%%%%%%%%5
 			\begin{matrix}
 				2 a_1 & 2 a_1 \\ \vdots & \vdots \\ 2 a_1 & 2 a_1
 			\end{matrix} & \mbox{\large $ 4 a_1 \cos \frac{i j \pi}{n}$}\bigg{|}_{1\leq i,j \leq n-1} & \begin{matrix}
 				2 a_1 \mathbb{J}_{n-1}  & 2 a_1 \mathbb{J}_{n-1} \end{matrix} & \mbox{\large $ 0 $} & \mbox{\large $ 0 $}\\
 			\hline
 			%%%%%%%%%%%%%%%%%%%%%%%%%%%%%%%%%%%%%%%%%%%%%%555
 			\begin{matrix}
 				a_1 & a_1 \\ a_1 & a_1
 			\end{matrix} & \begin{matrix}
 				2a_1 \mathbb{J}^T_{n-1} \\ 2a_1 \mathbb{J}^T_{n-1}
 			\end{matrix} & \begin{matrix}
 				\phantom{1}b_{1,+}\phantom{1} &\phantom{1}b_{1,-}\phantom{1} \\ b_{1,-} & b_{1,+}
 			\end{matrix} & \mbox{\large $ 0 $} & \begin{matrix}
 				\frac{i^n}{2 \sqrt{2}} & -\frac{i^n}{2 \sqrt{2}} \\ -\frac{i^n}{2 \sqrt{2}} & \frac{i^n}{2 \sqrt{2}}
 			\end{matrix} \\
 			\hline
 			%%%%%%%%%%%%%%%%%%%%%%%%%%%%%%%%%%%%%%%%%%%%%%%%%%555
 			\begin{matrix}
 				-2 a_2 & 2 a_2 \\ \vdots & \vdots \\ -2a_2 & 2 a_2
 			\end{matrix} & \mbox{\large $ 0 $} & \mbox{\large $ 0 $} & \mbox{\large $ 4 a_2 \cos \frac{ij\pi}{n+1}$}\bigg{|}_{1\leq i,j \leq n} & \begin{matrix}
 				2a_2 \mathbb{J}_{n} & 2a_2 \mathbb{J}_{n} \end{matrix}\\
 			\hline
 			%%%%%%%%%%%%%%%%%%%%%%%%%%%%%%%%%%%%%%%%%%%%%%%%%%%%%%%555555555
 			\begin{matrix}
 				- a_2 & a_2 \\ -a_2 & a_2
 			\end{matrix} & \mbox{\large $ 0 $} & \begin{matrix}
 				\frac{i^n}{2 \sqrt{2}} & -\frac{i^n}{2 \sqrt{2}} \\ -\frac{i^n}{2 \sqrt{2}} & \frac{i^n}{2 \sqrt{2}}
 			\end{matrix} & \begin{matrix}
 				2a_2 \mathbb{J}^T_{n} \\ 2a_2 \mathbb{J}^T_{n}
 			\end{matrix} & \begin{matrix}
 				\phantom{1}b_{2,+}\phantom{1} & \phantom{1}b_{2,-}\phantom{1}\\ b_{2,-} & b_{2,+}
 			\end{matrix}
 		\end{array}\right),
 		\\
 		&\widetilde{T}=e^{-\frac{2\pi i   }{24} } \mathrm{diag}\left[\exp\left(2 \pi i \left\{0,0,\frac{A^2}{(4n)}\bigg{|}_{A=1,\cdots,n-1},\frac{n}{4}^{\otimes 2}, \frac{B^2}{4(n+1)}\bigg{|}_{B=1,\cdots,n}, \frac{n+1}{4}^{\otimes 2}\right\}\right)\right].
 		\label{Haagerup modular data-2}
 	\end{split}
 \end{align}
They  are square matrices of dimension $(\frac{k+8}2)$. Here we define
 \begin{align}
 	\begin{split}
 		& n = \frac{k-2}{4}, \quad
 		\mathbb{J}^T_n = \left( -1, 1, -1, \cdots, (-1)^n \right),
 		\\ 
 		&b_{1, \pm} = (-1)^n a_1 \pm \frac{i^n}{2 \sqrt{2}}, \; \textrm{ and } \;b_{2,\pm} = (-1)^{n+1} a_2 \pm \frac{(-i)^n}{2 \sqrt{2}}.
 	\end{split}
 \end{align}
 For $k =  4m^2+4m+3$  ($m \in \mathbb{Z}_{\geq 0}$), the modular data, $(\widetilde{S}, \widetilde{T})$, is related to the generalized Haagerup-Izumi modular data   $\CD^{\omega=2}{\rm Hg}_{2m+1}$  \cite{Evans:2010yr}  by a Hecke (or Galois conjugate) transformation. The specific form of Galois conjugation \cite{Ng:2012ty, Harvey:2018rdc} is given by 
\begin{align}
	\begin{split}
		&\left( S \text{ of $\mathcal{D}^{\omega = 2} \text{Hg}_{2m+1}$ in \cite{Evans:2010yr}}\right) \cong \widetilde{T}^{\bar{p}} \widetilde{S}^{-1} \widetilde{T}^p \widetilde{S}\widetilde{T}^{\bar{p}} \widetilde{S}^2, \\
		&\left( T \text{ of $\mathcal{D}^{\omega=2} \text{Hg}_{2m+1}$ in \cite{Evans:2010yr}}\right) \cong \widetilde{T}^{\bar{p}}, \label{Hecke on modular data}
	\end{split}
\end{align}
up to permutations, for $p = 2$ and $\bar{p} = \frac{k^2-3}{2}$. When  $k=4m^2+4m+3$, the conductor $N$ of the modular matrices, i.e. the minimal positive integer such that $T^N=1$,  is $N=(k+2)(k-2)$ and $\bar{p}$ is a multiplcative inverse of $p$ in $\mathbb{Z}/N\mathbb{Z}$.  Especially when $m=1$, the $\CD^{\omega=0}{\rm Hg}_{2m+1}$ is the modular data of original Haagerup TQFT  with 12 simple objects. The Haggerup TQFT has drawn much attentions since it serve as an  example of 3D extoic unitary TQFTs, which can not be realized  as a Chern-Simons theory or its discrete gauging.  The modular data $\CD^{\omega}{\rm Hg}_{2m+1}$ provides a two-parameter generalization of the Haagerup TQFT's modular data, parametrized by $m \geq 0$ and $\omega \in \mathbb{Z}/({(2m+1)\mathbb{Z}} )$.  Our ${\rm TFT}_- [\CS_k]$ and ${\rm TFT}_- [\widetilde{\CS}_k]$ can be regarded as (non-unitary) generalized Haagerup TQFTs. The modular data of  ${\rm TFT}_- [\widetilde{\CS}_k]$ for $k=8$ is equivalent to one of rank 10 fermionic modular data  found in  \cite{Cho:2022kzf}.

\section{Generalized non-unitary Haagerup RCFTs and their characters} \label{sec 3}

Let us consider the Haagerup TQFT on a solid torus with a consistent holomorphic boundary condition $\mathcal{B}$. The solid torus is  equivalently
$D_2\times S^1$ where $D_2$ is a disk and $\partial D_2=\widetilde{S}^1$. 
A TQFT equipped with MTC has various loop operators associated with the anyons.  
We insert a loop operator ${\cal L}^\alpha$ running around the circle $S^1$ 
at the center of $D_2$. Upon the boundary condition, 
let us denote by $\{\chi_\alpha(\tau)\}$ the expectation value of the loop operators.
Here $\chi_0(\tau)$ is the partition function on $D_2\times S^1$ with no
loop operator insertion, and $\tau$ is the complex modulus
of the boundary two-torus $T^2= S^1 \times \widetilde{S}^1$. 
Given the modular data $S$ and $T$ \eqref{Modular S/T matrices} in the bulk, 
one can argue that $\{\chi_\alpha(\tau)\}$ should
obey the transformation rules below,
\begin{align}
\begin{split}
       \chi_\alpha( \tau+1) & = \sum_{\beta} T_{\alpha \beta} \chi_\beta(\tau) \ , 
       \\
       \chi_\alpha ( - 1/\tau) & = \sum_{\beta} S_{\alpha \beta} \chi_\beta(\tau)\ ,
\end{split}
\end{align}
In other words, $\{\chi_\alpha(\tau)\}$, which is referred to as 
the partition vectors, form a vector-valued modular form under the $SL(2,\mathbb{Z})$.

On the boundary of TQFT, there exists a chiral current that 
defines a certain chiral algebra.
Each partition vector $\chi_\alpha(\tau)$ 
can be often understood as the character of the chiral algebra. 
More precisely, the Hilbert space of the TQFT on $D_2$ 
becomes the vacuum representation of the chiral algebra.
When a loop operator $\cal L^\alpha$ is placed at the center of $D_2$,
the quantization then leads to the equivalence between the Hilbert space 
and a highest weight representation of the chiral algebra other than the vacuum. 
As a remark, this chirality is reflected in the 
the holomorphic dependence of $\{\chi_\alpha(\tau)\}$
on the complex modulus $\tau$. 

One expects that a chiral algebra living 
on the boundary includes the Virasoro algebra, 
and $\{\chi_\alpha(\tau)\}$ 
can be identified as the conformal characters of a putative 
rational CFT (RCFT). Indeed, the Haagerup TQFT 
is likely to have the Virasoro algebra on the boundary. 
This is because the holomorphic twist of the 3D supersymmetric 
theories gives a stress tensor on the boundary when 
the bulk theory is topological  \cite{Costello:2020ndc}. 
It is therefore interesting to see if 
the candidate characters $\{\chi_\alpha(\tau)\}$ obtained from the Haagerup TQFT 
can be regarded as the conformal characters of 2D RCFTs. We will refer them as to 
the Haagerup RCFTs, denoted by ${\cal R}_k$. 

Obviously, the consistent modular transformation of 
such candidate conformal characters $\{\chi_\alpha(\tau)\}$ does not guarantee 
that the putative RCFT exists. What else should $\{\chi_\alpha(\tau)\}$ 
satisfy to convince that the Haagerup RCFTs exist? 

In the limit $\tau \to i \infty$, each candidate conformal character 
$\chi_\alpha(\tau)$ can be expanded in powers of $q=e^{2\pi i \tau}$ 
as
\begin{align}\label{expansion}
    \chi_\alpha(q) = q^{h_\alpha - \frac{c}{24}} \sum_{n} a_\alpha(n) q^n \ , 
\end{align}
where the leading exponent can be identified as 
the central charge of the putative RCFT and the conformal weight $h_\alpha$
of the chiral primary. One can also argue that the conformal weight of the chiral primary  
equals to the statistical spin of the loop operator in the bulk modulo integer.
The salient feature of conformal characters of a consistent rational CFT 
is that each Fourier coefficient $a_\alpha(n)$ in \eqref{expansion}
is non-negative integer-valued. This is because 
$a_\alpha(n)$ counts the number of states of given conformal weight.

The integrality of Fourier coefficients $a_\alpha(n)$ 
requires that the set of candidate characters $\chi_\alpha(q)$ 
is a representation of $SL(2,\mathbb{Z}_N)$ for a certain $N$, 
which is known as the integrality theorem \cite{atkin1971modular,calegari2021unbounded} or the congruence property 
in the modular tensor category \cite{Ng:2012ty}. The integrality theorem plays 
a key role in classifying the space of bosonic RCFTs \cite{Kaidi:2021ent}, and it has been further 
generalized to the fermionic theories recently \cite{Bae:2021jkc,Duan:2022kxr}. 
As expected, one can show that the conformal weights 
together with modular matrices \eqref{Modular S/T matrices} are indeed consistent to a reducible representation 
of $SL(2,\mathbb{Z}_N)$. 
However, the non-negative coefficients are not guaranteed yet.

The present work aims to explicitly construct the conformal characters that
transform as a vector-valued modular form (vvmf) with the Haagerup modular data \eqref{Modular S/T matrices} and have
non-negative integer Fourier coefficients. 
Given such conformal characters, we strongly suspect that 
the Haagerup RCFTs ${\cal R}_k$ actually exist.  
Their existence also implies that  
the Haagreup TQFTs have consistent modular tensor 
categories beyond the mere modular data. In fact, one can 
view the full 2D Haagerup RCFTs ${\cal R}_k$ as the Haagerup TQFTs $\text{TFT}_-[{\cal S}_k]$
on an interval. 

\subsection{UV Abelian gauge theory and the half-index}

We can utilize the aforementioned UV abelian gauge theory  \eqref{S-fold from abelian} leading to the Haagerup 
TQFT to obtain the explicit $q$-expansion of a few 
candidate conformal characters.  

To do so, we first need to translate the consistent holomorphic boundary condition 
${\cal B}$ of the TQFT to a supersymmetric boundary condition  
in the UV abelian gauge theory ${\cal S}_k$ on $D_2\times S^1$. 
We propose that\footnote{Verifying whether the UV boundary condition flows to an IR SCFT boundary condition compatible with the topological twisting is a highly non-trivial task. Consistent RCFT characters are obtained, as we will see below, from the half-indices associated with the boundary condition $\mathcal{B}' = (D_c, \mathcal{D})$. This may imply that $\mathcal{B}'$ is compatible with the topological twisting.}  the boundary condition ${\cal B}$ in the IR can be described 
by a simple SUSY boundary condition in ${\cal S}_k$, which sets 
any chiral multiplet  $\{\phi, \psi_\pm, F\}$  to a deformed Dirichlet boundary condition ($D_c$), 
\begin{align}\label{boundary01}
    \phi|_{\partial} =c \ , \quad \psi_+|_{\partial} =0\ 
\end{align}
with a non-zero constant $c$, and any vector multiplet 
$\{ A_\mu, \lambda_\pm, \sigma, D\}$ to a Dirichlet boundary condition ($\CD$), 
\begin{align}\label{boundary02}
    A_{0}\pm A_1 |_{\partial}=0\;, \quad D|_{\partial} =0\;, \quad  \lambda_-|_{\partial} =0\; . 
\end{align}
Given the above SUSY boundary condition \eqref{boundary01} and \eqref{boundary02}, denoted by 
${\cal B}'=(D_c,\CD)$ collectively, 
the half-index can be defined as  \cite{Gadde:2013wq,Gadde:2013sca,Sugishita:2013jca,Dimofte:2017tpi}
\begin{align}
    \CI_{\CB'}(q, \eta, \nu)= \textrm{Tr}_{\CH} \Big[ (-1)^{R_\nu} q^{\frac{R_\nu}2 +j_3} \eta^A \Big]\;.
\end{align}
where ${\cal H}$ is the Hilbert space of a given SCFT on the disk, 
and the circle $S^1$ can be understood as the temporal circle. 
Specifically, the half-index of the SCFT ${\cal S}_k$ \eqref{S-fold from abelian} is 
given by \cite{Dimofte:2017tpi}
\begin{align}\label{halfindexvacuum}
    \CI_{\CB'} (q, \eta, \nu) = 
    \frac{1}{(q)_\infty^r}  \sum_{\mathfrak{m}}  
    q^{\frac12 \mathfrak{m}^T \cdot K \cdot \mathfrak{m}}
    \Big[ (-q^\frac12)^{\nu-1} \eta \Big]^{(r-1) m_1 + \sum\limits_{a=1}^r (a-1)m_a }
    \prod_{a=1}^r ( q^{-Q_a m_a+1 };q )_\infty,
\end{align}
where the magnetic fluxes $\mathfrak{m}=(m_1,m_2,\ldots, m_r)$ are quantized, i.e.,
$m_a\in \mathbb{Z} $ for all $a$. The $K$ matrix and the gauge charges $Q_a$
are given by \eqref{K-matrix} and \eqref{gauge charge}. We remark that 
the R-charges of chiral multiplets 
should be zero. Otherwise, the R-charge would not be 
compatible with \eqref{boundary01}. 
The Pochhammer symbols are defined by 
\begin{align}
	&(x; q)_k := \prod_{n=0}^{k-1} (1-q^n x), \quad (q)_k := (q;q)_k := \prod_{n=1}^k (1-q^n)\;.
\end{align}
One can further simplify the half-index \eqref{halfindexvacuum} into  
\begin{align}
    \CI_{\CB'} (q, \eta, \nu) = 
    \sum_{m_a \in \mathbb{Z}_{\geq0}}   
    \frac{q^{\frac12 \mathfrak{m}^T \cdot K \cdot \mathfrak{m}}}{(q)_{2m_1} (q)_{m_2}\ldots (q)_{m_r}}
    \Big[ (-q^\frac12)^{\nu-1} \eta \Big]^{-(r-1) m_1 - \sum\limits_{a=1}^r (a-1)m_a }\ .
\end{align}

Some of supersymmetric loop operators ${\cal L}^\alpha$ 
\eqref{SUSY loops} in the SCFT ${\cal S}_k$ flow 
to the loop operators in the Haagerup TQFT ${\rm TFT}_-[{\cal S}_k]$.  
The insertion of ${\cal L}^\alpha(S^1)$ placed at the center of $D_2$
corresponds to adding to ${\cal S}_k$ a certain static source at the center.  
In the presence of ${\cal L}^\alpha(S^1)$, the half-index 
thus becomes
\begin{align}\label{halfindexloop}
  \CI_{\CB'}^\alpha (q, \eta, \nu) = \text{Tr}_{\cal H^\alpha} \Big[ (-1)^{R_\nu} q^{\frac{R_\nu}2 +j_3} \eta^A \Big]\ ,  
\end{align}
where ${\cal H}^\alpha$ denotes the Hilbert space of ${\cal S}_k$ on $D_2$
with the source.  
Using the SUSY localization technique, one can compute 
the half-indices \eqref{halfindexloop} exactly:  
\begin{align}\label{halfindicesprimary}
\begin{split}
    \CI_{\CB'}^{\alpha=1} (q, \eta, \nu)
    &= \sum_{\mathfrak{m} \in {\cal Z}' } 
    \frac{q^{\frac12 \mathfrak{m}^T \cdot K \cdot \mathfrak{m}}}{(q)_{2m_1} (q)_{m_2}\ldots (q)_{m_r}}
    \Big[ (-q^\frac12)^{\nu-1} \eta \Big]^{-(r-1) m_1 - \sum\limits_{a=1}^r (a-1)m_a } \ ,
    \\ 
    \CI_{\CB'}^{\alpha=2} (q, \eta, \nu)
     &= \sum_{\mathfrak{m} \in {\cal Z} } 
    \frac{q^{\frac12 \mathfrak{m}^T \cdot K \cdot \mathfrak{m}}}{(q)_{2m_1} (q)_{m_2}\ldots (q)_{m_r}}
    \Big[ q (-q^\frac12)^{\nu-1} \eta \Big]^{-(r-1) m_1 - \sum\limits_{a=1}^r (a-1)m_a } \ ,  
    \\ 
    \CI_{\CB'}^{\alpha=3} (q, \eta, \nu)
     &= \sum_{\mathfrak{m} \in {\cal Z}' } 
    \frac{q^{\frac12 \mathfrak{m}^T \cdot K \cdot \mathfrak{m}}}{(q)_{2m_1} (q)_{m_2}\ldots (q)_{m_r}}
    \Big[ q (-q^\frac12)^{\nu-1} \eta \Big]^{-(r-1) m_1 - \sum\limits_{a=1}^r (a-1)m_a } \ . 
\end{split}
\end{align}
Here ${\cal Z}$ and ${\cal Z}'$ are defined as  
\begin{align}
    {\cal Z} =  \Big( \mathbb{Z}_{\geq 0} \Big)^{r}\ , \quad 
    {\cal Z}' = \Big( \mathbb{Z}_{\geq 0}+\frac12 \Big) \otimes  \Big( \mathbb{Z}_{\geq 0} \Big)^{r-1}\ .
\end{align}
The half-integer values of the $U(1)$ flux $m_1$ 
can be explained by the fact that the static sources  
corresponding to ${\cal L}^{\alpha=1,3}(S^1)$ carry 
the half-integral monopole charge. Note also that, 
since the chiral multiplet $\Phi_1$ has the $U(1)$ 
charge $Q_1=2$, the half-integral flux $m_1$ also obeys  
the Dirac quantization condition properly. 

We propose that the above half-indices \eqref{halfindexvacuum} and \eqref{halfindicesprimary} 
in the degenerate limit $\nu \to -1$ and $\eta \to +1$ agree with 
the first four characters 
$\chi_0(q)$, $\chi_1(q)$, $\chi_2(q)$, and $\chi_3(q)$ 
of the non-unitary Haagerup RCFT 
\begin{align}
\begin{split}
    \chi_0(q)  & = q^{\Delta_0 }  \sum_{\mathfrak{m} \in  \CZ}  \frac{q^{\frac12 \mathfrak{m}^T \cdot K  \cdot \mathfrak{m} +(r-1)m_1 +\sum\limits_{a=1}^{r} (a-1) m_{a} }}{(q)_{2m_1} (q)_{m_2}\ldots (q)_{m_r}}\;,
    \\
    \chi_1(q) & =  q^{\Delta_1 }  \sum_{\mathfrak{m}\in  \CZ'}  \frac{q^{\frac12 \mathfrak{m}^T \cdot K  \cdot \mathfrak{m} +(r-1)m_1 +\sum\limits_{a=1}^{r} (a-1) m_{a} - \frac{3(r-1)}{4}  } }{(q)_{2m_1} (q)_{m_2}\ldots (q)_{m_r}}\;,
    \\
    \chi_2(q) & =  q^{\Delta_2 }  \sum_{\mathfrak{m}\in  \CZ}  \frac{q^{\frac12 \mathfrak{m}^T \cdot K  \cdot \mathfrak{m}} }{(q)_{2m_1} (q)_{m_2}\ldots (q)_{m_r}}\;,
    \\
    \chi_3(q) & =  q^{\Delta_3 }  \sum_{\mathfrak{m}\in  \CZ'}  \frac{q^{\frac12 \mathfrak{m}^T \cdot K  \cdot \mathfrak{m}- \frac{(r-1)}{4} } }{(q)_{2m_1} (q)_{m_2}\ldots (q)_{m_r}}\;,
\end{split}
\label{eq:firstfourhalfindex}
\end{align}
up to the prefactors $q^{\Delta}$. The exponents $\Delta_0=-c/24$ and $\Delta_\alpha = h_\alpha-c/24$,
and they account for the ground state energies in the corresponding 
Hilbert spaces ${\cal H}$ and ${\cal H}^\alpha$.  
From the bulk computation of modular data in \eqref{Modular S/T matrices}, we can determine them only modulo $1$ and they are given by
\begin{align}
    \Big(\Delta_0, \Delta_1, \Delta_2 ,\Delta_3\Big) 
    = 
    \Big(\frac{6k+11}{24}, -\frac{1}{24}, -\frac{1}{24}, \frac{6k+11}{24}\Big) 
    \quad (\text{mod 1}) \;.
\end{align}

The connection between the half-indices of three-dimensional $\CN\geq 2$ gauge theories
and the conformal characters of two-dimensional CFTs has been explored in 
numerous recent studies. For instance, it was shown in  \cite{Dimofte:2017tpi} that 
the conformal characters of the $G_k$ WZW models with $k>0$ 
agrees with the half-indices of 3D $\CN=2$ pure $G_{k+h}$ Chern-Simons 
theories where $h$ denotes the dual Coxeter number of $G$. In fact, 
the 3D SUSY theories flow to the pure bosonic $G_k$ Chern-Simons theories
in the infrared limit.The relation between 
the characters of 2D logarithmic CFTs and the half-indices of 3D $\CN=2$ 
theories has been studied in \cite{Cheng:2018vpl,Cheng:2022rqr,Chung:2023qth}. 
More recently, Virasoro characters of the non-unitary minimal models are obtained
from the half-indices of 3D $\CN=4$ rank-0 SCFTs \cite{Gang:to-appear}. 
Our proposal is parallel to these ideas. 

Armed with the above four characters \eqref{eq:firstfourhalfindex},
we obtain in the next Section the explicit $q$-expansion of all conformal 
characters $\{ \chi_\alpha(q) \}$ ($\alpha=0,1,..,2k+1$) of the 
Haagerup RCFTs ${\cal R}_k$ with $3\leq k \leq 11$ 
using two well-known methods. 

Moreover, given such characters,
one can express the conformal characters $\widetilde{\chi}_\alpha(q)$
of the $\mathbb{Z}_2$ orbifold RCFT $\widetilde{\CR}_k$,  the boundary RCFT for the $\mathbb{Z}_2$ gauged TQFT ${\rm TFT}_-[\widetilde{\cal S}_k]$  \eqref{defTQFT02}, as follows: 
\paragraph{For $k\in 4\mathbb{Z}$}  ($\alpha=2,..,k/4$ and $\beta= (k+4)/4,..,k/2$)
\begin{align}
\begin{split}
    \widetilde{\chi}_0(q)  & = \chi_0(q) + \chi_1(q)  = q^{\Delta_0 }  \sum_{\mathfrak{m} \in  \CZ \oplus \CZ'}  \frac{q^{\frac12 \mathfrak{m}^T \cdot K  \cdot \mathfrak{m} +(r-1)m_1 +\sum\limits_{a=1}^{r} (a-1) m_{a} }}{(q)_{2m_1} (q)_{m_2}\ldots (q)_{m_r}}\;,  
    \\
    \widetilde{\chi}_1(q)  & = \chi_2(q) + \chi_3(q) =  q^{\Delta_2 }  \sum_{\mathfrak{m}\in  \CZ  \oplus \CZ'}  \frac{q^{\frac12 \mathfrak{m}^T \cdot K  \cdot \mathfrak{m}} }{(q)_{2m_1} (q)_{m_2}\ldots (q)_{m_r}}\  \ , 
    \\
    \widetilde{\chi}_\alpha(q) & = \chi_{2\alpha +1}(q) + \chi_{k+3-2\alpha}(q) \ ,
    \\ 
    \widetilde{\chi}_\beta(q) & = \chi_{2\beta+ \frac{k}{2} }(q)  + \chi_{\frac{5k}{2}+ 2-2\beta}(q)\  .
\end{split}
\end{align}

\paragraph{For $k\in 4\mathbb{Z}+2$} ($\alpha=2,..,(k-2)/4$ and $\beta= (k+10)/4,..,(k+2)/2$) 
\begin{align}\label{orbifoldk2}
	\begin{split}
	  \widetilde{\chi}_0(q)  & = \chi_0(q) + \chi_1(q)  = q^{\Delta_0 }  \sum_{\mathfrak{m} \in  \CZ \oplus \CZ'}  \frac{q^{\frac12 \mathfrak{m}^T \cdot K  \cdot \mathfrak{m} +(r-1)m_1 +\sum\limits_{a=1}^{r} (a-1) m_{a} }}{(q)_{2m_1} (q)_{m_2}\ldots (q)_{m_r}}\;,  
	\\
	\widetilde{\chi}_1(q)  & = \chi_2(q) + \chi_3(q)=  q^{\Delta_2 }  \sum_{\mathfrak{m}\in  \CZ  \oplus \CZ'}  \frac{q^{\frac12 \mathfrak{m}^T \cdot K  \cdot \mathfrak{m}} }{(q)_{2m_1} (q)_{m_2}\ldots (q)_{m_r}}\  \ , 
		\\
		\widetilde{\chi}_\alpha(q) & = \chi_{2\alpha +1}(q) + \chi_{k+3-2\alpha}(q) \ ,
		\\ 
		\widetilde{\chi}_{\frac{k}{4}+\frac12}(q) & = \widetilde{\chi}_{\frac{k}{4}+\frac32}(q) = \chi_{\frac{k}{2}+2}(q) \ ,
		\\
		\widetilde{\chi}_\beta(q) & = \chi_{\frac{k}{2} + 2\beta -3 }(q)  + \chi_{\frac{5k}{2}+ 5-2\beta}(q)\  ,
		\\
		\widetilde{\chi}_{\frac k2 +2}(q) & = \widetilde{\chi}_{\frac k2 +3}(q) = \chi_{\frac{3k}{2}+1}(q)\ .
	\end{split}
\end{align}
Note that  the characters  $\tilde{\chi}_0$ and $\tilde{\chi}_1$ above are expressed as a fermionic sum, also known as Nahm sums  \cite{Kedem:1993ze,Nahm:1994vas,Berkovich:1994es,Nahm:2004ch,welsh2005fermionic,Zagier:2007knq} after rescaling $m_1$ to $\frac{1}2 m_1$. 
It provide a new infinite series of modular forms in the Nahm sum representation. 

\paragraph{For $k\in 4\mathbb{Z} \pm 1$}  ($\alpha=2,..,(k-1)/2$ and $\beta= (k+1)/2,..,k$)
\begin{align}
\begin{split}
    \widetilde{\chi}_0(q)  & = \chi_0(q) \chi_1^\pm(q) + \chi_1(q) \chi_0^\pm(q)\ ,   
    \\
    \widetilde{\chi}_1(q)  & = \chi_2(q) \chi_0^\pm(q)  + \chi_3(q) \chi_1^\pm(q) \ , 
    \\
    \widetilde{\chi}_\alpha(q) & = \chi_{k+4-2\alpha}(q) \chi_0^\pm(q)  + \chi_{2\alpha}(q) \chi_1^\pm(q)  \ ,
    \\ 
    \widetilde{\chi}_\beta(q) & =  \chi_{3k+2-2\beta}(q) \chi_0^\pm(q) +  \chi_{2\beta}(q) \chi_1^\pm(q)\  ,
\end{split}
\end{align}
 Here $\chi_{0,1}^\pm(q)$ 
are the conformal characters of a RCFT corresponding to the bulk theory $\CA^{\pm }_{\mathbb{Z}_2}$ in \eqref{Sk/Z2}.

\subsection{Modular linear differential equation}

A simple and practical approach to construct 
a vector-valued modular form is the method of holomorphic modular boostrap, 
which is also known as the modular linear differential equation 
(MLDE). A general form of $d$-th order MLDE can be expressed as 
\begin{align}\label{MLDE}
    \Big[ {\cal D}^d + \sum_{s=0}^{d-1} \phi_s(\tau) {\cal D}^s \Big] f(\tau) = 0\ , 
\end{align}
where $\phi_s(\tau)$ are weakly holomorphic modular forms of weight $2(d-s)$. Here the $s$-th 
order derivative ${\cal D}^s$ is defined as 
\begin{align}
    {\cal D}^s = {\cal D}_{2s-2} {\cal D}_{2s-4} \cdots {\cal D}_0    
\end{align}
where ${\cal D}_t$ is the Serre derivative 
\begin{align}
    {\cal D}_t = \frac{1}{2\pi i}\frac{d}{d\tau} - \frac{t}{12} E_2(\tau)\ , 
\end{align}
that maps a modular form of weight $t$ to a modular form of weight $t+2$. 
$E_2(\tau)$ denotes the Eisenstein series of weight two. 
Since \eqref{MLDE} is invariant under the modular 
transformation, one can argue that its $d$ independent 
solutions have to transform as a $d$-dimensional vector-valued 
modular function under $SL(2,\mathbb{Z})$. 

There are two useful parameters that provides 
a organizing principle to explore the space of solutions 
to MDLEs. One of them is the number of independent 
solutions $d$, and the other is the so-called Wronskian index $l$ 
that constrains the number of allowed 
singularities of the the coefficients $\phi_k(\tau)$ 
in the fundamental domain. For a given $l$, one can 
determine $\phi_k(\tau)$ with finitely 
many unknown parameters, which makes the MLDE method  
highly practical. Note that one can also 
desrcribe the Wronskian index $l$ as 
\begin{align}
    l = \frac{d(d-1)}{2} -  6\sum_{i=0}^{d-1} \Delta_i , 
\end{align}
where $\Delta_i$ are the leading exponents of the solutions in the limit $q\to 0$. 
For more details, please see \cite{Mukhi:2019xjy} and the reference therein.

The question is then how to pin down the MLDEs whose solutions 
can be identified as the conformal characters of the Haagerup 
RCFTs. To this end, we demand that the above four conformal 
characters \eqref{eq:firstfourhalfindex} are 
solutions to the MLDE. Since the Fourier coefficients 
of such characters are all specified, the finitely many 
parameters fixing the coefficients $\phi_p(\tau)$
become rigid unless the value of the Wronskian index $l$
changes.  

However the MLDE approach to our problem has a small drawback.
There is no physical input to fix the value of the Wronskian index $l$ a priori. 
The bulk description, either the Haagerup TQFTs 
or their SCFTs, only provides the values of $\Delta_i$
modulo integer. We thus have to examine the space of possible MLDEs 
by scanning over different values of $l$ until the solutions give rise to
a vvmf that only admits non-negative integer Fourier coefficients. 
One can then regard this vvmf as the conformal characters associated with 
the bulk TQFTs. 

\paragraph{Example} 

To demonstrate this procedure, let us construct the conformal 
characters of the orbifold Haagerup RCFT  $\widetilde{\CR}_k$ with $k=6$. 

Since ${\rm TFT}_-[\widetilde{\cal S}_{k=6}]$ is a non-spin TQFT, the putative RCFT 
of our interest has to be bosonic. It has seven conformal characters 
$\widetilde{\chi}_\alpha(\tau)$ \eqref{orbifoldk2}, but only five  
are independent. This is because the charge conjugation matrix $\widetilde{{\cal C}}=\widetilde{S}^2$ 
where $\widetilde{S}$ is the modular S-matrix  \eqref{Haagerup modular data-2}
\begin{align}\label{modularS:ex}
    \widetilde{S}
    = \left(
		\begin{array}{ccccccc}
			\frac{\sqrt2 + 1}{4}  & \frac{\sqrt2 -1}{4}  & \frac{\sqrt2}{4} & \frac{\sqrt2}{4} & -\frac{1}{2} & -\frac{1}{4} & -\frac{1}{4} \\
			\frac{\sqrt2 - 1}{4}  & \frac{\sqrt2 +1 }{4}  & \frac{1}{2 \sqrt{2}} & \frac{1}{2 \sqrt{2}} & \frac{1}{2} & \frac{1}{4} & \frac{1}{4} \\
			\frac{1}{2 \sqrt{2}} & \frac{1}{2 \sqrt{2}} & -\frac{1-i}{2\sqrt{2}} & -\frac{1+i}{2\sqrt{2}} & 0 & \frac{i}{2 \sqrt{2}} & -\frac{i}{2 \sqrt{2}} \\
			\frac{1}{2 \sqrt{2}} & \frac{1}{2 \sqrt{2}} & -\frac{1+i}{2\sqrt{2}} & -\frac{1-i}{2\sqrt{2}} & 0 & -\frac{i}{2 \sqrt{2}} & \frac{i}{2 \sqrt{2}} \\
			-\frac{1}{2} & \frac{1}{2} & 0 & 0 & 0 & -\frac{1}{2} & -\frac{1}{2} \\
			-\frac{1}{4} & \frac{1}{4} & \frac{i}{2 \sqrt{2}} & -\frac{i}{2 \sqrt{2}} & -\frac{1}{2} & \frac{1-i \sqrt{2}}{4}  & \frac{1+i \sqrt{2}}{4}  \\
			-\frac{1}{4} & \frac{1}{4} & -\frac{i}{2 \sqrt{2}} & \frac{i}{2 \sqrt{2}} & -\frac{1}{2} & \frac{1+i \sqrt{2}}{4}  & \frac{1-i \sqrt{2}}{4}  \\
		\end{array}
		\right)\ ,  
\end{align}
exchanges $\widetilde{\chi}_2 \leftrightarrow \widetilde{\chi}_3$ and 
$\widetilde{\chi}_5 \leftrightarrow \widetilde{\chi}_6$, and thus 
those characters become degenerate, $\widetilde{\chi}_2(\tau)=\widetilde{\chi}_3(\tau)$
and $\widetilde{\chi}_5(\tau)=\widetilde{\chi}_6(\tau)$.

From \eqref{orbifoldk2}, one can also read off the explicit $q$-expansion 
of the first two characters 
\begin{align}\label{sol1}
\begin{split}
    \widetilde{\chi}_0 & = q^{\widetilde{\Delta}_0} \sum_{\mathfrak{m}\in {\cal Z}\oplus {\cal Z}'}\frac{q^{\frac{1}2 \mathfrak{m}^T.  K . \mathfrak{m}+ 4m_1+m_2+2m_3 + 3 m_4+ 4m_5 }}{(q)_{2m_1} (q)_{m_2} (q)_{m_3} (q)_{m_4} (q)_{m_5}}\ , 
    \\
    & = q^{\widetilde{\Delta}_0} \Big(1+ q^{2}+2q^3+3q^{4}+4q^5+7 q^{6}+8 q^7+14 q^{8}+18 q^9+\ldots \Big)\ , 
\end{split}
\end{align}
and 
\begin{align}\label{sol2}
\begin{split}
    \widetilde{\chi}_1 & = q^{\widetilde{\Delta}_1} \sum_{\mathfrak{m}\in {\cal Z}\oplus {\cal Z}'}\frac{q^{\frac{1}2 \mathfrak{m}^T. K  . \mathfrak{m}}}{(q)_{2m_1} (q)_{m_2} (q)_{m_3} (q)_{m_4} (q)_{m_5}}\ , 
    \\
    & = q^{\tilde{\Delta}_1} \Big(1+2 q+4 q^2 + 6 q^3+ 11 q^4+16 q^5+25 q^6 + 36 q^7+ 54 q^8+76 q^9 +\ldots \Big)\ . 
\end{split}
\end{align}
where the $K$ matrix \eqref{K-matrix} is 
\begin{align}
    K = \left(
	\begin{array}{ccccc}
		8 & 0 & 2 & 4 & 6 \\
		0 & 2 & 2 & 2 & 2 \\
		2 & 2 & 4 & 4 & 4 \\
		4 & 2 & 4 & 6 & 6 \\
		6 & 2 & 4 & 6 & 8 \\
	\end{array}
	\right)\ . 
\end{align}
Here the exponents are  $\widetilde{\Delta}_0 = \Delta_0$
and $\widetilde{\Delta}_1= \Delta_2$.
 Note that the descendants of each character only carry integral conformal weights relative to 
that of the primary. This is consistent with the fact that the orbifold with $k=6$ is bosonic. 

To construct the rests of the conformal characters, 
we have to determine an MLDE. Since there are five independent
conformal characters, the order of the MLDE is fixed by five, $d=5$. 
Let us start  the exploration with the lowest value of the Wronskian 
index $l=0$, and see if the solutions have non-negative 
integer Fourier coefficients. The most general MLDE 
with $d=5$ and $l=0$ can be described as
\begin{align}\label{MLDEex}
	\bigg[ {\cal D}^5 + \mu_1 E_4(\tau) {\cal D}^3 
	+ \mu_2 E_6(\tau) {\cal D}^2 + \mu_3 E_4^2(\tau) {\cal D} 
	+ \mu_4 E_4(\tau) E_6(\tau) \bigg] f(\tau) = 0\ , 
\end{align}
where $E_4(\tau)$ and $E_6(\tau)$ are the Eisenstein series of weight four and six. 
We then require that either \eqref{sol1} or \eqref{sol2} is a solution to the MLDE \eqref{MLDEex},
which fixes the parameters uniquely as 
\begin{align}
	\mu_1= - \frac{24}{144}, \quad \mu_2= \frac{65}{2304}, 
	\quad \mu_3 = -\frac{965}{331776}, \quad \mu_4= \frac{1265}{3981312}\ .  
\end{align}
Having fixed the MLDE completely, we solve \eqref{MLDEex} using the Frobenius method 
to obtain the candidate conformal characters in powers of $q$, 
\begin{align}\label{chex}
    \widetilde{\chi}_0(q) & = q^{\frac{23}{24}} \Big(1+ q^{2}+2q^3+3q^{4}+4q^5+7 q^{6}+8 q^7+14 q^{8}+18 q^9+\ldots \Big)\ ,
    \nonumber \\
    \widetilde{\chi}_1(q) & = q^{-\frac{1}{24}} \Big(1+2 q+4 q^2 + 6 q^3+ 11 q^4+16 q^5+25 q^6 + 36 q^7+ 54 q^8+76 q^9 +\ldots \Big)\ , \nonumber
    \\
    \widetilde{\chi}_2 (q)  &= \widetilde{\chi}_3 (q) = 
    q^{\frac{5}{24}} \Big( 1+q +3 q^2+ 4 q^3+7 q^4+10 q^5+17 q^6+23 q^7+ 35 q^8+\ldots \Big)\ ,
	\nonumber \\
	\widetilde{\chi}_4(q) & = q^{\frac{1}{12}} \Big( 1+2 q+3 q^2+ 6 q^3+ 9 q^4+ 14 q^5+ 22 q^6+32 q^7+ 46 q^8+\ldots \Big)\ ,
	\nonumber \\
	\widetilde{\chi}_5(q) & = \widetilde{\chi}_6 (q)  = q^{\frac{11}{24}} 
    \Big( 1+ q+2 q^2+ 3 q^3+ 6 q^4+ 8 q^5+ 13 q^6+18 q^7+ 27 q^8+\ldots \Big)\ .   \nonumber 
\end{align}
We can see that all exponents are rational-valued and agree with \eqref{Haagerup modular data-2}, 
and that Fourier coefficients are non-negative integer-valued. One can also 
show numerically that, under the modular transformation $\tau \to - 1/\tau$,
\begin{align}
    \widetilde{\chi}_\alpha (-1/\tau) = \sum_{\beta=0}^6 \widetilde{S}_{\alpha \beta}
    \widetilde{\chi}_\beta(\tau) 
\end{align}
with the modular S-matrix $\widetilde{S}$ \eqref{modularS:ex}. In the numerical check, we choose the value of $\tau$ near $i$ so that both $q:=e^{2\pi i \tau}$ and $\tilde{q}:=e^{2\pi i (-1/\tau)}$ are small. 
Thus, we can conclude that \eqref{chex} are the very conformal characters of 
the orbifold. 

Several remarks are in order. Relying on our experiences with the lower values of $k $, 
we propose that the candidate characters of the orbifold $\widetilde{\CR}_k$ for $k\in 4\mathbb{Z}_{\geq 1}+2$ have the leading exponents 
as follows,
\begin{align}
    \widetilde{\Delta}_\alpha = \Big( 
    \frac{k-2}{4} , ~ 0 ,  \hspace*{-0.7cm} \underbrace{\frac{n^2}{k-2}}_\text{$n=1,2,..,(k-6)/4$} \hspace*{-0.8cm} , 
    ~\frac{k-2}{16}, ~\frac{k-2}{16}, \hspace*{-0.7cm} \underbrace{\frac{m^2}{k+2}}_\text{$m=1,2,..,(k-2)/4$} \hspace*{-0.8cm}, ~\frac{k+2}{16}, ~\frac{k+2}{16} 
    \Big) - \frac{1}{24}\ . 
\end{align}
It implies that, as $k$ increases, the values of $d$ and $l$ grow rapidly, making the MLDE approach 
less practical. In addition, the MLDE technique applied to the Haagerup RCFTs 
shares the same challenges with their $\mathbb{Z}_2$ orbifolds. Notice that the $c_{\rm eff} := -24\times \textrm{min}_\alpha \{\tilde{\Delta}_\alpha\} $ is always $1$, which is  expected to be also true for the $\CR_k$s.

To address this challeges, we shall employ an alternative method 
to construct a vvmf based on a given modular data 
and four characters \eqref{eq:firstfourhalfindex} in what follows.

\subsection{Riemann-Hilbert method}

It was shown in \cite{Bantay:2005vk,Bantay:2007zz, Gannon:2013jua, Cheng:2020srs} 
that there exists a characteristic matrix $\Xi(\tau)$ that describes the 
space of vvmfs for a given modular data such as $S$ and $T$ matrices. 
More precisely, each column of $\Xi(\tau)$ becomes a generator that 
transforms as a given $d$-dimensional representation of $SL(2,\mathbb{Z})$. 
Any such vvmf $\{\xi_\alpha(\tau)\}$ can thus be expressed as 
\begin{align}\label{hjk}
    \xi_\alpha(\tau) = \Xi_{\alpha \beta }(\tau) P_\beta\big(j(\tau) \big) \ , 
\end{align}
where $P_\beta\big(j(\tau)\big)$ are polynomials of the Klein 
$j$-function,
\begin{align}
    j(\tau)  =  \frac{1728 E_4^3(\tau)}{E_4^3(\tau) - E^2_6(\tau)}\ .
\end{align}
This approach is known as the theory of Bantay and Gannon \cite{Bantay:2005vk}, 
which is also referred to as the Riemann-Hilbert method in the literature.

The analytic construction of the matrix $\Xi(\tau)$ needs two constant 
matrices, denoted by $\Lambda$ and $\chi$. We first briefly  
discuss the constraints on the pair of two matrices. 
$\Lambda$ is a diagonal matrix with 
\begin{align}\label{LambdaT}
    \exp{\Big[ 2\pi i \Lambda \Big]} = T \ ,
\end{align}
and satisfies the relation below 
\begin{align}
	\Tr \Lambda = - \frac{7 d}{12} + \frac{1}{4} \Tr S + \frac{2}{3\sqrt{3}} \Re \left[ e^{-\frac{\pi i}{6}} \Tr\left(S T^{-1}\right)\right]\ .
	\label{eq:trLambda}
\end{align}
Here $S$ and $T$ are given $d\times d$ modular S- and T-matrices. 
We define two $d \times d$ matrices $B_2$ and $B_3$,
\begin{align}
\begin{split}    
    B_2 & =  -\frac{31}{72} \Lambda - \frac{1}{1728}\Big(\chi + \left[\Lambda,\chi \right]\Big)
    \\
    B_3 & = -\frac{41}{72} \Lambda + \frac{1}{1728} \Big(\chi + \left[\Lambda,\chi \right]\Big)\ ,
\end{split}
\end{align}
to describe the conditions that the other 
constant matrix $\chi$ should obey, 
\begin{align}\label{RHcondition}
    B_2 \Big( B_2 - \frac12 \Big) = 0 \ , \quad 
    B_3 \Big( B_3 - \frac13  \Big) \Big( B_3 - \frac23 \Big) = 0\ . 
\end{align}
Given two constant matrices $\Lambda$ and $\chi$, let a $d\times d$ matrix  
$\Xi(\tau)$ be expanded in powers of $q$ as 
\begin{align}\label{expansionxi}
    \Xi(\tau) = q^\Lambda \sum_{n=0}^\infty  \Xi[n] q^n   
\end{align}
with $\Xi[0] = {\bf 1}_d$ and $\Xi[1]= \chi$.

One can show that 
any vvmf transforming as a given representation of $SL(2,\mathbb{Z})$ 
can be generated by the columns of $\Xi(\tau)$, when the other matrices $\Xi[n]$ in 
\eqref{expansionxi} satisfy the recursion relation below, 
\begin{align}
	n \Xi[n] + \big[ \Lambda , \Xi[n] \big] = \sum_{i=0}^{n-1} \Xi[i] \bigg\{
	f_{n-i} \Lambda + g_{n-i} \big( \chi + \big[\Lambda, \chi \big] \big)   \bigg\}
	, 
 \label{eq:recursiverelation}
\end{align}
where $f_i$ and $g_i$ are defined as  
\begin{align}
	\begin{split}
		\Big( j(\tau) - 984\Big) \frac{\Delta(\tau)}{E_{10} (\tau)} &= \sum_{n=0}^\infty f_n q^n,\\
		\frac{\Delta(\tau)}{E_{10} (\tau)} &= \sum_{n=0}^\infty g_n q^n\ .
	\end{split}   
\end{align}
Here $\Delta(\tau)$  is the modular discriminant, 
\begin{align}
    \Delta(\tau)  = \frac{E_4^3(\tau) - E_6^2(\tau)  }{1728}\ .
\end{align}

For the sake of self-containedness, let us  briefly sketch the key ideas of the Bantay-Gannon  method.
When a given modular representation admits a diagonal $T$ matrix, it was proven 
that there always exists the bijective exponent $\Lambda$ that must obey 
the trace formula \eqref{eq:trLambda}. By bijective exponent, we mean 
that any vvmf $\chi_\alpha(\tau)$ for the given modular representation can be expressed 
in powers of $q$ as
\begin{align}
    \chi_{\alpha}(\tau) = q^{\Lambda_{ii}} \sum_{\substack{n\in\mathbb{Z}\\n\gg-\infty}} b_\alpha(n) q^n
\end{align}
where the principal part $b_\alpha(n)$ with $n\leq 0$ uniquely determines 
the rest of the coefficients $b_\alpha(n)$ with $n>0$. Suppose that each column of $\Xi(\tau)$ 
defined in \eqref{expansionxi} transforms as a vvmf for the given modular matrices. 
One can then argue that 
\begin{align}\label{qwer}
    \frac{E_4(\tau) E_6(\tau)}{\Delta(\tau)} {\cal D} \Xi(\tau) 
    = \Xi(\tau) \mathfrak{P}\big(j(\tau)\big)\ ,
\end{align}
where the polynomial matrix $\mathfrak{P}$ is 
\begin{align}
    \mathfrak{P}\big(j(\tau)\big) = \big( j(\tau) - 984 \big) \Lambda + \chi + \big[ \Lambda, \chi]\ . 
\end{align}
Notice that the both sides of \eqref{qwer} are vvmfs that share the same principal part. 
The bijective exponent $\Lambda$ guarantees that they should be equal. The recursion relation \eqref{eq:recursiverelation} follows directly from \eqref{qwer}. Since each column of 
$\Xi(\tau)$ has a simple principal part, one can easily read off the polynomials of \eqref{hjk} such that 
the principal part of $\xi_\alpha(\tau)$ matches that of a character $\chi_\alpha(\tau)$ of our interest. Again, the bijective $\Lambda$ 
implies that $\chi_\alpha(\tau)$ has to agree with $\xi_\alpha(\tau)$.

For a given modular data, it is rather non-trivial to determine 
the two constant matrices $\Lambda$ and $\chi$ solving 
\eqref{RHcondition} in general. As we will see shortly, 
the known explicit $q$-expansions of four characters 
\eqref{eq:firstfourhalfindex} plays a crucial role in further constraining 
the matrix $\chi$, and make the Riemann-Hilbert method higly effective.  
It eventually leads us to determine 
the characteristic matrix $\Xi(\tau)$ completely. Once $\Xi(\tau)$ is obtained,
it is straightforward to construct the conformal characters 
of the Haagerup RCFTs.

\paragraph{the Haagerup RCFT $\CR_{k=4}$} 

To illustrate, let us describe the procedure 
for determining the conformal characters of the Haagerup RCFT ${\cal R}_k$
with $k=4$ via the Riemann-Hilbert method. 

Given the modular matrices \eqref{Modular S/T matrices}, 
the conformal characters $\{\chi_\alpha(\tau)\}$ of $\CR_{k=4}$ 
transform as a ten-dimensional representation of $SL(2,\mathbb{Z})$. 
Thus, one can describe them in terms of a 
$10 \times 10$ characteristic matrix $\Xi(\tau)$. 
To generate $\Xi(\tau)$, it is essential to 
construct two constant matrices $\Lambda$ and $\chi$. 
We elaborate on how to determine them in what follows. 

The constant matrix $\Lambda$ should be compatible with 
the modular T-matrix \eqref{LambdaT} and its trace is constrained to 
be equal to $-5$ due to \eqref{eq:trLambda}.  
Our choice of $\Lambda$ is 
\begin{align}\label{Lambdachoice}
    \Lambda = \diag \Big(-\frac{13}{24},-\frac{1}{24},-\frac{1}{24},-\frac{13}{24},-\frac{11}{12},-1,-\frac{7}{8},-\frac{2}{3},-\frac{3}{8},0 \Big).
\end{align}
Each eigenvalue of $\Lambda$ is equal to the leading exponent $\Delta_\alpha$
of each character $\chi_\alpha(\tau)$ modulo integer. 
In fact, solving the modular crossing equation numerically, 
\begin{align}\label{crossing}
    \chi_\alpha\big(-1/\tau \big) = \sum_\beta S_{\alpha\beta} \chi_\beta\big(\tau\big)\ , 
\end{align}
one can read off the exponent $\Delta_\alpha$ of each character. 
In particular, the exponents of the first four characters are given by
\begin{align}\label{datak6}
    \Delta_0 = \frac{11}{24}\ , \quad \Delta_1= \frac{47}{24}\ , \quad
    \Delta_2 = -\frac{1}{24}\ , \quad \Delta_3 = \frac{11}{24}\ . 
 \end{align}

The next is to determine another constant matrix $\chi$ 
that solves the conditions \eqref{RHcondition}. 
To this end, we first note that there exists two pairs of conformal characters, 
$\big(\chi_0(\tau),\chi_3(\tau)\big)$ or $\big(\chi_1(\tau),\chi_2(\tau)\big)$,
whose conformal weights differ by an integer. 
For each pair, any linear combination of two characters 
remains an eigenvector of the modular T-matrix $T$. 
One can also see that the combinations 
\begin{align}
    \chi_0(\tau) + \chi_3(\tau) \ , \quad \chi_1(\tau) + \chi_2(\tau)   
\end{align}
together with $\chi_4(\tau)$ transform among themselves and decouple from the rests under the S
transformation. This implies that 
the modular S-matrix can be block-diagonalized as follows 
\begin{align}
    S = U^{-1}\begin{pmatrix} S_{1}  & & 
    \\ & S_2 
    \end{pmatrix} U    \ , 
\end{align}
where each block becomes a congruence representation of $SL(2,\mathbb{Z})$,
\begin{align}
    S_1  = 
        \begin{pmatrix}
            \frac12 & \frac12 & -\frac{1}{\sqrt2} \\
            \frac12 & \frac12 & \frac{1}{\sqrt2}\\
            -\frac{1}{\sqrt2} & \frac{1}{\sqrt2} & 0
        \end{pmatrix}  \ ,       
\end{align}
and
\begin{align}
    S_2  = 
         \begin{pmatrix}
            \frac{1}{2\sqrt3} & \frac{1}{2\sqrt3}  & - \frac{1}{\sqrt6} & - \frac{1}{\sqrt6}  & - \frac{1}{\sqrt6} & - \frac{1}{\sqrt6} & - \frac{1}{\sqrt6} \\
            \frac{1}{2\sqrt3} & \frac{1}{2\sqrt3}  &  \frac{1}{\sqrt6} & - \frac{1}{\sqrt6}  &  \frac{1}{\sqrt6} & - \frac{1}{\sqrt6} &  \frac{1}{\sqrt6} \\
            - \frac{1}{\sqrt6} &   \frac{1}{\sqrt6} & \frac12 & \frac{1}{2\sqrt3} & 0 & - \frac{1}{2\sqrt3} & -\frac12 \\
            - \frac{1}{\sqrt6} & - \frac{1}{\sqrt6}  & \frac{1}{2\sqrt3} & - \frac{1}{2\sqrt3}
            & -\frac{1}{\sqrt3} & - \frac{1}{2\sqrt3} & \frac{1}{2\sqrt3} \\ 
            - \frac{1}{\sqrt6} &   \frac{1}{\sqrt6} & 0 & -\frac{1}{\sqrt3} & 0 &  \frac{1}{\sqrt3} & 0 \\
            - \frac{1}{\sqrt6} & - \frac{1}{\sqrt6}  & -\frac{1}{2\sqrt3} & - \frac{1}{2\sqrt3}
            & \frac{1}{\sqrt3} & - \frac{1}{2\sqrt3} & -\frac{1}{2\sqrt3} \\ 
            - \frac{1}{\sqrt6} &   \frac{1}{\sqrt6} & - \frac12 & \frac{1}{2\sqrt3} & 0 & - \frac{1}{2\sqrt3} & \frac12 \\
        \end{pmatrix}     \ .  
\end{align}
Hence, $\{\chi_\alpha(\tau)\}$ of our interest should be in 
a reducible representation rather than an irreducible representation. 
Accordingly, the constant matrix $\chi$ can be described as 
\begin{align}
    \chi = U^{-1} \begin{pmatrix} \chi_1 & \\ & \chi_2 \end{pmatrix} U\ ,    
\end{align}
where $\chi_1$ and $\chi_2$ are $3\times 3$ and $7\times 7$ matrices. 
One can express $\chi$ explicitly as 
\begin{align}\label{parametrization02}
    \chi = \left(\begin{array}{cccccccccc}
\chi_{1,1} & \chi_{1,2} & \chi_{1,3} & \chi_{4,1} & \chi_{1,5} & \chi_{1,6} & \chi_{1,7} & \chi_{1,8} & \chi_{1,9} & \chi_{1,10} \\
\chi_{2,1} & \chi_{2,2} & \chi_{3,2} & \chi_{3,1} & \chi_{2,5} & \chi_{2,6} & \chi_{2,7} & \chi_{2,8} & \chi_{2,9} & \chi_{2,10} \\
\chi_{3,1} & \chi_{3,2} & \chi_{2,2} & \chi_{2,1} & \chi_{2,5} & -\chi_{2,6} & -\chi_{2,7} & -\chi_{2,8} & -\chi_{2,9} & -\chi_{2,10} \\
\chi_{4,1} & \chi_{1,3} & \chi_{1,2} & \chi_{1,1} & \chi_{1,5} & -\chi_{1,6} & -\chi_{1,7} & -\chi_{1,8} & -\chi_{1,9} & -\chi_{1,10} \\
\chi_{5,1} & \chi_{5,2} & \chi_{5,2} & \chi_{5,1} & \chi_{5,5} & 0 & 0 & 0 & 0 & 0 \\
\chi_{6,1} & \chi_{6,2} & -\chi_{6,2} & -\chi_{6,1} & 0 & \chi_{6,6} & \chi_{6,7} & \chi_{6,8} & \chi_{6,9} & \chi_{6,10} \\
\chi_{7,1} & \chi_{7,2} & -\chi_{7,2} & -\chi_{7,1} & 0 & \chi_{7,6} & \chi_{7,7} & \chi_{7,8} & \chi_{7,9} & \chi_{7,10} \\
\chi_{8,1} & \chi_{8,2} & -\chi_{8,2} & -\chi_{8,1} & 0 & \chi_{8,6} & \chi_{8,7} & \chi_{8,8} & \chi_{8,9} & \chi_{8,10} \\
\chi_{9,1} & \chi_{9,2} & -\chi_{9,2} & -\chi_{9,1} & 0 & \chi_{9,6} & \chi_{9,7} & \chi_{9,8} & \chi_{9,9} & \chi_{9,10} \\
\chi_{10,1} & \chi_{10,2} & -\chi_{10,2} & -\chi_{10,1} & 0 & \chi_{10,6} & \chi_{10,7} & \chi_{10,8} & \chi_{10,9} & \chi_{10,10} 
\end{array}\right).
\end{align}

For the sake of later convenience, we demand that 
the conformal characters $\{\chi_\alpha(\tau)\}$ is identified as the 
the third column of $\Xi(\tau)$. 
Based on the parameterization \eqref{parametrization02} with 
the exponents \eqref{datak6} estimated numerically, 
this requirement fixes the four unknown parameters by the four characters  \eqref{eq:firstfourhalfindex},
\begin{align}
    \chi_{1,3} = 1\, \quad \chi_{3,2}=0\ , \quad \chi_{2,2}=1\ , \quad \chi_{1,2}=1\ . 
\end{align}
Given \eqref{Lambdachoice} and \eqref{parametrization02}, one can express the solution 
to \eqref{RHcondition} as 
\begin{align}
\chi &=\resizebox{0.9\hsize}{!}{$\left(\begin{array}{cccccccccccc}
221 & 1 & 1 & 52 & \chi_{1,5} & -\frac{32x}{\chi_{6,2}} & \chi_{1,7} & \chi_{1,8} & \chi_{1,9} & 0 \\
52\left(25+\frac{756}{x}\right) & 1 & 0 & 52\left(25-\frac{756}{x}\right) & -22\chi_{1,5} & \frac{44928}{\chi_{6,2}} & \frac{5616\chi_{1,7}}{5x} & -\frac{3564\chi_{1,8}}{5x} & \frac{297}{\chi_{9,2}} & 0 \\
52\left(25-\frac{756}{x}\right) & 0 & 1 & 52\left(25+\frac{756}{x}\right) & -22\chi_{1,5} & -\frac{44928}{\chi_{6,2}} & -\frac{5616\chi_{1,7}}{5x} & \frac{3564\chi_{1,8}}{5x} & -\frac{297}{\chi_{9,2}} & 0 \\
52 & 1 & 1 & 221 & \chi_{1,5} & \frac{32x}{\chi_{6,2}} & -\chi_{1,7} & -\chi_{1,8} & -\chi_{1,9} & 0 \\
\frac{53248}{\chi_{1,5}} & -\frac{2048}{\chi_{1,5}} & -\frac{2048}{\chi_{1,5}} & \frac{53248}{\chi_{1,5}} & -22 & 0 & 0 & 0 & 0 & 0 \\
-\frac{756\chi_{6,2}}{x} & \chi_{6,2} & -\chi_{6,2} & \frac{756\chi_{6,2}}{x} & 0 & -12 & \frac{24\chi_{1,7}\chi_{6,2}}{35x} & \frac{27\chi_{1,8}\chi_{6,2}}{16x} & -\frac{10\chi_{6,2}}{\chi_{9,2}} & \chi_{6,10} \\
\frac{12285}{\chi_{1,7}} & \frac{35x}{2\chi_{1,7}} & -\frac{35x}{2\chi_{1,7}} & -\frac{12285}{\chi_{1,7}} & 0 & \frac{1120x}{\chi_{1,7}\chi_{6,2}} & -63 & \frac{189\chi_{1,8}}{\chi_{1,7}} & -\frac{210\chi_{1,9}}{\chi_{1,7}} & 0 \\
\frac{24960}{\chi_{1,8}} & -\frac{160x}{27\chi_{1,8}} & \frac{160x}{27\chi_{1,8}} & -\frac{24960}{\chi_{1,8}} & 0 & \frac{160x}{\chi_{1,8}\chi_{6,2}} & \frac{256\chi_{1,7}}{\chi_{1,8}} & 4 & -\frac{320\chi_{1,9}}{\chi_{1,8}} & 0 \\
\frac{19656}{\chi_{1,9}} & \chi_{9,2} & -\chi_{9,2} & -\frac{19656}{\chi_{1,9}} & 0 & -\frac{5760\chi_{9,2}}{\chi_{6,2}} & -\frac{832\chi_{1,7}}{5x} & -\frac{1404\chi_{1,8}}{5x} & -87 & 0 \\
-\frac{94068\chi_{10,2}}{x} & \chi_{10,2} & -\chi_{10,2} & \frac{94068\chi_{10,2}}{x} & 0 & -\frac{98370\chi_{10,2}}{\chi_{6,2}} & \frac{6952\chi_{1,7}\chi_{10,2}}{5x} & -\frac{297\chi_{1,8}\chi_{10,2}}{40x} & -\frac{330\chi_{10,2}}{\chi_{9,2}} & 0 \\
\end{array}
\right)$} \ , \nonumber
\end{align}
where $x = \chi_{1,9} \chi_{9,2}$. One can then compute the characteristic matrix $\Xi(\tau)$ using the recursion relation \eqref{eq:recursiverelation}, which leads to 
the conformal characters

\begin{align}
\begin{split}
\allowdisplaybreaks
    \chi_0(\tau) &= q^{\frac{11}{24}} \left( 1 + \frac{54-x}{54} q + \frac{108-x}{54} q^2 + \cdots \right), \\
\chi_1(\tau) &= q^{\frac{47}{24}} \left(1 + q + 2q^2 + 3q^3 + \cdots \right),\\
\chi_2(\tau) &= q^{-\frac{1}{24}} \left(1 + q + 3q^2 + 4q^3 + 7q^4 + \cdots \right), \\
\chi_3(\tau) &= q^{\frac{11}{24}} \left(1 + \frac{54+x}{54}q + \frac{108+x}{54} q^2 + \cdots \right),\\
\chi_4(\tau) &= - \frac{2048}{\chi_{1,5}} q^\frac{1}{12} \left(1 + 2q + 3q^2 + 6q^3 + 9q^4 + \cdots \right), \\
\chi_5(\tau) &= -\chi_{6,2} \left(1 + 2q + 4q^2 + 6q^3 + \cdots\right) - \chi_{6,10}\chi_{10,2}\left( q + 2q^2 + 3q^3 + 5q^4 + \cdots\right),\\
\chi_6(\tau) &= - \frac{35x}{2 \chi_{1,7}} q^{\frac{1}{8}} \left(1 + q + 2q^2 + 3q^3 + 6q^4 + \cdots \right),\\
\chi_7(\tau) &= \frac{160 x}{27 \chi_{1,8}} q^\frac{1}{3} \left(1 + q + 2q^2 + 4q^3 + 6q^4 + \cdots\right), \\
\chi_8(\tau) &= - \chi_{9,2} q^\frac{5}{8} \left(1 + q + 3q^2 + 4q^3 + 7q^4 + \cdots \right),\\
\chi_9(\tau) &= - \chi_{10,2} q \left(1 + 2q + 3q^2 + 5q^3 + \cdots \right)\ .
 \end{split}
\end{align}

Since the first four characters should agree with \eqref{eq:firstfourhalfindex},
one can easily fix a parameter $x$ as $x=54$. Demanding all $q$-expansion 
coefficients are non-negative integers, the 
left-over parameters in $\chi$ can be fixed completely 
by solving the modular crossing equation \eqref{crossing}
numerically. To be concrete, the constant matrix $\chi$ is
\begin{align}
    \chi = \left(
\begin{array}{cccccccccc}
221 & 1 & 1 & 52 & -2048 & 1728 & -945 & 320 & -54 & 0 \\
2028 & 1 & 0 & 572 & 45056 & -44928 & -19656 & -4224 & -297 & 0 \\
572 & 0 & 1 & 2028 & 45056 & 44928 & 19656 & 4224 & 297 & 0 \\
52 & 1 & 1 & 221 & -2048 & -1728 & 945 & -320 & 54 & 0 \\
-26 & 1 & 1 & -26 & -22 & 0 & 0 & 0 & 0 & 0 \\
14 & -1 & 1 & -14 & 0 & -12 & 12 & -10 & -10 & -1 \\
-13 & -1 & 1 & 13 & 0 & 64 & -63 & -64 & -12 & 0 \\
78 & -1 & 1 & -78 & 0 & -27 & -756 & 4 & 54 & 0 \\
-364 & -1 & 1 & 364 & 0 & -5760 & -2912 & 1664 & -87 & 0 \\
1742 & -1 & 1 & -1742 & 0 & -98370 & 24332 & 44 & -330 & 0 
\end{array}
\right),
\end{align}
and thus the conformal characters of the Haagerup RCFT $\CR_{k=4}$ are
\begin{align}
\begin{split}
    \chi_0(\tau) &= q^{\frac{11}{24}} \left(1 + q^2 + q^3 + 3q^4 + 3q^5 + 6q^6 + \cdots\right), \\
    \chi_1(\tau) &= q^{\frac{47}{24}} \left( 1 + q + 2q^2 + 3q^3 + 4q^4 + 6q^5 + 10q^6 + \cdots
 \right),\\
    \chi_2(\tau) &= q^{-\frac{1}{24}}\left(1+q+3q^2 + 4q^3 + 7q^4 + 10q^5 + 17q^6 + \cdots \right), \\
    \chi_3(\tau) &= q^{\frac{11}{24}}  \left(1 + 2q + 3q^2 + 5q^3 + 9q^4 + 13 q^5 + 20 q^6 + \cdots \right),\\
    \chi_4(\tau) &= q^\frac{1}{12}  \left( 1 + 2q + 3q^2 + 6q^3 + 9q^4 + 14 q^5 + 22 q^6 + \cdots \right), \\
    \chi_5(\tau) &= q^0 \left(1 + q + 2q^2 + 3q^3 + 5q^4 + 8q^5 + 12 q^6 + \cdots\right),\\
    \chi_6(\tau) &= q^\frac{1}{8} \left(1+q+2q^2+3q^3+6q^4+8q^5+13q^6 + \cdots \right),\\
    \chi_7(\tau) &= q^\frac{1}{3} \left( 1 + q + 2q^2+4q^3+6q^4+9q^5+14q^6+\cdots \right), \\
    \chi_8(\tau) &= q^\frac{5}{8} \left(1+q+3q^2+4q^3+7q^4+10q^5+16q^6+\cdots\right),\\
    \chi_9(\tau) &= q\left(1+2q+3q^2+5q^3+8q^4+12q^5+18q^6+\cdots\right)\ .
\end{split}  
\end{align}
The characters coincide with characters of bosonoic theory of supersymmetric $\CN=1$ minimal model $SM(2,12)$:
\begin{align}
\begin{split}
&\chi^{SM(2,12)}_{(1,1)} (q) = \chi_{0} (q)+\chi_1 (q)\;,
\\
&\chi^{SM(2,12)}_{(1,3)} (q) = \chi_6 (q) +\chi_8(q)\;,
\\
&\chi^{SM(2,12)}_{(1,5)} (q) = \chi_2 (q)+\chi_3 (q)\;.
\end{split}
\end{align}
NS characters of $\CN=1$ supersymmetric Virasoro minimal model $SM(P,Q)$ are  given as ($1\leq r<P,1\leq s<Q$ with $r-s\in 2\mathbb{Z}$) 
\begin{align}
\begin{split}
&\chi^{SM(P,Q)}_{(r,s)} = q^{h-c/24} \frac{(-q^{1/2};q)_\infty}{(q)_\infty}  \sum_{n\in \mathbb{Z}} \left( q^{(n^2 P Q+n (Qr-Ps))/2} - q^{(nP+r)(nQ+s)/2}\right)\;,
\\
&\textrm{with }h=\frac{(Qr-Ps)^2-(P-Q)^2}{8PQ}\;, \quad c=\frac{3}2\left( 1-\frac{2(P-Q)^2}{PQ}\right)\;.
\end{split}
\end{align}
 
As a final remark, we propose that  $\chi_4(\tau)$, $\chi_5(\tau)$, $\chi_7(\tau)$, and $\chi_9(\tau)$
can be summarized as 
\begin{align}
\begin{split}
    \chi_4(\tau) & = \frac{1}{2} \sum_{\mathfrak{m}\in {\cal Z}\oplus {\cal Z}'} \frac{q^{\frac{1}{2} \mathfrak{m}^T \cdot K \cdot \mathfrak{m} + (1, -1, -1) \cdot \mathfrak{m} +\frac{1}{12}}}{(q)_{2m_1} (q)_{m_2} (q)_{m_3}}\ , 
    \\
    \chi_5(\tau) & = \sum_{\mathfrak{m} \in {\cal Z} } \frac{q^{\frac{1}{2} \mathfrak{m}^T\cdot K \cdot \mathfrak{m} + (1, 0, 0)\cdot \mathfrak{m} }}{(q)_{2m_1} (q)_{m_2} (q)_{m_3}}\ ,
    \\
    \chi_7(\tau) & = \sum_{\mathfrak{m}\in {\cal Z}\oplus {\cal Z}'} \frac{q^{\frac{1}{2} \mathfrak{m}^T \cdot K \cdot \mathfrak{m} + (1, 1, 2) \cdot \mathfrak{m} +1/3}}{(q)_{2m_1} (q)_{m_2} (q)_{m_3}}\ ,
    \\
    \chi_9(\tau) & =  \sum_{\mathfrak{m} \in {\cal Z}' } \frac{q^{\frac{1}{2} \mathfrak{m}^T \cdot K \cdot \mathfrak{m} + (1, 0, 0)\cdot \mathfrak{m} }}{(q)_{2m_1} (q)_{m_2} (q)_{m_3}}\ .
\end{split}
\end{align}
The matrix $K$ is given in \eqref{K-matrix} with $k=4$. It is noteworthy here that five among six modular forms studied in \cite{Zagier:2007knq} with 
$A = \resizebox{0.06\hsize}{!}{$\left(\begin{array}{ccc}
    4 &2 & 1  \\
     2& 2 & 0 \\
     1 & 0 & 1
\end{array}\right)$}$
are present as the above conformal characters $\chi_\alpha(\tau)$ \
with the summation range slightly modified.

To present the results for other values of $k$ concisely, we only provide
essential data $\Lambda$, $\chi$ that generate the characteristic matrix $\Xi(\tau)$.
One can always read off the conformal characters from the third column of $\Xi(\tau)$ and confirm that their  $q$-expansion coefficients are always non-negative integers.

\paragraph{the Haagerup RCFT ${\cal R}_{k=3}$} The two constant matrices are given by 
\begin{align}
    \Lambda =\textrm{diag}  \left\{-\frac{19}{24},-\frac{1}{24},-\frac{1}{24},-\frac{19}{24},-\frac{119}{120},-\frac{101}{120},-\frac{71}{120},-\frac{29}{120}\right\}  \;    \ , 
\end{align}
and
\begin{align}
\begin{split}
\chi= \left(
\begin{array}{cccccccc}
	-171 & 1 & 1 & -76 & 210 & -120 & 45 & -10 \\
	19304 & 2 & 1 & 7448 & -46760 & -17030 & -2520 & -65 \\
	7448 & 1 & 2 & 19304 & 46760 & 17030 & 2520 & 65 \\
	-76 & 1 & 1 & -171 & -210 & 120 & -45 & 10 \\
	19 & -1 & 1 & -19 & 21 & -6 & -12 & -8 \\
	-56 & -1 & 1 & 56 & 56 & -131 & -56 & 1 \\
	494 & -1 & 1 & -494 & -780 & -1014 & 277 & 12 \\
	-4256 & -1 & 1 & 4256 & -23200 & 3585 & 672 & -61 \\
\end{array}
\right)\;.
\end{split}
\end{align}
The third column of $\Xi(\tau)$ gives conformal characters of $\CR_{k=3}$, which agree perfectly with the conformal characters of a product of two non-unitary minimal models $M(3,5)\otimes M(2,5)$, 
\begin{align}
	\allowdisplaybreaks
	\begin{split}
		\chi_0(\tau)  &=q^{\frac{5}{24}} \sum_{m_1 \in \mathbb{Z}_{\geq 0}, m_2 \in \mathbb{Z}_{\geq 0 }} \frac{q^{m_1^2+m_1 + m_2^2 +m_2}}{(q)_{2m_1} (q)_{m_2}}\;,
		\\
		\chi_1(\tau)  &=q^{\frac{23}{24}} \sum_{m_1 \in \mathbb{Z}_{\geq 0}+\frac{1}2, m_2 \in \mathbb{Z}_{\geq 0 }} \frac{q^{m_1^2+m_1 + m_2^2 +m_2- \frac{3}4}}{(q)_{2m_1} (q)_{m_2}}\;,
		\\
		\chi_2(\tau)  &=q^{-\frac{1}{24}} \sum_{m_1 \in \mathbb{Z}_{\geq 0}, m_2 \in \mathbb{Z}_{\geq 0 }} \frac{q^{m_1^2+ m_2^2}}{(q)_{2m_1} (q)_{m_2}}\;,
		\\
		\chi_3(\tau)  &=q^{\frac{5}{24}} \sum_{m_1 \in \mathbb{Z}_{\geq 0} +\frac{1}2, m_2 \in \mathbb{Z}_{\geq 0 }} \frac{q^{m_1^2 + m_2^2-\frac{1}4}}{(q)_{2m_1} (q)_{m_2}}\;,
		\\
		\chi_4(\tau)  &=q^{\frac{1}{120}} \sum_{m_1 \in \mathbb{Z}_{\geq 0}, m_2 \in \mathbb{Z}_{\geq 0 }} \frac{q^{m_1^2+m_1 + m_2^2}}{(q)_{2m_1} (q)_{m_2}}\;,
		\\
		\chi_5(\tau)  &=q^{\frac{19}{120}} \sum_{m_1 \in \mathbb{Z}_{\geq 0}, m_2 \in \mathbb{Z}_{\geq 0 }} \frac{q^{m_1^2 + m_2^2+m_2}}{(q)_{2m_1} (q)_{m_2}}\;,
		\\
		\chi_6(\tau)  &=q^{\frac{49}{120}} \sum_{m_1 \in \mathbb{Z}_{\geq 0}+\frac{1}2, m_2 \in \mathbb{Z}_{\geq 0 }} \frac{q^{m_1^2 + m_2^2+m_2-\frac{1}4}}{(q)_{2m_1} (q)_{m_2}}\;,
		\\
		\chi_7(\tau) & =q^{\frac{91}{120}} \sum_{m_1 \in \mathbb{Z}_{\geq 0}+\frac{1}2, m_2 \in \mathbb{Z}_{\geq 0 }} \frac{q^{m_1^2 +m_1+ m_2^2-\frac{3}4}}{(q)_{2m_1} (q)_{m_2}}\;.
	\end{split}
	\label{eq:k3character}
\end{align}

\vspace{0.2cm}
\paragraph{the Haagerup RCFT ${\cal R}_{k=5}$} The data are given by 
\begin{align}
    \Lambda = \resizebox{0.8\hsize}{!}{$\diag \left\{-\frac{7}{24},-\frac{1}{24},-\frac{1}{24},-\frac{7}{24},-\frac{23}{24},-\frac{17}{24},-\frac{169}{168},-\frac{151}{168},-\frac{121}{168},-\frac{79}{168}-\frac{25}{168}-\frac{127}{168}\right\}$}\ , \nonumber
\end{align}
\begin{align}
    \chi = \resizebox{0.8\hsize}{!}{$\left(\begin{array}{cccccccccccc}
-62 & 1 & 1 & -15 & -11178 & 2430 & 9555 & -5152 & 1701 & -280 & 7 & -2184 \\
134 & 1 & 0 & 34 & 48600 & 8262 & -43316 & -21350 & -5950 & -714 & -14 & -7735 \\
34 & 0 & 1 & 134 & 48600 & 8262 & 43316 & 21350 & 5950 & 714 & 14 & 7735 \\
-15 & 1 & 1 & -62 & -11178 & 2430 & -9555 & 5152 & -1701 & 280 & -7 & 2184 \\
-7 & 1 & 1 & -7 & 23 & -34 & 0 & 0 & 0 & 0 & 0 & 0 \\
20 & 1 & 1 & 20 & -552 & -187 & 0 & 0 & 0 & 0 & 0 & 0 \\
5 & -1 & 1 & -5 & 0 & 0 & 13 & -2 & -1 & -10 & -3 & -22 \\
-2 & -1 & 1 & 2 & 0 & 0 & 52 & -27 & -50 & -25 & -2 & 26 \\
12 & -1 & 1 & -12 & 0 & 0 & 156 & -450 & -152 & 46 & 5 & -78 \\
-30 & -1 & 1 & 30 & 0 & 0 & -1300 & -2976 & 798 & 153 & -10 & -338 \\
75 & -1 & 1 & -75 & 0 & 0 & -28900 & -5250 & 6375 & -850 & 6 & 5746 \\
-14 & 0 & 0 & 14 & 0 & 0 & -468 & 225 & -50 & -6 & 6 & -179 \\
\end{array}
\right)$}\; , \nonumber
\end{align}
 and the characters are
 \begin{align}
 \begin{split}
 {\chi}_0 &= q^{\frac{17}{24}}\left( 1+ q^2 + q^3 + 2q^4 + 2q^5 + 5q^6 + \cdots \right), \\
 {\chi}_1 &= q^{\frac{71}{24}} \left( 1 + q + 2q^2 + 3q^3 + 4q^4 + 6q^5 + 9q^6 + \cdots   \right),\\
 {\chi}_2 &= q^{-\frac{1}{24}} \left( 1+q+2q^2+4q^3+6q^4+9q^5 +14 q^6 + \cdots   \right), \\
 {\chi}_3 &= q^{\frac{17}{24}}\left(1+2q+3q^2+5q^3+8q^4+12q^5 + 19 q^6 +  \cdots\right), \\
 {\chi}_4 &= q^\frac{1}{24} \left( 1 + q + 3q^2 + 4q^3 + 8q^4 + 11q^5 + 18 q^6 +  \cdots \right),\\
 {\chi}_5 &= q^\frac{7}{24} \left( 1 + 2q+3q^2+5q^3+8q^4+13q^5 +19 q^6 +  \cdots  \right), \\
 {\chi}_6 &= q^{-\frac{1}{168}} \left( 1+q+2q^2+3q^3 + 5q^4 + 7q^5 + 12 q^6 + \cdots \right), \\
%  &= q^{-\frac{1}{168}}  \sum_{\substack{m_1 \in \mathbf{Z}_{\geq 0}, \\ m_{a: 2 \leq a \leq 4} \in \mathbf{Z}_{\geq 0 }}} \frac{q^{\frac{1}2 \mathbf{m}^T \cdot K^{\rm eff}\cdot  \mathbf{m} + m_1 }}{(q)_{2m_1} (q)_{m_2} (q)_{m_3}(q)_{m_4}} ,\\
 {\chi}_7 &= q^\frac{17}{168} \left( 1 + q + 2q^2 + 3q^3 + 5 q^4 + 8q^5 + 12 q^6 + \cdots\right),\\
 {\chi}_8 &= q^{\frac{47}{168}} \left(1 + q + 2q^2 + 3q^3 + 6 q^4 + 8q^5 + 13q^6 +  \cdots   \right), \\
 {\chi}_9 &= q^\frac{89}{168} \left( 1 + q + 2q^2 + 4q^3 +6q^4 + 9q^5 + 14 q^6 + \cdots \right), \\
 {\chi}_{10} &= q^\frac{143}{168} \left(1+q+3q^2 + 4q^3 + 7q^4 + 10 q^5 + 16 q^6 + \cdots  \right), \\
 {\chi}_{11} &= q^\frac{209}{168} \left(1 + 2q + 3q^2 + 5q^3 + 8q^4 + 12q^5 + 18 q^6 + \cdots  \right). %\\
 %&=q^\frac{209}{168} \sum_{\substack{m_1 \in \mathbf{Z}_{\geq 0} + \frac{1}{2}, \\ m_{a: 2 \leq a \leq 4} \in \mathbf{Z}_{\geq 0 }}} \frac{q^{\frac{1}2 \mathbf{m}^T \cdot K^{\rm eff}\cdot  \mathbf{m} + m_1 -\frac{5}{4}}}{(q)_{2m_1} (q)_{m_2} (q)_{m_3}(q)_{m_4}}.
 \end{split}
 \end{align}

\vspace{0.2cm}
\paragraph{the Haagerup RCFT ${\cal R}_{k=6}$} We present two constant matrices below, 
\begin{align}
    \Lambda = \resizebox{0.8\hsize}{!}{$\diag \left\{ -\frac{1}{24},-\frac{1}{24},-\frac{1}{24},-\frac{1}{24},-\frac{47}{48},-\frac{19}{24},-\frac{23}{48},-\frac{97}{96},-\frac{11}{12},-\frac{73}{96},-\frac{13}{24},-\frac{25}{96},-\frac{11}{12},-\frac{49}{96},  \right\}$}\ , \nonumber
\end{align}
\begin{align}
    \chi = \resizebox{0.8\hsize}{!}{$\left(\begin{array}{cccccccccccccccc}
1 & 1 & 1 & 0 & -48128 & 13376 & -1024 & 41600 & -22464 & 7488 & -1300 & 64 & -22592 & 960 \\
1 & 1 & 0 & 1 & 48128 & 13376 & 1024 & -41600 & -22592 & -7488 & -1300 & -64 & -22464 & -960 \\
1 & 0 & 1 & 1 & 48128 & 13376 & 1024 & 41600 & 22592 & 7488 & 1300 & 64 & 22464 & 960 \\
0 & 1 & 1 & 1 & -48128 & 13376 & -1024 & -41600 & 22464 & -7488 & 1300 & -64 & 22592 & -960 \\
-1 & 1 & 1 & -1 & 0 & 0 & -23 & 0 & 0 & 0 & 0 & 0 & 0 & 0 \\
1 & 1 & 1 & 1 & 0 & -247 & 0 & 0 & 0 & 0 & 0 & 0 & 0 & 0 \\
-1 & 1 & 1 & -1 & -4371 & 0 & 253 & 0 & 0 & 0 & 0 & 0 & 0 & 0 \\
1 & -1 & 1 & -1 & 0 & 0 & 0 & -2 & 16 & -14 & 0 & -7 & -16 & -17 \\
0 & -1 & 1 & 0 & 0 & 0 & 0 & 48 & -7 & -40 & -26 & -8 & -15 & 8 \\
1 & -1 & 1 & -1 & 0 & 0 & 0 & 147 & -240 & -182 & 0 & 12 & 240 & -35 \\
-1 & -1 & 1 & 1 & 0 & 0 & 0 & 0 & -2048 & 0 & 273 & 0 & -2048 & 0 \\
1 & -1 & 1 & -1 & 0 & 0 & 0 & -8625 & -9200 & 4575 & 0 & -69 & 9200 & 322 \\
-1 & 0 & 0 & 1 & 0 & 0 & 0 & -48 & -15 & 40 & -26 & 8 & -7 & -8 \\
0 & 0 & 0 & 0 & 0 & 0 & 0 & -3675 & 1792 & -441 & 0 & 21 & -1792 & 256 \\
\end{array}
\right)$}\;,  \nonumber
\end{align}
 and characters are
 \begin{align}
 \begin{split}
 \chi_0 &= q^\frac{23}{24} \left( 1+q^2+q^3+2q^4+2q^5+4q^6+\cdots \right), \\
 \chi_1 &= q^\frac{95}{24} \left( 1+q+2q^2+3q^3+4q^4+6q^5+9q^6+\cdots \right), \\
 \chi_2 &= q^{-\frac{1}{24}} \left(1 + q + 2 q^2 + 3 q^3 + 6 q^4 + 8 q^5 + 13 q^6 + \cdots \right),\\
 \chi_3 &= q^\frac{23}{24} \left(1 + 2 q + 3 q^2 + 5 q^3 + 8 q^4 + 12 q^5 + 18 q^6 + \cdots \right),\\
 \chi_4 &= q^\frac{1}{48} \left(1 + q + 2 q^2 + 4 q^3 + 6 q^4 + 10 q^5 + 15 q^6 + \cdots \right),\\
 \chi_5 &= q^\frac{5}{24} \left(1 + q + 3 q^2 + 4 q^3 + 7 q^4 + 10 q^5 + 17 q^6+ \cdots \right),\\
 \chi_6 &= q^\frac{25}{48} \left(1 + 2 q + 3 q^2 + 5 q^3 + 8 q^4 + 12 q^5 + 18 q^6+ \cdots \right),\\
 \chi_7 &= q^{-\frac{1}{96}} \left(1 + q + 2 q^2 + 3 q^3 + 5 q^4 + 7 q^5 + 11 q^6+ \cdots \right),\\
 \chi_8 &= q^\frac{1}{12} \left(1 + q + 2 q^2 + 3 q^3 + 5 q^4 + 7 q^5 + 12 q^6+ \cdots \right),\\
 \chi_9 &= q^\frac{23}{96} \left(1 + q + 2 q^2 + 3 q^3 + 5 q^4 + 8 q^5 + 12 q^6+ \cdots \right),\\
 \chi_{10} &= q^\frac{11}{24} \left(1 + q + 2 q^2 + 3 q^3 + 6 q^4 + 8 q^5 + 13 q^6 + \cdots \right),\\
 \chi_{11} &= q^\frac{71}{96} \left(1 + q + 2 q^2 + 4 q^3 + 6 q^4 + 9 q^5 + 14 q^6+ \cdots \right),\\
 \chi_{12} &= q^\frac{13}{12} \left(1 + q + 3 q^2 + 4 q^3 + 7 q^4 + 10 q^5 + 16 q^6 + \cdots \right),\\
 \chi_{13} &= q^\frac{143}{96} \left(1 + 2 q + 3 q^2 + 5 q^3 + 8 q^4 + 12 q^5 + 18 q^6 + \cdots \right).
 \end{split}
 \end{align}

\vspace{0.2cm}
\paragraph{the Haagerup RCFT ${\cal R}_{k=7}$} Two constant matrices are 
\begin{align}
        \Lambda =\resizebox{0.8\hsize}{!}{$ \diag \left\{ -\frac{19}{24}, -\frac{1}{24},-\frac{1}{24},-\frac{19}{24},-\frac{119}{120},-\frac{101}{120},-\frac{71}{120},-\frac{29}{120},-\frac{73}{72},-\frac{67}{72},-\frac{19}{24},-\frac{43}{72},-\frac{25}{72},-\frac{1}{24},-\frac{49}{72},-\frac{19}{72}\right\}$}\ , \nonumber
\end{align}
\begin{align}
    \chi = \resizebox{0.8\hsize}{!}{$\left(\begin{array}{cccccccccccccccc}
-86 & 0 & 0 & -9 & -210 & 120 & -45 & 10 & 162 & -135 & 85 & -36 & 9 & -1 & 45 & -9 \\
10380 & 1 & 0 & 1476 & 46760 & 17030 & 2520 & 65 & -40176 & -23301 & -8924 & -1953 & -180 & -1 & -3816 & -63 \\
1476 & 0 & 1 & 10380 & 46760 & 17030 & 2520 & 65 & 40176 & 23301 & 8924 & 1953 & 180 & 1 & 3816 & 63 \\
-9 & 0 & 0 & -86 & -210 & 120 & -45 & 10 & -162 & 135 & -85 & 36 & -9 & 1 & 45 & 9 \\
-19 & 1 & 1 & -19 & 21 & -6 & -12 & -8 & 0 & 0 & 0 & 0 & 0 & 0 & 0 & 0 \\
56 & 1 & 1 & 56 & 56 & -131 & -56 & 1 & 0 & 0 & 0 & 0 & 0 & 0 & 0 & 0 \\
-494 & 1 & 1 & -494 & -780 & -1014 & 277 & 12 & 0 & 0 & 0 & 0 & 0 & 0 & 0 & 0 \\
4256 & 1 & 1 & 4256 & -23200 & 3585 & 672 & -61 & 0 & 0 & 0 & 0 & 0 & 0 & 0 & 0 \\
7 & -1 & 1 & -7 & 0 & 0 & 0 & 0 & 7 & -5 & 7 & -8 & -5 & -1 & -20 & -7 \\
-20 & -1 & 1 & 20 & 0 & 0 & 0 & 0 & 32 & -2 & -20 & -29 & -12 & -1 & -8 & 1 \\
85 & -1 & 1 & -85 & 0 & 0 & 0 & 0 & 162 & -135 & -162 & -36 & 9 & 1 & 45 & -9 \\
-344 & -1 & 1 & 344 & 0 & 0 & 0 & 0 & 304 & -1279 & -344 & 195 & 40 & -1 & -384 & 2 \\
1735 & -1 & 1 & -1735 & 0 & 0 & 0 & 0 & -2138 & -7285 & 1735 & 764 & -100 & -1 & 1618 & 13 \\
-8924 & -1 & 1 & 8924 & 0 & 0 & 0 & 0 & -40176 & -23301 & 17828 & -1953 & -180 & 2 & -3816 & -63 \\
189 & 0 & 0 & -189 & 0 & 0 & 0 & 0 & -735 & 112 & 189 & -140 & 35 & 0 & -31 & 20 \\
-2808 & 0 & 0 & 2808 & 0 & 0 & 0 & 0 & -19760 & 9984 & -2808 & 260 & 40 & 0 & 2080 & -61 \\
\end{array}\right)$}\;, \nonumber
\end{align}
 and the characters are
  \begin{align}
      \begin{split}
      {\chi}_{0} &= q^\frac{29}{24} \left( 1 + q^2 + q^3 + 2q^4 + 2q^5  + \cdots \right),   \\
 {\chi}_{1} &= q^{\frac{119}{24}} \left(1 + q + 2q^2 + 3q^3 + 4q^4 + 6q^5 + 9q^6 + \cdots \right), \\
 {\chi}_{2} &= q^{-\frac{1}{24}}\left( 1 + q + 2q^2 + 3q^3 + 5q^4 + 8q^5 + 12 q^6 + \cdots \right), \\
  {\chi}_{3} &= q^\frac{29}{24} \left(1 + 2q + 3q^2 + 5q^3 + 8q^4 + 12q^5 + 18 q^6 + \cdots\right), \\
{\chi}_{4} &= q^\frac{1}{120} \left( 1 + q + 2q^2 + 3q^3 + 6q^4 + 8q^5 + 14 q^6 + \cdots
  \right),  \\
 {\chi}_{5} &= q^\frac{19}{120} \left( 1 + q + 2q^2 + 4q^3 + 6q^4 + 9q^5 + 14 q^6 + \cdots \right), \\
 {\chi}_{6} &=q^\frac{49}{120} \left( 1 + q + 3q^2 + 4q^3 + 7q^4 + 10q^5 + 16 q^6 + \cdots \right),  \\
 {\chi}_{7} &= q^\frac{91}{120} \left( 1 + 2q + 3q^2 + 5q^3 + 8q^4 + 12q^5 + 18 q^6 + \cdots \right), \\
 {\chi}_{8} &=q^{-\frac{1}{72}} \left( 1 + q + 2q^2 + 3q^3 + 5q^4 + 7q^5 + 11q^6 + \cdots \right),  \\
% &= q^{-\frac{1}{72}} \sum_{\substack{m_1 \in \mathbf{Z}_{\geq 0}, \\ m_{a: 2 \leq a \leq 6} \in \mathbf{Z}_{\geq 0 }}} \frac{q^{\frac{1}2 \mathbf{m}^T \cdot K^{\rm eff}\cdot  \mathbf{m} + m_1 }}{(q)_{2m_1} (q)_{m_2} (q)_{m_3}(q)_{m_4}(q)_{m_5}(q)_{m_6}} \\
 {\chi}_{9} &= q^\frac{5}{72} \left( 1 + q + 2q^2 + 3q^3 + 5q^4 + 7q^5 + 11q^6 + \cdots \right), \\
 {\chi}_{10} &=q^\frac{5}{24} \left( 1 + q + 2q^2 + 3q^3 + 5q^4 + 7q^5 + 12q^6 + \cdots \right), \\
 {\chi}_{11} &= q^\frac{29}{72} \left( 1 + q + 2q^2 + 3q^3 + 5q^4 + 8q^5 + 12 q^6 + \cdots \right),  \\
 {\chi}_{12} &= q^\frac{47}{72} \left( 1 + q + 2q^2 + 3q^3 + 6q^4 + 8q^5 + 13 q^6 + \cdots \right), \\
 {\chi}_{13} &= q^\frac{23}{24} \left( 1 + q + 2q^2 + 4q^3 + 6q^4 + 9q^5 + 14q^6 + \cdots \right), \\
 {\chi}_{14} &= q^\frac{95}{72} \left( 1 + q + 3q^2 + 4q^3 + 7q^4 + 10 q^5 + 16 q^6 + \cdots  \right),  \\
 {\chi}_{15} &= q^\frac{125}{72} \left( 1 + 2q + 3q^2 + 5q^3 + 8q^4 + 12q^5 + 18 q^6 + \cdots \right). %\\
%&=q^{\frac{125}{72}} \sum_{\substack{m_1 \in \mathbf{Z}_{\geq 0}, \\ m_{a: 2 \leq a \leq 6} \in \mathbf{Z}_{\geq 0 }}} \frac{q^{\frac{1}2 \mathbf{m}^T \cdot K^{\rm eff}\cdot  \mathbf{m} + m_1 -\frac{7}{4}}}{(q)_{2m_1} (q)_{m_2} (q)_{m_3}(q)_{m_4}(q)_{m_5}(q)_{m_6}} 
      \end{split}
     \end{align}

\vspace{0.2cm}
\paragraph{the Haagerup RCFT ${\cal R}_{k=8}$} The seed data $\Lambda$ and $\chi$ are 
\begin{align}
    \Lambda &= \resizebox{0.9\hsize}{!}{$\diag \left\{ -\frac{13}{24},-\frac{1}{24}, -\frac{1}{24},-\frac{13}{24},  -1, -\frac{7}{8}, -\frac{2}{3}, -\frac{3}{8}, 0, -\frac{61}{60}, -\frac{113}{120}, -\frac{49}{60}, -\frac{77}{120}, -\frac{5}{12}, -\frac{17}{120}, -\frac{49}{60}, -\frac{53}{120}, -\frac{1}{60}
\right\}$}\ , \nonumber
\end{align}
\begin{align}
    \chi &=\resizebox{0.8\hsize}{!}{$\left(
\begin{array}{cccccccccccccccccccc}
148 & 0 & 0 & 21 & -1728 & 945 & -320 & 54 & 0 & 1440 & -1035 & 560 & -210 & 48 & -5 & 560 & -70 & 0 \\
1302 & 1 & 0 & 154 & 44928 & 19656 & 4224 & 297 & 0 & -38720 & -23870 & -10080 & -2695 & -352 & -10 & -10080 & -440 & 0 \\
154 & 0 & 1 & 1302 & 44928 & 19656 & 4224 & 297 & 0 & 38720 & 23870 & 10080 & 2695 & 352 & 10 & 10080 & 440 & 0 \\
21 & 0 & 0 & 148 & -1728 & 945 & -320 & 54 & 0 & -1440 & 1035 & -560 & 210 & -48 & 5 & -560 & 70 & 0 \\
-14 & 1 & 1 & -14 & -12 & 12 & -10 & -10 & -1 & 0 & 0 & 0 & 0 & 0 & 0 & 0 & 0 & 0 \\
13 & 1 & 1 & 13 & 64 & -63 & -64 & -12 & 0 & 0 & 0 & 0 & 0 & 0 & 0 & 0 & 0 & 0 \\
-78 & 1 & 1 & -78 & -27 & -756 & 4 & 54 & 0 & 0 & 0 & 0 & 0 & 0 & 0 & 0 & 0 & 0 \\
364 & 1 & 1 & 364 & -5760 & -2912 & 1664 & -87 & 0 & 0 & 0 & 0 & 0 & 0 & 0 & 0 & 0 & 0 \\
-1742 & 1 & 1 & -1742 & -98370 & 24332 & 44 & -330 & 0 & 0 & 0 & 0 & 0 & 0 & 0 & 0 & 0 & 0 \\
8 & -1 & 1 & -8 & 0 & 0 & 0 & 0 & 0 & -10 & 14 & -7 & 0 & -6 & -2 & -7 & -14 & -1 \\
-7 & -1 & 1 & 7 & 0 & 0 & 0 & 0 & 0 & 32 & 4 & -16 & -22 & -16 & -3 & -16 & -4 & 0 \\
28 & -1 & 1 & -28 & 0 & 0 & 0 & 0 & 0 & 111 & -54 & -133 & -56 & -1 & 2 & 112 & 2 & -1 \\
-62 & -1 & 1 & 62 & 0 & 0 & 0 & 0 & 0 & 384 & -770 & -448 & 77 & 64 & 2 & -448 & -63 & 0 \\
224 & -1 & 1 & -224 & 0 & 0 & 0 & 0 & 0 & -45 & -4970 & 20 & 880 & 2 & -10 & 20 & 210 & 0 \\
-777 & -1 & 1 & 777 & 0 & 0 & 0 & 0 & 0 & -13344 & -21987 & 9744 & 1078 & -496 & 4 & 9744 & -504 & 0 \\
20 & 0 & 0 & -20 & 0 & 0 & 0 & 0 & 0 & -123 & -68 & 119 & -56 & 5 & 4 & -126 & 16 & 1 \\
-210 & 0 & 0 & 210 & 0 & 0 & 0 & 0 & 0 & -4928 & 1254 & 672 & -616 & 160 & -6 & 672 & 73 & 0 \\
1260 & 0 & 0 & -1260 & 0 & 0 & 0 & 0 & 0 & -84843 & 44100 & -13475 & 1848 & 15 & -4 & 13524 & -784 & 1 \\
\end{array}\right)$} \; , \nonumber
\end{align}
and characters are
 \begin{align}
 \begin{split}
 \chi_0 &= q^\frac{35}{24} \left(1 + q^2 + q^3 + 2q^4+2q^5+4q^6 + \cdots \right),\\
 \chi_1 &= q^\frac{143}{24} (1 + q + 2 q^2 + 3 q^3 + 4 q^4 + 6 q^5 + 9 q^6 + \cdots) \\
  \chi_2 &= q^{-\frac{1}{24}} \left(1 + q + 2 q^2 + 3 q^3 + 5 q^4 + 7 q^5 + 12 q^6 + \cdots\right),\\
 \chi_3 &= q^\frac{35}{24} (1 + 2 q + 3 q^2 + 5 q^3 + 8 q^4 + 12 q^5 + 18 q^6 + \cdots ),\\
 \chi_4 &= 1 + q + 2 q^2 + 3 q^3 + 5 q^4 + 8 q^5 + 12 q^6 + \cdots,\\
 \chi_5 &= q^\frac{1}{8} (1 + q + 2 q^2 + 3 q^3 + 6 q^4 + 8 q^5 + 13 q^6 + \cdots),\\
 \chi_6 &= q^\frac{1}{3} (1 + q + 2 q^2 + 4 q^3 + 6 q^4 + 9 q^5 + 14 q^6 + \cdots ),\\
 \chi_7 &= q^\frac{5}{8} (1 + q + 3 q^2 + 4 q^3 + 7 q^4 + 10 q^5 + 16 q^6 + \cdots ),\\
 \chi_8 &= q (1 + 2 q + 3 q^2 + 5 q^3 + 8 q^4 + 12 q^5 + 18 q^6+\cdots),\\
 \chi_9 &= q^{-\frac{1}{60}} \left( 1 + q + 2 q^2 + 3 q^3 + 5 q^4 + 7 q^5 + 11 q^6 +\cdots \right),\\
 \chi_{10} &= q^\frac{7}{120} (1 + q + 2 q^2 + 3 q^3 + 5 q^4 + 7 q^5 + 11 q^6  +\cdots ),\\
 \chi_{11} &= q^\frac{11}{60} ( 1 + q + 2 q^2 + 3 q^3 + 5 q^4 + 7 q^5 + 11 q^6 + \cdots),\\
 \chi_{12} &= q^\frac{43}{120} ( 1 + q + 2 q^2 + 3 q^3 + 5 q^4 + 7 q^5 + 12 q^6 + \cdots),\\
 \chi_{13} &= q^\frac{7}{12} ( 1 + q + 2 q^2 + 3 q^3 + 5 q^4 + 8 q^5 + 12 q^6  + \cdots),\\
 \chi_{14} &= q^\frac{103}{120} (  1 + q + 2 q^2 + 3 q^3 + 6 q^4 + 8 q^5 + 13 q^6 + \cdots),\\
 \chi_{15} &= q^\frac{71}{60} ( 1 + q + 2 q^2 + 4 q^3 + 6 q^4 + 9 q^5 + 14 q^6  + \cdots),\\
 \chi_{16} &= q^\frac{187}{120} ( 1 + q + 3 q^2 + 4 q^3 + 7 q^4 + 10 q^5 + 16 q^6  + \cdots),\\
 \chi_{17} &= q^\frac{119}{60} (  1 + 2 q + 3 q^2 + 5 q^3 + 8 q^4 + 12 q^5 + 18 q^6 + \cdots).\\
 \end{split}
 \end{align}

\vspace{0.2cm}
\paragraph{the Haagerup RCFT ${\cal R}_{k=9}$} Two constant matrices are presented below, 
\begin{align}
    \Lambda = \resizebox{0.9\hsize}{!}{$\diag \left\{ -\frac{7}{24}, -\frac{1}{24}, -\frac{1}{24},-\frac{7}{24}, -\frac{169}{168}, 
-\frac{151}{168}, -\frac{121}{168}, -\frac{79}{168}, -\frac{25}{168}, -\frac{127}{168}, 
-\frac{269}{264}, -\frac{251}{264}, -\frac{221}{264}, -\frac{179}{264}, -\frac{125}{264}, 
-\frac{59}{264}, -\frac{245}{264}, -\frac{155}{264}, -\frac{53}{264}, -\frac{203}{264} \right\}$} \ , \nonumber
\end{align}
\begin{align}
    \chi = \resizebox{0.8\hsize}{!}{$\left(\begin{array}{cccccccccccccccccccc}
-43 & 0 & 0 & -4 & -9555 & 5152 & -1701 & 280 & -7 & 2184 & 8184 & -5610 & 2860 & -1012 & 220 & -22 & 4895 & -550 & 11 & -1892 \\
92 & 1 & 0 & 8 & 43316 & 21350 & 5950 & 714 & 14 & 7735 & -37532 & -24101 & -11154 & -3410 & -594 & -33 & -20658 & -1617 & -22 & -6798 \\
8 & 0 & 1 & 92 & 43316 & 21350 & 5950 & 714 & 14 & 7735 & 37532 & 24101 & 11154 & 3410 & 594 & 33 & 20658 & 1617 & 22 & 6798 \\
-4 & 0 & 0 & -43 & -9555 & 5152 & -1701 & 280 & -7 & 2184 & -8184 & 5610 & -2860 & 1012 & -220 & 22 & -4895 & 550 & -11 & 1892 \\
-5 & 1 & 1 & -5 & 13 & -2 & -1 & -10 & -3 & -22 & 0 & 0 & 0 & 0 & 0 & 0 & 0 & 0 & 0 & 0 \\
2 & 1 & 1 & 2 & 52 & -27 & -50 & -25 & -2 & 26 & 0 & 0 & 0 & 0 & 0 & 0 & 0 & 0 & 0 & 0 \\
-12 & 1 & 1 & -12 & 156 & -450 & -152 & 46 & 5 & -78 & 0 & 0 & 0 & 0 & 0 & 0 & 0 & 0 & 0 & 0 \\
30 & 1 & 1 & 30 & -1300 & -2976 & 798 & 153 & -10 & -338 & 0 & 0 & 0 & 0 & 0 & 0 & 0 & 0 & 0 & 0 \\
-75 & 1 & 1 & -75 & -28900 & -5250 & 6375 & -850 & 6 & 5746 & 0 & 0 & 0 & 0 & 0 & 0 & 0 & 0 & 0 & 0 \\
14 & 0 & 0 & 14 & -468 & 225 & -50 & -6 & 6 & -179 & 0 & 0 & 0 & 0 & 0 & 0 & 0 & 0 & 0 & 0 \\
3 & -1 & 1 & -3 & 0 & 0 & 0 & 0 & 0 & 0 & 11 & 2 & 0 & -4 & -2 & -4 & -17 & -10 & -5 & -18 \\
-2 & -1 & 1 & 2 & 0 & 0 & 0 & 0 & 0 & 0 & 28 & 12 & -14 & -19 & -14 & -7 & -38 & -7 & -2 & 14 \\
7 & -1 & 1 & -7 & 0 & 0 & 0 & 0 & 0 & 0 & 114 & -36 & -91 & -64 & -14 & 2 & 127 & 28 & -1 & -62 \\
-8 & -1 & 1 & 8 & 0 & 0 & 0 & 0 & 0 & 0 & 352 & -457 & -416 & -33 & 64 & 9 & -104 & -133 & -8 & 125 \\
23 & -1 & 1 & -23 & 0 & 0 & 0 & 0 & 0 & 0 & 658 & -3220 & -780 & 656 & 131 & -14 & -2513 & 188 & 16 & -206 \\
-52 & -1 & 1 & 52 & 0 & 0 & 0 & 0 & 0 & 0 & -3672 & -16284 & 3536 & 2431 & -376 & -31 & 12512 & 612 & -32 & -766 \\
3 & 0 & 0 & -3 & 0 & 0 & 0 & 0 & 0 & 0 & -27 & -38 & 39 & -4 & -11 & 6 & 13 & -18 & 4 & 12 \\
-18 & 0 & 0 & 18 & 0 & 0 & 0 & 0 & 0 & 0 & -1268 & -201 & 702 & -366 & 54 & 9 & -1242 & 202 & 2 & -396 \\
50 & 0 & 0 & -50 & 0 & 0 & 0 & 0 & 0 & 0 & -24507 & 8010 & 1365 & -2100 & 615 & -42 & 11100 & -210 & -11 & 5180 \\
-10 & 0 & 0 & 10 & 0 & 0 & 0 & 0 & 0 & 0 & -372 & 255 & -130 & 46 & -10 & 1 & 110 & -45 & 10 & -141 
\end{array}\right)$}\ . \nonumber
\end{align}
and characters are
 \begin{align}
 \begin{split}
 \chi_{0} &= q^{\frac{41}{24}} \left(1 + q^2 + q^3 + 2 q^4 + 2 q^5 + 4 q^6 +  \cdots \right),\\
 \chi_{1} &=q^{\frac{167}{24}} \left(1 + q + 2 q^2 + 3 q^3 + 4 q^4 + 6 q^5 + 9 q^6 + \cdots \right),\\
 \chi_{2} &=  q^{-\frac{1}{24}} \left(1 + q + 2 q^2 + 3 q^3 + 5 q^4 + 7 q^5 + 11 q^6+ \cdots \right),\\
 \chi_{3} &= q^{\frac{41}{24}} \left(1 + 2 q + 3 q^2 + 5 q^3 + 8 q^4 + 12 q^5 + 18 q^6+ \cdots \right),\\
 \chi_{4} &= {q^{-\frac{1}{168}}} \left(1 + q + 2 q^2 + 3 q^3 + 5 q^4 + 7 q^5 + 12 q^6+ \cdots \right),\\
 \chi_{5} &=q^{\frac{17}{168}} \left(1 + q + 2 q^2 + 3 q^3 + 5 q^4 + 8 q^5 + 12 q^6 + \cdots \right),\\
 \chi_{6} &= q^{\frac{47}{168}} \left(1 + q + 2 q^2 + 3 q^3 + 6 q^4 + 8 q^5 + 13 q^6+ \cdots \right),\\
 \chi_{7} &=q^{\frac{89}{168}} \left(1 + q + 2 q^2 + 4 q^3 + 6 q^4 + 9 q^5 + 14 q^6 + \cdots \right),\\
 \chi_{8} &= q^{\frac{143}{168}} \left(1 + q + 3 q^2 + 4 q^3 + 7 q^4 + 10 q^5 + 16 q^6+ \cdots \right),\\
 \chi_{9} &= q^{\frac{209}{168}} \left(1 + 2 q + 3 q^2 + 5 q^3 + 8 q^4 + 12 q^5 + 18 q^6 + \cdots \right),\\
 \chi_{10} &= {q^{-\frac{5}{264}}} \left(1 + q + 2 q^2 + 3 q^3 + 5 q^4 + 7 q^5 + 11 q^6+ \cdots \right),\\
 \chi_{11} &= q^{\frac{13}{264}} \left(1 + q + 2 q^2 + 3 q^3 + 5 q^4 + 7 q^5 + 11 q^6+ \cdots \right),\\
 \chi_{12} &= q^{\frac{43}{264}} \left(1 + q + 2 q^2 + 3 q^3 + 5 q^4 + 7 q^5 + 11 q^6+ \cdots \right),\\
 \chi_{13} &= q^{\frac{85}{264}} \left(1 + q + 2 q^2 + 3 q^3 + 5 q^4 + 7 q^5 + 11 q^6+ \cdots \right),\\
 \chi_{14} &=q^{\frac{139}{264}} \left(1 + q + 2 q^2 + 3 q^3 + 5 q^4 + 7 q^5 + 12 q^6+ \cdots \right),\\
 \chi_{15} &= q^{\frac{205}{264}} \left(1 + q + 2 q^2 + 3 q^3 + 5 q^4 + 8 q^5 + 12 q^6+ \cdots \right),\\
 \chi_{16} &= q^{\frac{283}{264}} \left(1 + q + 2 q^2 + 3 q^3 + 6 q^4 + 8 q^5 + 13 q^6+ \cdots \right),\\
 \chi_{17} &= q^{\frac{373}{264}} \left(1 + q + 2 q^2 + 4 q^3 + 6 q^4 + 9 q^5 + 14 q^6+ \cdots \right),\\
 \chi_{18} &= q^{\frac{475}{264}} \left(1 + q + 3 q^2 + 4 q^3 + 7 q^4 + 10 q^5 + 16 q^6+ \cdots \right),\\
 \chi_{19} &= q^{\frac{589}{264}} \left(1 + 2 q + 3 q^2 + 5 q^3 + 8 q^4 + 12 q^5 + 18 q^6+ \cdots \right).
 \end{split}
 \end{align}

\vspace{0.2cm}
\paragraph{the Haagerup RCFT ${\cal R}_{k=10}$} For $k=10$, $\Lambda$ and $\chi$ 
become
\begin{align}
    \Lambda = \resizebox{0.9\hsize}{!}{$\diag \left\{-\frac{1}{24}, -\frac{1}{24}, -\frac{1}{24}, -\frac{1}{24}, -\frac{97}{96}, -\frac{11}{12}, -\frac{73}{96}, -\frac{13}{24}, -\frac{25}{96}, -\frac{11}{12}, -\frac{49}{96}, -\frac{49}{48}, -\frac{23}{24}, -\frac{41}{48}, -\frac{17}{24}, -\frac{25}{48}, -\frac{7}{24}, -\frac{1}{48}, -\frac{17}{24}, -\frac{17}{48}, -\frac{23}{24}, -\frac{25}{48}\right\}$}\ , \nonumber
\end{align}
\begin{align}
    \chi = \resizebox{0.8\hsize}{!}{$\left( \begin{array} {cccccccccccccccccccccc}1 &0 &0 &0 &-41600 &22464 &-7488 &1300 &-64 &22592 &-960 &36288 &-24300 &12000 &-4104 &864 &-84 &0 &-4158 &160 &-24300 &864 \\0 &1 &0 &0 &41600 &22592 &7488 &1300 &64 &22464 &960 &-36288 &-24300 &-12000 &-4158 &-864 &-84 &0 &-4104 &-160 &-24300 &-864 \\0 &0 &1 &0 &41600 &22592 &7488 &1300 &64 &22464 &960 &36288 &24300 &12000 &4158 &864 &84 &0 &4104 &160 &24300 &864 \\0 &0 &0 &1 &-41600 &22464 &-7488 &1300 &-64 &22592 &-960 &-36288 &24300 &-12000 &4104 &-864 &84 &0 &4158 &-160 &24300 &-864 \\-1 &1 &1 &-1 &-2 &16 &-14 &0 &-7 &-16 &-17 &0 &0 &0 &0 &0 &0 &0 &0 &0 &0 &0 \\0 &1 &1 &0 &48 &-7 &-40 &-26 &-8 &-15 &8 &0 &0 &0 &0 &0 &0 &0 &0 &0 &0 &0 \\-1 &1 &1 &-1 &147 &-240 &-182 &0 &12 &240 &-35 &0 &0 &0 &0 &0 &0 &0 &0 &0 &0 &0 \\1 &1 &1 &1 &0 &-2048 &0 &273 &0 &-2048 &0 &0 &0 &0 &0 &0 &0 &0 &0 &0 &0 &0 \\-1 &1 &1 &-1 &-8625 &-9200 &4575 &0 &-69 &9200 &322 &0 &0 &0 &0 &0 &0 &0 &0 &0 &0 &0 \\1 &0 &0 &1 &-48 &-15 &40 &-26 &8 &-7 &-8 &0 &0 &0 &0 &0 &0 &0 &0 &0 &0 &0 \\0 &0 &0 &0 &-3675 &1792 &-441 &0 &21 &-1792 &256 &0 &0 &0 &0 &0 &0 &0 &0 &0 &0 &0 \\1 &-1 &1 &-1 &0 &0 &0 &0 &0 &0 &0 &-7 &0 &-2 &10 &-13 &0 &-1 &-10 &-10 &0 &-13 \\0 &-1 &1 &0 &0 &0 &0 &0 &0 &0 &0 &21 &0 &-1 &-12 &-19 &-7 &-1 &-22 &-5 &23 &6 \\1 &-1 &1 &-1 &0 &0 &0 &0 &0 &0 &0 &81 &0 &-77 &-54 &-27 &0 &0 &54 &3 &0 &-27 \\0 &-1 &1 &0 &0 &0 &0 &0 &0 &0 &0 &320 &-276 &-352 &-87 &32 &20 &0 &-100 &-32 &-276 &32 \\1 &-1 &1 &-1 &0 &0 &0 &0 &0 &0 &0 &770 &-2048 &-1001 &330 &209 &0 &-1 &-330 &44 &2048 &-66 \\-1 &-1 &1 &1 &0 &0 &0 &0 &0 &0 &0 &0 &-11178 &0 &2430 &0 &-77 &0 &2430 &0 &-11178 &0 \\1 &-1 &1 &-1 &0 &0 &0 &0 &0 &0 &0 &-21252 &-47104 &18952 &4554 &-1819 &0 &0 &-4554 &-253 &47104 &506 \\-1 &0 &0 &1 &0 &0 &0 &0 &0 &0 &0 &-320 &-276 &352 &-100 &-32 &20 &0 &-87 &32 &-276 &-32 \\0 &0 &0 &0 &0 &0 &0 &0 &0 &0 &0 &-7371 &0 &2925 &-1728 &351 &0 &0 &1728 &-74 &0 &351 \\-1 &0 &0 &1 &0 &0 &0 &0 &0 &0 &0 &-21 &23 &1 &-22 &19 &-7 &1 &-12 &5 &0 &-6 \\0 &0 &0 &0 &0 &0 &0 &0 &0 &0 &0 &-3045 &2048 &-999 &320 &-53 &0 &1 &-320 &54 &-2048 &222 \\ \end{array} \right)$}\;, \nonumber 
\end{align}
and the characters are
\begin{align}
 \begin{split}
 \chi_{0} &= q^{\frac{47}{24}} \left(1 + q^2 + q^3 + 2 q^4 + 2 q^5 + 4 q^6+ \cdots \right),\\
 \chi_{1} &= q^{\frac{191}{24}} \left(1 + q + 2 q^2 + 3 q^3 + 4 q^4 + 6 q^5 + 9 q^6+ \cdots \right),\\
 \chi_{2} &= {q^{-\frac{1}{24}}} \left(1 + q + 2 q^2 + 3 q^3 + 5 q^4 + 7 q^5 + 11 q^6 + \cdots \right),\\
 \chi_{3} &= q^{\frac{47}{24}} \left(1 + 2 q + 3 q^2 + 5 q^3 + 8 q^4 + 12 q^5 + 18 q^6  + \cdots \right),\\
 \chi_{4} &= {q^{-\frac{1}{96}}} \left(1 + q + 2 q^2 + 3 q^3 + 5 q^4 + 7 q^5 + 11 q^6+ \cdots \right),\\
 \chi_{5} &=q^{\frac{1}{12}} \left(1 + q + 2 q^2 + 3 q^3 + 5 q^4 + 7 q^5 + 12 q^6 + \cdots \right),\\
 \chi_{6} &= q^{\frac{23}{96}} \left(1 + q + 2 q^2 + 3 q^3 + 5 q^4 + 8 q^5 + 12 q^6+ \cdots \right),\\
 \chi_{7} &=q^{\frac{11}{24}} \left(1 + q + 2 q^2 + 3 q^3 + 6 q^4 + 8 q^5 + 13 q^6 + \cdots \right),\\
 \chi_{8} &= q^{\frac{71}{96}} \left(1 + q + 2 q^2 + 4 q^3 + 6 q^4 + 9 q^5 + 14 q^6+ \cdots \right),\\
 \chi_{9} &= q^{\frac{13}{12}} \left(1 + q + 3 q^2 + 4 q^3 + 7 q^4 + 10 q^5 + 16 q^6+ \cdots \right),\\
 \chi_{10} &= q^{\frac{143}{96}} \left(1 + 2 q + 3 q^2 + 5 q^3 + 8 q^4 + 12 q^5 + 18 q^6+ \cdots \right),\\
 \chi_{11} &={q^{-\frac{1}{48}}} \left(1 + q + 2 q^2 + 3 q^3 + 5 q^4 + 7 q^5 + 11 q^6 + \cdots \right),\\
 \chi_{12} &= q^{\frac{1}{24}} \left(1 + q + 2 q^2 + 3 q^3 + 5 q^4 + 7 q^5 + 11 q^6+ \cdots \right),\\
 \chi_{13} &=q^{\frac{7}{48}} \left(1 + q + 2 q^2 + 3 q^3 + 5 q^4 + 7 q^5 + 11 q^6+ \cdots \right),\\
 \chi_{14} &= q^{\frac{7}{24}} \left(1 + q + 2 q^2 + 3 q^3 + 5 q^4 + 7 q^5 + 11 q^6+ \cdots \right),\\
 \chi_{15} &= q^{\frac{23}{48}} \left(1 + q + 2 q^2 + 3 q^3 + 5 q^4 + 7 q^5 + 11 q^6+ \cdots \right),\\
 \chi_{16} &=q^{\frac{17}{24}} \left(1 + q + 2 q^2 + 3 q^3 + 5 q^4 + 7 q^5 + 12 q^6 + \cdots \right),\\
 \chi_{17} &=q^{\frac{47}{48}} \left(1 + q + 2 q^2 + 3 q^3 + 5 q^4 + 8 q^5 + 12 q^6 + \cdots \right),\\
 \chi_{18} &= q^{\frac{31}{24}} \left(1 + q + 2 q^2 + 3 q^3 + 6 q^4 + 8 q^5 + 13 q^6+ \cdots \right),\\
 \chi_{19} &= q^{\frac{79}{48}} \left(1 + q + 2 q^2 + 4 q^3 + 6 q^4 + 9 q^5 + 14 q^6+ \cdots \right),\\
 \chi_{20} &= q^{\frac{49}{24}} \left(1 + q + 3 q^2 + 4 q^3 + 7 q^4 + 10 q^5 + 16 q^6+ \cdots \right),\\
 \chi_{21} &=q^{\frac{119}{48}} \left(1 + 2 q + 3 q^2 + 5 q^3 + 8 q^4 + 12 q^5 + 18 q^6 + \cdots \right).
 \end{split}
 \end{align}
\paragraph{the Haagerup RCFT ${\cal R}_{k=11}$} Finally, the two constant 
matrices for $k=11$ are 
\begin{align}
    \Lambda = \resizebox{0.9\hsize}{!}{$\diag \left\{-\frac{19}{24}, -\frac{1}{24}, -\frac{1}{24}, -\frac{19}{24}, -\frac{73}{72}, -\frac{67}{72}, -\frac{19}{24}, -\frac{43}{72}, -\frac{25}{72}, -\frac{1}{24}, -\frac{49}{72}, -\frac{19}{72}, -\frac{319}{312}, -\frac{301}{312}, -\frac{271}{312}, -\frac{229}{312}, -\frac{175}{312}, -\frac{109}{312}, -\frac{31}{312}, -\frac{253}{312}, -\frac{151}{312}, -\frac{37}{312}, -\frac{223}{312}, -\frac{85}{312}\right\}$}\ , \nonumber
\end{align}
\begin{align}
    \chi &= \resizebox{0.9\hsize}{!}{$\left(
\begin{array}{*{24}{c}}
-70 & 0 & 0 & -7 & -162 & 135 & -85 & 36 & -9 & 1 & -45 & 9 & 141 & -112 & 86 & -58 & 28 & -8 & 1 & -64 & 16 & -2 & 44 & -8 \\
8132 & 1 & 0 & 772 & 40176 & 23301 & 8924 & 1953 & 180 & 1 & 3816 & 63 & -35300 & -24304 & -12772 & -4825 & -1204 & -150 & -4 & -8480 & -584 & -5 & -4180 & -59 \\
772 & 0 & 1 & 8132 & 40176 & 23301 & 8924 & 1953 & 180 & 1 & 3816 & 63 & 35300 & 24304 & 12772 & 4825 & 1204 & 150 & 4 & 8480 & 584 & 5 & 4180 & 59 \\
-7 & 0 & 0 & -70 & -162 & 135 & -85 & 36 & -9 & 1 & -45 & 9 & -141 & 112 & -86 & 58 & -28 & 8 & -1 & 64 & -16 & 2 & -44 & 8 \\
-7 & 1 & 1 & -7 & 7 & -5 & 7 & -8 & -5 & -1 & -20 & -7 & 0 & 0 & 0 & 0 & 0 & 0 & 0 & 0 & 0 & 0 & 0 & 0 \\
20 & 1 & 1 & 20 & 32 & -2 & -20 & -29 & -12 & -1 & -8 & 1 & 0 & 0 & 0 & 0 & 0 & 0 & 0 & 0 & 0 & 0 & 0 & 0 \\
-85 & 1 & 1 & -85 & 162 & -135 & -162 & -36 & 9 & 1 & 45 & -9 & 0 & 0 & 0 & 0 & 0 & 0 & 0 & 0 & 0 & 0 & 0 & 0 \\
344 & 1 & 1 & 344 & 304 & -1279 & -344 & 195 & 40 & -1 & -384 & 2 & 0 & 0 & 0 & 0 & 0 & 0 & 0 & 0 & 0 & 0 & 0 & 0 \\
-1735 & 1 & 1 & -1735 & -2138 & -7285 & 1735 & 764 & -100 & -1 & 1618 & 13 & 0 & 0 & 0 & 0 & 0 & 0 & 0 & 0 & 0 & 0 & 0 & 0 \\
8924 & 1 & 1 & 8924 & -40176 & -23301 & 17828 & -1953 & -180 & 2 & -3816 & -63 & 0 & 0 & 0 & 0 & 0 & 0 & 0 & 0 & 0 & 0 & 0 & 0 \\
-189 & 0 & 0 & -189 & -735 & 112 & 189 & -140 & 35 & 0 & -31 & 20 & 0 & 0 & 0 & 0 & 0 & 0 & 0 & 0 & 0 & 0 & 0 & 0 \\
2808 & 0 & 0 & 2808 & -19760 & 9984 & -2808 & 260 & 40 & 0 & 2080 & -61 & 0 & 0 & 0 & 0 & 0 & 0 & 0 & 0 & 0 & 0 & 0 & 0 \\
9 & -1 & 1 & -9 & 0 & 0 & 0 & 0 & 0 & 0 & 0 & 0 & 6 & -2 & 5 & -4 & -1 & -4 & -1 & -6 & -11 & -2 & -11 & -6 \\
-4 & -1 & 1 & 4 & 0 & 0 & 0 & 0 & 0 & 0 & 0 & 0 & 18 & 7 & -2 & -11 & -14 & -10 & -2 & -20 & -12 & -1 & 10 & 1 \\
35 & -1 & 1 & -35 & 0 & 0 & 0 & 0 & 0 & 0 & 0 & 0 & 84 & 0 & -47 & -56 & -28 & -6 & 0 & 48 & 15 & 0 & -12 & -8 \\
-108 & -1 & 1 & 108 & 0 & 0 & 0 & 0 & 0 & 0 & 0 & 0 & 262 & -147 & -278 & -133 & 14 & 19 & 2 & 24 & -44 & -4 & -66 & 4 \\
386 & -1 & 1 & -386 & 0 & 0 & 0 & 0 & 0 & 0 & 0 & 0 & 790 & -1316 & -982 & 80 & 203 & 30 & -2 & -812 & -17 & 4 & 374 & -12 \\
-1408 & -1 & 1 & 1408 & 0 & 0 & 0 & 0 & 0 & 0 & 0 & 0 & 1212 & -7472 & -1572 & 1769 & 364 & -83 & -4 & 1824 & 312 & -3 & -1716 & 4 \\
5534 & -1 & 1 & -5534 & 0 & 0 & 0 & 0 & 0 & 0 & 0 & 0 & -6096 & -34104 & 6652 & 6508 & -1064 & -190 & 5 & 7240 & -848 & -4 & 6160 & 16 \\
-52 & 0 & 0 & 52 & 0 & 0 & 0 & 0 & 0 & 0 & 0 & 0 & -74 & -140 & 114 & 12 & -42 & 12 & 2 & -113 & 12 & 3 & -66 & -2 \\
624 & 0 & 0 & -624  & 0 & 0 & 0 & 0 & 0 & 0 & 0 & 0 & -2304 & -1104 & 1832 & -684 & -36 & 64 & -3 & 192 & 145 & -4 & 297 & 24 \\
-5044 & 0 & 0 & 5044 & 0 & 0 & 0 & 0 & 0 & 0 & 0 & 0 & -33638 & 3794 & 9558 & -6456 & 1554 & -81 & -2 & 10152 & -852 & 7 & 738 & -68 \\
117 & 0 & 0 & -117 & 0 & 0 & 0 & 0 & 0 & 0 & 0 & 0 & -531 & 294 & -47 & -72 & 63 & -22 & 3 & -206 & 27 & 2 & -98 & 18 \\
-2236 & 0 & 0 & 2236 & 0 & 0 & 0 & 0 & 0 & 0 & 0 & 0 & -16890 & 11116 & -5318 & 1720 & -322 & 20 & 2 & -1766 & 292 & -4 & 2398 & -57
\end{array}\right)$}\;, \nonumber
\end{align}
and the characters are
\begin{align}
 \begin{split}
 \chi_{0} &= q^{\frac{53}{24}} \left(1 + q^2 + q^3 + 2 q^4 + 2 q^5 + 4 q^6 + \cdots \right),\\
 \chi_{1} &= q^{\frac{215}{24}} \left(1 + q + 2 q^2 + 3 q^3 + 4 q^4 + 6 q^5 + 9 q^6+ \cdots \right),\\
 \chi_{2} &={q^{-\frac{1}{24}}} \left(1 + q + 2 q^2 + 3 q^3 + 5 q^4 + 7 q^5 + 11 q^6 + \cdots \right),\\
 \chi_{3} &= q^{\frac{53}{24}} \left(1 + 2 q + 3 q^2 + 5 q^3 + 8 q^4 + 12 q^5 + 18 q^6+ \cdots \right),\\
 \chi_{4} &={q^{-\frac{1}{72}}} \left(1 + q + 2 q^2 + 3 q^3 + 5 q^4 + 7 q^5 + 11 q^6 + \cdots \right),\\
 \chi_{5} &=q^{\frac{5}{72}} \left(1 + q + 2 q^2 + 3 q^3 + 5 q^4 + 7 q^5 + 11 q^6 + \cdots \right),\\
 \chi_{6} &=q^{\frac{5}{24}} \left(1 + q + 2 q^2 + 3 q^3 + 5 q^4 + 7 q^5 + 12 q^6 + \cdots \right),\\
 \chi_{7} &=q^{\frac{29}{72}} \left(1 + q + 2 q^2 + 3 q^3 + 5 q^4 + 8 q^5 + 12 q^6 + \cdots \right),\\
 \chi_{8} &=q^{\frac{47}{72}} \left(1 + q + 2 q^2 + 3 q^3 + 6 q^4 + 8 q^5 + 13 q^6 + \cdots \right),\\
 \chi_{9} &=q^{\frac{23}{24}} \left(1 + q + 2 q^2 + 4 q^3 + 6 q^4 + 9 q^5 + 14 q^6 + \cdots \right),\\
 \chi_{10} &=q^{\frac{95}{72}} \left(1 + q + 3 q^2 + 4 q^3 + 7 q^4 + 10 q^5 + 16 q^6 + \cdots \right),\\
 \chi_{11} &=  q^{\frac{125}{72}} \left(1 + 2 q + 3 q^2 + 5 q^3 + 8 q^4 + 12 q^5 + 18 q^6+ \cdots \right),\\
 \chi_{12} &={q^{-\frac{7}{312}}} \left(1 + q + 2 q^2 + 3 q^3 + 5 q^4 + 7 q^5 + 11 q^6  + \cdots \right),\\
 \chi_{13} &= q^{\frac{11}{312}} \left(1 + q + 2 q^2 + 3 q^3 + 5 q^4 + 7 q^5 + 11 q^6+ \cdots \right),\\
 \chi_{14} &=q^{\frac{41}{312}} \left(1 + q + 2 q^2 + 3 q^3 + 5 q^4 + 7 q^5 + 11 q^6 + \cdots \right),\\
 \chi_{15} &= q^{\frac{83}{312}} \left(1 + q + 2 q^2 + 3 q^3 + 5 q^4 + 7 q^5 + 11 q^6+ \cdots \right),\\
 \chi_{16} &=q^{\frac{137}{312}} \left(1 + q + 2 q^2 + 3 q^3 + 5 q^4 + 7 q^5 + 11 q^6 + \cdots \right),\\
 \chi_{17} &=q^{\frac{203}{312}} \left(1 + q + 2 q^2 + 3 q^3 + 5 q^4 + 7 q^5 + 11 q^6  + \cdots \right),\\
 \chi_{18} &=q^{\frac{281}{312}} \left(1 + q + 2 q^2 + 3 q^3 + 5 q^4 + 7 q^5 + 12 q^6 + \cdots \right),\nonumber
 \end{split}
 \end{align}
 \begin{align}
 \begin{split}
  \chi_{19} &=q^{\frac{371}{312}} \left(1 + q + 2 q^2 + 3 q^3 + 5 q^4 + 8 q^5 + 12 q^6 + \cdots \right), 
  \\
 \chi_{20} &= q^{\frac{473}{312}} \left(1 + q + 2 q^2 + 3 q^3 + 6 q^4 + 8 q^5 + 13 q^6+ \cdots \right),\\
 \chi_{21} &= q^{\frac{587}{312}} \left(1 + q + 2 q^2 + 4 q^3 + 6 q^4 + 9 q^5 + 14 q^6+ \cdots \right),\\
 \chi_{22} &=q^{\frac{713}{312}} \left(1 + q + 3 q^2 + 4 q^3 + 7 q^4 + 10 q^5 + 16 q^6 + \cdots \right),\\
 \chi_{23} &=q^{\frac{851}{312}} \left(1 + 2 q + 3 q^2 + 5 q^3 + 8 q^4 + 12 q^5 + 18 q^6 + \cdots \right). \nonumber
 \end{split}
 \end{align}

	\section{Conclusion and Future works}
	
	We have explored the non-unitary version of bulk-boundary correspondence with concrete examples. For those examples, we have confirmed that some basic dictionaries  work well even in the non-unitary cases. By applying the correspondence, we obtain a new class of non-unitary RCFTs from known bulk rank-0 theories. By applying the correspondence in the reverse direction,  one may construct new classes of 3D rank-0 SCFTs from known non-unitary RCFTs as studied in \cite{Gang:to-appear}.  
	
	Here we list a set of related problems which we hope to address in the near future. 
	\paragraph{Chiral algebra of $\CR_k$} In this paper, we analyze the correspondence using modular data and   characters. Another important mathematical object  of a 2D RCFT is its underlying chiral algebra.  Each character is associated to an irreducible representation of the algebra. It would be interesting to identify the chiral algebra of the $\CR_k$ by analyzing the boundary operator algebra of the twisted S-fold SCFT using the techniques developed in \cite{Gaiotto:2016wcv, Gaiotto:2017euk, Costello:2018fnz, Costello:2018swh, Costello:2020ndc,Creutzig:2021ext, Ballin:2022rto, Garner:2022vds, Garner:2023pmt, Yoshida:2023wyt, Beem:2023dub}. For $k=3$, the $\CR_k$ is a product of two Virasoro minimal models, $M(3,5)$ and $M(2,5)$, and the chiral algebra is expected to be two copies of Virasoro algebra.  
	
    \paragraph{Hecke transformation to unitary RCFTs} Hecke transformation acts on a vvmf to another vvmf. Applying the transformation, we may hope to find other admissible RCFT characters from the characters of $\CR_k$ we obtained. For some cases, the Hecke transform relates non-unitary RCFTs to unitary ones. It would be interesting to see if one can generate a new class of   admissible characters of unitary RCFTs by  applying Heck transformations to the characters of $\CR_k$.

	\paragraph{More general classes of Haagerup RCFTs} When $k =  4m^2+4m+3$  ($m \in \mathbb{Z}_{\geq 0}$), the modular data of $\CS_k$ is related to  an unitary Haagerup-Izumi modular data $\CD^{\omega}{\rm Hg}_{2m+1}$  with $\omega = \pm 2$ by a Hecke transformation (modulo decoupled $U(1)_{\pm 2}$). One may also consider non-unitary modular data   obtained by acting a Hecke transformation on  $\CD^{\omega}{\rm Hg}_{2m+1}$ with $\omega \neq \pm 2$. It would be interesting to find 3D rank-0 SCFTs which realize  non-unitary TQFTs with the modular data via a topological twisting. Since the existence of non-unitary 3D TQFTs with the modular data is obscure, there is no guarantee that such rank-0 SCFTs should exit.

	%%%%%%%%%%%%%%%%%%%%%%%%%%%%%%%%%%%%%%%%%%%%%%%%%%%
	\section*{Acknowledgements}
	We would like to thank  Heeyeon Kim, Kimyeong Lee, and Brandon Rayhaun for the useful discussion.
	The work of DG and DK is supported in part by the National Research Foundation of Korea grant  NRF-2022R1C1C1011979. DG also acknowledges support by Creative-Pioneering Researchers Program through Seoul National University. SL is supported in part by KIAS Grant PG056502.

\newpage 

	\appendix
	
		\section{Dual Abelian description of $\CS_k$ from 3D-3D correspondence } \label{App : dual of Sk}
	
    3D-3D correspondence predicts following  IR duality  \cite{Terashima:2011qi,Gang:2013sqa, Gang:2021hrd}
	\begin{align}
	\begin{split}
	&\left( \widetilde{\CS}_k  \textrm{ in \eqref{S-fold from T[SU(2)]} and \eqref{Sk/Z2}} \right)
	\\
	&\simeq \CT_{DGG}[(\Sigma_{1,1}\times S^1)_{\varphi = ST^k \simeq LR^{k-2}}; A = \lambda ]	\;. \label{3D-3D correpondence}
	\end{split}
	\end{align}
    Here $\CT_{DGG}[N;A]$ is the 3D $\CN=2$  gauge theory labeled by an ideally triangulated  3-manifold $N$ with a torus boundary proposed in \cite{Dimofte:2011ju}.  The theory also depends on  a choice of primitive boundary  1-cycle $A \in H_1 (\partial N, \mathbb{Z})$.  Here  $(\Sigma_{1,1} \times S^1)_{\varphi }$ denotes the once-punctured torus bundle with a monodromy $\varphi \in SL(2,\mathbb{Z})$ (mapping class group of $\Sigma_{1,1}$, once-punctured torus):
    \begin{align}
    \begin{split}
    &(\Sigma_{1,1} \times S^1)_{\varphi} = (\Sigma_{1,1} \times [0,1])/\sim 
    \\
     &\textrm{ with the equivalence relation  } (x \in \Sigma_{1,1}, 0) \sim (\varphi(x),1)\;.
    \end{split}
    \end{align}
     $A=\lambda$ is a boundary 1-cycle circling the puncture on  $\Sigma_{1,1}$. The $SL(2,\mathbb{Z})$ matrices $S,T=R$ and $L$ are
    \begin{align}
    	S = \begin{pmatrix}
    		0 & 1 \\
    		-1 & 0 
    	\end{pmatrix}\;, \quad  	T =R= \begin{pmatrix}
    	1 & 0 \\
    	1 & 1 
    \end{pmatrix}\;,  \quad L = \begin{pmatrix}
    1 & 1 \\
    0 & 1 
\end{pmatrix}\;.
    \end{align}
    Topology of the torus bundle depends only on the conjugacy class of $\varphi$ and note that $ST^{k} \simeq  LR^{k-2}$ up to a conjugation. 
	
	One convenient way to identify a field theory description of $\CT_{DGG}[N;A]$ is calculating  $SL(2,\mathbb{C})$  Chern-Simons theory partition functions  $\CZ^{SL(2,\mathbb{C}) \textrm{ on }(N,A)}_{(k,\sigma)}(M_A)$ on the 3-manifold $N$ with various $k\in \mathbb{Z}$:
	\begin{align}
	\begin{split}
    &\CZ^{SL(2,\mathbb{C}) \textrm{ on }(N,A)}_{(k,\sigma)}(M_A) 
    \\
    &= \int \frac{[D\CA][D\bar{\CA}]}{(\rm gauge)}\bigg{|}_{b.c} \exp \left[ i  \left(\int_N \frac{k+\sigma}{8\pi} \textrm{Tr} (\CA \wedge d \CA + \frac{2}3 \CA^3) +\frac{k-\sigma}{8\pi} \textrm{Tr} (\bar{\CA} \wedge d \bar{\CA} + \frac{2}3 \bar{\CA}^3)  \right) \right]\;,
    \\
	 &\textrm{with the  boundary condition}\;:\; P \exp\left( \oint_A  \CA \right)\bigg{|}_{\partial N} \sim \begin{pmatrix} e^{M_A /2} & * \\ 0 & e^{-M_A /2}\end{pmatrix} \in SL(2,\mathbb{C})\;. \label{SL(2,C) CS ptn}
	 \end{split}
	\end{align}
	According to 3D-3D relations \cite{Terashima:2011qi,Dimofte:2011ju,Dimofte:2011py,Lee:2013ida,Yagi:2013fda,Cordova:2013cea,Dimofte:2014zga}, the CS partition functions with  $k (\in \mathbb{Z})$  and $\sigma (\in \mathbb{R} \;\textrm{ or } i \mathbb{R})$ compute various SUSY partition functions of $\CT_{DGG}[N;A]$. 
	Especially when $\textrm{with } k=1$ and  $\sigma= \frac{1-b^2}{1+b^2} $,  the  CS partition function is related  to the squashed 3-sphere partition function of $\CT_{DGG}[N;A]$ in the following way
	\begin{align}
	\begin{split}
	&\left( \CZ_{S^3_b} (m,\nu)  \textrm{ of } \CT_{DGG}[N;A] \right) =\sqrt{ \frac{|H_1 (N, \mathbb{Z}_2)|}2} \CZ_{(1, \frac{1-b^2}{1+b^2})}^{SL(2,\mathbb{C}) \textrm{ on } (N,A)} (M_A )\;,
	\\
	&\textrm{with }  m+(i \pi +\frac{\hbar}2) \nu = \begin{cases}
		 \frac{M_A}2\;, \quad A \in \textrm{Ker}\left(i_* : H_1 (\partial N, \mathbb{Z}) \rightarrow H_1 (N, \mathbb{Z}_2)\right)
		 \\
		 M_A \;, \quad \textrm{otherwise}
	\end{cases}\;.
	\end{split} \label{3D-3D correpondence-2}
	\end{align}
	The $\CT_{DGG}[N;A]$ has a $U(1)_A$ symmetry associated to the torus boundary and $(m,\nu)$ is the (rescaled real mass, R-symmetry  mixing parameter) of the flavor symmetry. The boundary 1-cycle  $\lambda$ of $N=(\Sigma_{1,1}\times S^1)_\varphi$  belongs to the  kernel of the map $i_* : H_1 (\partial N, \mathbb{Z}) \rightarrow H_1 (N, \mathbb{Z}_2)$. As we will see below, the Chern-Simons partition function can be given as a finite dimensional integral which is identical to a squashed 3-sphere partition function localization integral  of a 3D $\CN=2$ gauge theory, which can be identified as $\CT_{DGG}[N,A]$.
	
\begin{figure}[h!]
		\centering
		\includegraphics[width=0.60\linewidth,height=0.46\linewidth]{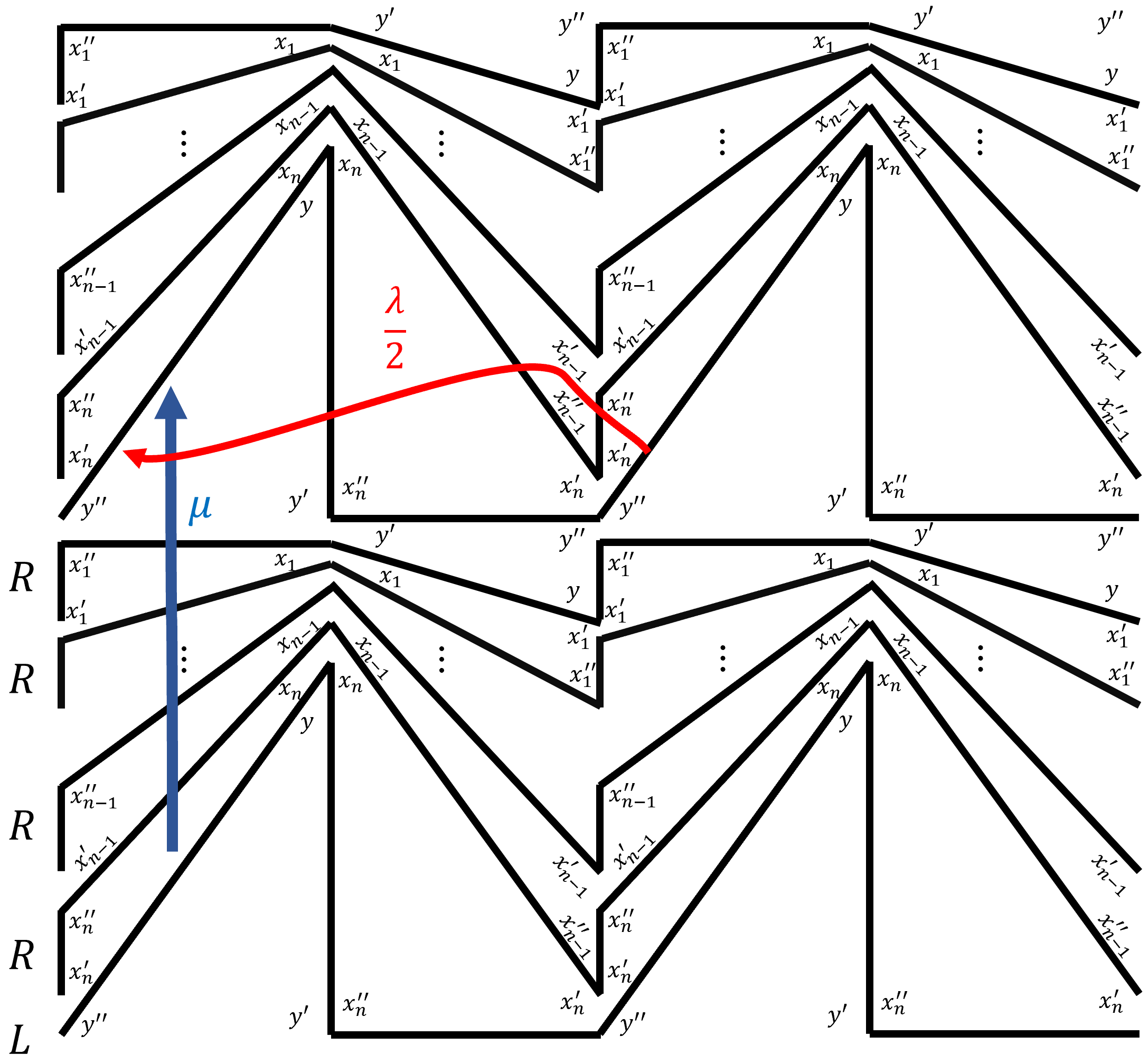}
		\caption{Farey tessellation of once punctured torus bundle with monodromy $LR^n$.}
		\label{fig:tessellation}
	\end{figure}

			The Chern-Simons partition function can be computed using the so-called state-integral model based on an ideal triangulation of $N$ \cite{HIKAMI_2001,hikami2007generalized,Dimofte:2009yn,EllegaardAndersen:2011vps,Dimofte:2012qj,Dimofte:2011gm,Dimofte:2014zga}. 
	There is a canonical way to ideally triangulate the once-puncture torus bundle $(\Sigma_{1,1}\times S^1)_{\varphi = LR^{n=k-2}}$ with  $(n+1)$ ideal tetrahedra  \cite{2004math......6242G} .  Farey tessellation from the triangulation is given as figure \ref{fig:tessellation}.
	From the farey tessellation, we can read off the following gluing data
	\begingroup
	\allowdisplaybreaks
	\begin{align}
		\begin{split}
			M_{\mu} &= Y'' - \sum_{i=1}^n X_i, \\
		    \frac{M_\lambda}{2} &= X_n ' + X_{n-1}'' - X_n - Y = (i \pi + \frac{\hbar}{2}) - 2 X_n - X_n '' + X_{n-1}'' - Y, \\
			C_1 &= 2 Y'' + 2 X_n' + X_1'' + X_{n-1}'' - 2(i \pi + \frac{\hbar}{2}) = 2 Y'' - 2 X_n + X_1'' + X_{n-1}'' - 2 X_n'',\\
			C_2 &= Y + 2 X_1' + X_2 '' -2(i \pi + \frac{\hbar}{2})= Y - 2 X_1 - 2 X_1'' + X_2'', \\
			&\vdots \\
			C_{n-1} &= X_{n-3}'' + 2 X_{n-2}' + X_{n-1}'' - 2 (i \pi + \frac{\hbar}{2}) = - 2 X_{n-2} + X_{n-3}'' - 2 X_{n-2}'' + X_{n-1}'',\\
			C_n &= X_{n-2}'' + 2 X_{n-1}' + X_n '' - 2(i \pi + \frac{\hbar}{2})= - 2 X_{n-1} + X_{n-2}''-2 X_{n-1}'' + X_n '',\\
			C_{n+1} &= 2 Y' + Y + X_n'' +2 \sum_{i=1}^n X_i - 2(i \pi + \frac{\hbar}{2}) = - Y - 2 Y'' + X_n'' + 2 \sum_{i=1}^n X_i.
		\end{split}
	\end{align}
\endgroup
In the construction of $\CT_{DGG}$ \cite{Dimofte:2011ju}, the `easy' internal edges, $C_1, \ldots, C_{n}$, correspond to chiral primary operators appearing in the superpotential \eqref{monopole operators}.  
  Generally, the gluing data is known to have a symplectic structure  \cite{neumann1985volumes} and one  consider following  affine symplectic transformation 
	\begin{align}
		\begin{pmatrix}
			\frac{M_\lambda}{2} \\ C_1 \\ \vdots \\ C_n \\ - M_\mu \\ \Gamma_1 \\ \vdots \\ \Gamma_n
		\end{pmatrix} = 
		\left(\begin{array}{ccc|ccc}
			& & & & & \\ 
			&\mbox{\Huge $A$}& & &\mbox{\Huge $B$}& \\ 
			& & & & & \\ \hline 
			& & & & & \\ 
			&\mbox{\Huge $C$}& & &\mbox{\Huge $D$}& \\ 
			& & & & & \\ 
		\end{array}\right) \begin{pmatrix}
			Y \\ X_1 \\ \vdots \\ X_n \\ Y'' \\ X_1'' \\ \vdots \\ X_n''
		\end{pmatrix} - (i \pi + \hbar/2) \begin{pmatrix}
			-1 \\ 0 \\ \vdots \\ 0 \\ 0 \\ 0 \\ \vdots \\ 0
		\end{pmatrix},
		\label{eq:gluing_data}
	\end{align}
	where two block matrices $A, B$ are
	\begin{align}
		\begin{split}
			A_{(n+1)\times(n+1)} = \begin{pmatrix}
				-1 & 0 &0 & \cdots &  0 & -2 \\
				0 & 0 & 0 & \cdots & 0 & -2 \\
				1 & -2 & 0 & \cdots & 0 & 0 \\
				0 & 0 & -2 & 0 & \cdots & 0 \\
				& & \vdots & & & \\
				0 & 0 & \cdots & 0 & -2 & 0
			\end{pmatrix},  \quad B_{(n+1)\times(n+1)} = \begin{pmatrix}
				0 & 0 & \cdots & 0 & 0& 1 & -1 \\
				2 & 1 & 0 & \cdots & 0 & 1 & -2 \\
				0 & -2 & 1  & 0 & 0 & \cdots & 0 \\
				0 & 1 & -2 & 1 & 0 & \cdots & 0 \\
				& & & \vdots  & & & \\
				0 & \cdots & 0 & 1 & -2 & 1 & 0\\
				0 & \cdots & 0 & 0 & 1 & -2 & 1
			\end{pmatrix}.
		\end{split}
		\label{eq:AandB}
	\end{align}
	for $n \geq 3$.\footnote{When $n$ is 1 or 2, the exact expression for $A$ and $B$ is different from \eqref{eq:AandB}. However, the final results agrees with \eqref{eq:chernsimonslevel} and \eqref{eq:stateintegral} for the two cases.} Using the symplectic gluing data,  the state integral can be  given as \cite{Dimofte:2012qj}
	\begin{align}
		\begin{split}
			&\CZ_{(1, \frac{1-b^2}{1+b^2})}^{SL(2,\mathbb{C}) \textrm{ on }(N,A)} (M_\lambda)
			\\
			&= \frac{1}{\sqrt{\det B}} \int \frac{d^{n+1}\mathbf{Z}}{(2 \pi \hbar)^\frac{n+1}{2}} \exp \left[ \frac{1}{2 \hbar} \left( \mathbf{X}^T D B^{-1} \mathbf{X} - 2 \mathbf{Z}^T  B^{-1} \mathbf{X}+ \mathbf{Z}^T  B^{-1} A \mathbf{Z}  \right)\right] \prod_{i=1}^{n+1} \psi_\hbar (Z_i), \\
			&= \frac{e^{f(M_\lambda, \hbar)}}{\sqrt{2}}  \int \frac{d^{n+1}\mathbf{Z}}{(2\pi\hbar)^{\frac{n+1}{2}}} \exp \bigg{[} \frac{1}{2\hbar}  \mathbf{Z}^T B^{-1}A\mathbf{Z} + \frac{1}{\hbar } (X- i \pi - \frac{\hbar }2)  ( \frac{n}{2} Z_1 + \sum_{k=2}^{n} kZ_{k+1}) \bigg{]} \prod_{i=1}^{n+1} \psi_\hbar (Z_i)\;.
		\end{split}
	\end{align}
    Here $e^{f(M_\lambda, \hbar )}$ is a  factor independent of $\mathbf{Z}$ which can be interpreted as  a contribution from background mixed Chern-Simons levels of $U(1)_A\times U(1)_R$ symmetries in the field theory $\CT_{DGG}[N;A]$.  In the last line, we let
    \begin{align}
    	\mathbf{X}= \left(\frac{M_\lambda}2 - i \pi -\frac{\hbar}2, C_1, \ldots, C_n \right)\bigg{|}_{C_{1\leq i\leq n} =0}\;.
    \end{align}
  The matrix $B^{-1} A$ is 
	\begin{align}
	B^{-1} A = \begin{pmatrix}
		\frac{n}{2} & 0 & 1 & 2 & 3 & \ldots & n-1\\
		0 & 2 & 2 & 2 & 2& \ldots &  2\\
		1 & 2 & 4 & 4 & 4 & \ldots  & 4\\
		2 & 2 & 4 & 6 & 6 & \ldots  & 6\\
		3 & 2 & 4 & 6 & 8 & \ldots &8  \\
		\ldots & \ldots & \ldots & \ldots & \ldots & \ldots & \ldots \\
		n-1 & 2 & 4 & 6 & 8 & \ldots & 2n
	\end{pmatrix}.
	\label{eq:chernsimonslevel}
\end{align}
Rescaling the $Z_1 $ to $2 Z_1$, we have (ignoring the  factor $e^{f(M_\lambda, \hbar)}$)
	\begin{align}
		\begin{split}
			&\CZ_{(1, \frac{1-b^2}{1+b^2})}^{SL(2,\mathbb{C}) \textrm{ on }(N,A)} (M_\lambda)
             \\
			&= \sqrt{2} \int \frac{d^{n+1}\mathbf{Z}}{(2\pi\hbar)^{\frac{n+1}{2}}} \exp \bigg{[} \frac{1}{2\hbar}  \mathbf{Z}^T K \mathbf{Z}   + \frac{1}\hbar (\frac{M_\lambda}2- i \pi -\frac{\hbar}2)  ( n Z_1 + \sum_{k=2}^{n} kZ_{k+1} ) \bigg{]}  \prod_{i=1}^{n+1} \psi_\hbar (Z_i)\;. \label{eq:stateintegral}
		\end{split}
	\end{align}
Here $K$ is the mixed Chern-Simons level in \eqref{K-matrix}. Then combining the 3D-3D relations in \eqref{3D-3D correpondence} and \eqref{3D-3D correpondence-2} and the following facts,
\begin{align}
&\left( \CZ_{S^3_b} \textrm{ of }\CS_k^{/\mathbb{Z}_2}\right) =  \begin{cases}
	2\times  \left( \CZ_{S^3_b} \textrm{ of }\CS_k \right), \quad k\in 2\mathbb{Z}
	\\
	\sqrt{2} \times \left( \CZ_{S^3_b} \textrm{ of }\CS_k \right), \quad   k\in 2\mathbb{Z}+1
\end{cases}
\\
&H_1 \left((\Sigma_{1,1}\times S^1)_{\varphi = LR^{n=k-2}}, \mathbb{Z}_2 \right) =   \begin{cases}
     \mathbb{Z}_2\times \mathbb{Z}_2\;, \quad k\in 2\mathbb{Z}
	\\
	\mathbb{Z}_2\;, \quad   k\in 2\mathbb{Z}+1
\end{cases}\;,
\end{align}
we have 
\begin{align}
\left( \CZ_{S^3_b}  (m, \nu) \textrm{ of }\CS_k \textrm{ in \eqref{S-fold from T[SU(2)]}}\right) =  \left(\CZ_{S^3_b}(m, \nu) \textrm{ in }\eqref{3-sphere of Sk}\right)\;.
\end{align} 
It strongly supports the proposed IR duality between \eqref{S-fold from T[SU(2)]} and \eqref{S-fold from abelian}.

	\bibliographystyle{ytphys}
	\bibliography{ref}

\providecommand{\href}[2]{#2}\begingroup\raggedright\begin{thebibliography}{10}

\bibitem{zbMATH04092352}
E.~Witten, ``Quantum field theory and the {Jones} polynomial,''
  \href{http://dx.doi.org/10.1007/BF01217730}{{\em Commun. Math. Phys.}
  {\bfseries 121} no.~3, (1989) 351--399}.

\bibitem{moore1989lectures}
G.~Moore and N.~Seiberg, ``Lectures on rcft (rational conformal field
  theory),'' tech. rep., Institute for Advanced Study, Princeton, NJ (USA);
  Yale Univ., New Haven, CT~…, 1989.

\bibitem{moore1989classical}
G.~Moore and N.~Seiberg, ``Classical and quantum conformal field theory,'' {\em
  Communications in Mathematical Physics} {\bfseries 123} (1989) 177--254.

\bibitem{turaev1992modular}
V.~G. Turaev, ``Modular categories and 3-manifold invariants,'' {\em
  International Journal of Modern Physics B} {\bfseries 6} no.~11n12, (1992)
  1807--1824.

\bibitem{Gang:2018huc}
D.~Gang and M.~Yamazaki, ``{Three-dimensional gauge theories with supersymmetry
  enhancement},'' \href{http://dx.doi.org/10.1103/PhysRevD.98.121701}{{\em
  Phys. Rev. D} {\bfseries 98} no.~12, (2018) 121701},
  \href{http://arxiv.org/abs/1806.07714}{{\ttfamily arXiv:1806.07714
  [hep-th]}}.

\bibitem{Gang:2021hrd}
D.~Gang, S.~Kim, K.~Lee, M.~Shim, and M.~Yamazaki, ``{Non-unitary TQFTs from 3D
  $ \mathcal{N} $ = 4 rank 0 SCFTs},''
  \href{http://dx.doi.org/10.1007/JHEP08(2021)158}{{\em JHEP} {\bfseries 08}
  (2021) 158}, \href{http://arxiv.org/abs/2103.09283}{{\ttfamily
  arXiv:2103.09283 [hep-th]}}.

\bibitem{Gang:2022kpe}
D.~Gang and D.~Kim, ``{Generalized non-unitary Haagerup-Izumi modular data from
  3D S-fold SCFTs},'' \href{http://dx.doi.org/10.1007/JHEP03(2023)185}{{\em
  JHEP} {\bfseries 03} (2023) 185},
  \href{http://arxiv.org/abs/2211.13561}{{\ttfamily arXiv:2211.13561
  [hep-th]}}.

\bibitem{Bantay:2005vk}
P.~Bantay and T.~Gannon, ``{Conformal characters and the modular
  representation},''
  \href{http://dx.doi.org/10.1088/1126-6708/2006/02/005}{{\em JHEP} {\bfseries
  02} (2006) 005}, \href{http://arxiv.org/abs/hep-th/0512011}{{\ttfamily
  arXiv:hep-th/0512011}}.

\bibitem{Ballin:2022rto}
A.~Ballin and W.~Niu, ``{3d Mirror Symmetry and the $\beta\gamma$ VOA},''
  \href{http://arxiv.org/abs/2202.01223}{{\ttfamily arXiv:2202.01223
  [hep-th]}}.

\bibitem{Gaiotto:2008ak}
D.~Gaiotto and E.~Witten, ``{S-Duality of Boundary Conditions In N=4 Super
  Yang-Mills Theory},''
  \href{http://dx.doi.org/10.4310/ATMP.2009.v13.n3.a5}{{\em Adv. Theor. Math.
  Phys.} {\bfseries 13} no.~3, (2009) 721--896},
  \href{http://arxiv.org/abs/0807.3720}{{\ttfamily arXiv:0807.3720 [hep-th]}}.

\bibitem{Assel:2018vtq}
B.~Assel and A.~Tomasiello, ``{Holographic duals of 3d S-fold CFTs},''
  \href{http://dx.doi.org/10.1007/JHEP06(2018)019}{{\em JHEP} {\bfseries 06}
  (2018) 019}, \href{http://arxiv.org/abs/1804.06419}{{\ttfamily
  arXiv:1804.06419 [hep-th]}}.

\bibitem{Garozzo:2018kra}
I.~Garozzo, G.~Lo~Monaco, and N.~Mekareeya, ``{The moduli spaces of $S$-fold
  CFTs},'' \href{http://dx.doi.org/10.1007/JHEP01(2019)046}{{\em JHEP}
  {\bfseries 01} (2019) 046}, \href{http://arxiv.org/abs/1810.12323}{{\ttfamily
  arXiv:1810.12323 [hep-th]}}.

\bibitem{Garozzo:2019hbf}
I.~Garozzo, G.~Lo~Monaco, and N.~Mekareeya, ``{Variations on $S$-fold CFTs},''
  \href{http://dx.doi.org/10.1007/JHEP03(2019)171}{{\em JHEP} {\bfseries 03}
  (2019) 171}, \href{http://arxiv.org/abs/1901.10493}{{\ttfamily
  arXiv:1901.10493 [hep-th]}}.

\bibitem{Garozzo:2019ejm}
I.~Garozzo, G.~Lo~Monaco, N.~Mekareeya, and M.~Sacchi, ``{Supersymmetric
  Indices of 3d $S$-fold SCFTs},''
  \href{http://dx.doi.org/10.1007/JHEP08(2019)008}{{\em JHEP} {\bfseries 08}
  (2019) 008}, \href{http://arxiv.org/abs/1905.07183}{{\ttfamily
  arXiv:1905.07183 [hep-th]}}.

\bibitem{Beratto:2020qyk}
E.~Beratto, N.~Mekareeya, and M.~Sacchi, ``{Marginal operators and
  supersymmetry enhancement in 3d $S$-fold SCFTs},''
  \href{http://dx.doi.org/10.1007/JHEP12(2020)017}{{\em JHEP} {\bfseries 12}
  (2020) 017}, \href{http://arxiv.org/abs/2009.10123}{{\ttfamily
  arXiv:2009.10123 [hep-th]}}.

\bibitem{Arav:2021gra}
I.~Arav, J.~P. Gauntlett, M.~M. Roberts, and C.~Rosen, ``{Marginal deformations
  and RG flows for type IIB S-folds},''
  \href{http://dx.doi.org/10.1007/JHEP07(2021)151}{{\em JHEP} {\bfseries 07}
  (2021) 151}, \href{http://arxiv.org/abs/2103.15201}{{\ttfamily
  arXiv:2103.15201 [hep-th]}}.

\bibitem{Bobev:2021yya}
N.~Bobev, F.~F. Gautason, and J.~van Muiden, ``{The holographic conformal
  manifold of 3d $ \mathcal{N} $ = 2 S-fold SCFTs},''
  \href{http://dx.doi.org/10.1007/JHEP07(2021)221}{{\em JHEP} {\bfseries 07}
  no.~221, (2021) 221}, \href{http://arxiv.org/abs/2104.00977}{{\ttfamily
  arXiv:2104.00977 [hep-th]}}.

\bibitem{Terashima:2011qi}
Y.~Terashima and M.~Yamazaki, ``{SL(2,R) Chern-Simons, Liouville, and Gauge
  Theory on Duality Walls},''
  \href{http://dx.doi.org/10.1007/JHEP08(2011)135}{{\em JHEP} {\bfseries 08}
  (2011) 135}, \href{http://arxiv.org/abs/1103.5748}{{\ttfamily arXiv:1103.5748
  [hep-th]}}.

\bibitem{Dimofte:2011ju}
T.~Dimofte, D.~Gaiotto, and S.~Gukov, ``{Gauge Theories Labelled by
  Three-Manifolds},'' \href{http://dx.doi.org/10.1007/s00220-013-1863-2}{{\em
  Commun. Math. Phys.} {\bfseries 325} (2014) 367--419},
  \href{http://arxiv.org/abs/1108.4389}{{\ttfamily arXiv:1108.4389 [hep-th]}}.

\bibitem{Gang:2013sqa}
D.~Gang, E.~Koh, S.~Lee, and J.~Park, ``{Superconformal Index and 3d-3d
  Correspondence for Mapping Cylinder/Torus},''
  \href{http://dx.doi.org/10.1007/JHEP01(2014)063}{{\em JHEP} {\bfseries 01}
  (2014) 063}, \href{http://arxiv.org/abs/1305.0937}{{\ttfamily arXiv:1305.0937
  [hep-th]}}.

\bibitem{Hama:2010av}
N.~Hama, K.~Hosomichi, and S.~Lee, ``{Notes on SUSY Gauge Theories on
  Three-Sphere},'' \href{http://dx.doi.org/10.1007/JHEP03(2011)127}{{\em JHEP}
  {\bfseries 03} (2011) 127}, \href{http://arxiv.org/abs/1012.3512}{{\ttfamily
  arXiv:1012.3512 [hep-th]}}.

\bibitem{Hama:2011ea}
N.~Hama, K.~Hosomichi, and S.~Lee, ``{SUSY Gauge Theories on Squashed
  Three-Spheres},'' \href{http://dx.doi.org/10.1007/JHEP05(2011)014}{{\em JHEP}
  {\bfseries 05} (2011) 014}, \href{http://arxiv.org/abs/1102.4716}{{\ttfamily
  arXiv:1102.4716 [hep-th]}}.

\bibitem{Gang:2019jut}
D.~Gang and M.~Yamazaki, ``{Expanding 3d $ \mathcal{N} $ = 2 theories around
  the round sphere},'' \href{http://dx.doi.org/10.1007/JHEP02(2020)102}{{\em
  JHEP} {\bfseries 02} (2020) 102},
  \href{http://arxiv.org/abs/1912.09617}{{\ttfamily arXiv:1912.09617
  [hep-th]}}.

\bibitem{Jafferis:2010un}
D.~L. Jafferis, ``{The Exact Superconformal R-Symmetry Extremizes Z},''
  \href{http://dx.doi.org/10.1007/JHEP05(2012)159}{{\em JHEP} {\bfseries 05}
  (2012) 159}, \href{http://arxiv.org/abs/1012.3210}{{\ttfamily arXiv:1012.3210
  [hep-th]}}.

\bibitem{Faddeev:1993rs}
L.~D. Faddeev and R.~M. Kashaev, ``{Quantum Dilogarithm},''
  \href{http://dx.doi.org/10.1142/S0217732394000447}{{\em Mod. Phys. Lett. A}
  {\bfseries 9} (1994) 427--434},
  \href{http://arxiv.org/abs/hep-th/9310070}{{\ttfamily arXiv:hep-th/9310070}}.

\bibitem{Closset:2017zgf}
C.~Closset, H.~Kim, and B.~Willett, ``{Supersymmetric partition functions and
  the three-dimensional A-twist},''
  \href{http://dx.doi.org/10.1007/JHEP03(2017)074}{{\em JHEP} {\bfseries 03}
  (2017) 074}, \href{http://arxiv.org/abs/1701.03171}{{\ttfamily
  arXiv:1701.03171 [hep-th]}}.

\bibitem{Closset:2018ghr}
C.~Closset, H.~Kim, and B.~Willett, ``{Seifert fibering operators in 3d
  $\mathcal{N}=2$ theories},''
  \href{http://dx.doi.org/10.1007/JHEP11(2018)004}{{\em JHEP} {\bfseries 11}
  (2018) 004}, \href{http://arxiv.org/abs/1807.02328}{{\ttfamily
  arXiv:1807.02328 [hep-th]}}.

\bibitem{Gukov:2015sna}
S.~Gukov and D.~Pei, ``{Equivariant Verlinde formula from fivebranes and
  vortices},'' \href{http://dx.doi.org/10.1007/s00220-017-2931-9}{{\em Commun.
  Math. Phys.} {\bfseries 355} no.~1, (2017) 1--50},
  \href{http://arxiv.org/abs/1501.01310}{{\ttfamily arXiv:1501.01310
  [hep-th]}}.

\bibitem{Benini:2015noa}
F.~Benini and A.~Zaffaroni, ``{A topologically twisted index for
  three-dimensional supersymmetric theories},''
  \href{http://dx.doi.org/10.1007/JHEP07(2015)127}{{\em JHEP} {\bfseries 07}
  (2015) 127}, \href{http://arxiv.org/abs/1504.03698}{{\ttfamily
  arXiv:1504.03698 [hep-th]}}.

\bibitem{Benini:2016hjo}
F.~Benini and A.~Zaffaroni, ``{Supersymmetric partition functions on Riemann
  surfaces},'' {\em Proc. Symp. Pure Math.} {\bfseries 96} (2017) 13--46,
  \href{http://arxiv.org/abs/1605.06120}{{\ttfamily arXiv:1605.06120
  [hep-th]}}.

\bibitem{Closset:2016arn}
C.~Closset and H.~Kim, ``{Comments on twisted indices in 3d supersymmetric
  gauge theories},'' \href{http://dx.doi.org/10.1007/JHEP08(2016)059}{{\em
  JHEP} {\bfseries 08} (2016) 059},
  \href{http://arxiv.org/abs/1605.06531}{{\ttfamily arXiv:1605.06531
  [hep-th]}}.

\bibitem{Gukov:2020lqm}
S.~Gukov, P.-S. Hsin, H.~Nakajima, S.~Park, D.~Pei, and N.~Sopenko,
  ``{Rozansky-Witten geometry of Coulomb branches and logarithmic knot
  invariants},'' \href{http://dx.doi.org/10.1016/j.geomphys.2021.104311}{{\em
  J. Geom. Phys.} {\bfseries 168} (2021) 104311},
  \href{http://arxiv.org/abs/2005.05347}{{\ttfamily arXiv:2005.05347
  [hep-th]}}.

\bibitem{Creutzig:2021ext}
T.~Creutzig, T.~Dimofte, N.~Garner, and N.~Geer, ``{A QFT for non-semisimple
  TQFT},'' \href{http://arxiv.org/abs/2112.01559}{{\ttfamily arXiv:2112.01559
  [hep-th]}}.

\bibitem{Cho:2020ljj}
G.~Y. Cho, D.~Gang, and H.-C. Kim, ``{M-theoretic Genesis of Topological
  Phases},'' \href{http://dx.doi.org/10.1007/JHEP11(2020)115}{{\em JHEP}
  {\bfseries 11} (2020) 115}, \href{http://arxiv.org/abs/2007.01532}{{\ttfamily
  arXiv:2007.01532 [hep-th]}}.

\bibitem{Choi:2022dju}
S.~Choi, D.~Gang, and H.-C. Kim, ``{Infrared phases of 3D class R theories},''
  \href{http://dx.doi.org/10.1007/JHEP11(2022)151}{{\em JHEP} {\bfseries 11}
  (2022) 151}, \href{http://arxiv.org/abs/2206.11982}{{\ttfamily
  arXiv:2206.11982 [hep-th]}}.

\bibitem{Gang:to-appear}
D.~Gang, H.~Kim, and S.~Stubbs, ``{Three-Dimensional Topological Field Theories
  and Non-Unitary Minimal Models},''
  \href{http://arxiv.org/abs/2310.09080}{{\ttfamily arXiv:2310.09080
  [hep-th]}}.

\bibitem{Dedushenko:2018bpp}
M.~Dedushenko, S.~Gukov, H.~Nakajima, D.~Pei, and K.~Ye, ``{3d TQFTs from
  Argyres\textendash{}Douglas theories},''
  \href{http://dx.doi.org/10.1088/1751-8121/abb481}{{\em J. Phys. A} {\bfseries
  53} no.~43, (2020) 43LT01}, \href{http://arxiv.org/abs/1809.04638}{{\ttfamily
  arXiv:1809.04638 [hep-th]}}.

\bibitem{Jolicoeur:2014isa}
T.~Jolicoeur, T.~Mizusaki, and P.~Lecheminant, ``{Absence of a gap in the
  Gaffnian state},'' \href{http://dx.doi.org/10.1103/PhysRevB.90.075116}{{\em
  Phys. Rev. B} {\bfseries 90} no.~7, (2014) 075116},
  \href{http://arxiv.org/abs/1406.5891}{{\ttfamily arXiv:1406.5891
  [cond-mat.mes-hall]}}.

\bibitem{Hsin:2018vcg}
P.-S. Hsin, H.~T. Lam, and N.~Seiberg, ``{Comments on One-Form Global
  Symmetries and Their Gauging in 3d and 4d},''
  \href{http://dx.doi.org/10.21468/SciPostPhys.6.3.039}{{\em SciPost Phys.}
  {\bfseries 6} no.~3, (2019) 039},
  \href{http://arxiv.org/abs/1812.04716}{{\ttfamily arXiv:1812.04716
  [hep-th]}}.

\bibitem{Gang:2018wek}
D.~Gang and K.~Yonekura, ``{Symmetry enhancement and closing of knots in 3d/3d
  correspondence},'' \href{http://dx.doi.org/10.1007/JHEP07(2018)145}{{\em
  JHEP} {\bfseries 07} (2018) 145},
  \href{http://arxiv.org/abs/1803.04009}{{\ttfamily arXiv:1803.04009
  [hep-th]}}.

\bibitem{Evans:2010yr}
D.~E. Evans and T.~Gannon, ``{The exoticness and realisability of twisted
  Haagerup-Izumi modular data},''
  \href{http://dx.doi.org/10.1007/s00220-011-1329-3}{{\em Commun. Math. Phys.}
  {\bfseries 307} (2011) 463--512},
  \href{http://arxiv.org/abs/1006.1326}{{\ttfamily arXiv:1006.1326 [math.QA]}}.

\bibitem{Ng:2012ty}
S.-H. Ng and X.~Lin, ``{Congruence Property In Conformal Field Theory},''
  \href{http://dx.doi.org/10.2140/ant.2015.9.2121}{{\em Alg. Numb. Theor.}
  {\bfseries 9} no.~9, (2015) 2121--2166},
  \href{http://arxiv.org/abs/1201.6644}{{\ttfamily arXiv:1201.6644 [math.QA]}}.

\bibitem{Harvey:2018rdc}
J.~A. Harvey and Y.~Wu, ``{Hecke Relations in Rational Conformal Field
  Theory},'' \href{http://dx.doi.org/10.1007/JHEP09(2018)032}{{\em JHEP}
  {\bfseries 09} (2018) 032}, \href{http://arxiv.org/abs/1804.06860}{{\ttfamily
  arXiv:1804.06860 [hep-th]}}.

\bibitem{Cho:2022kzf}
G.~Y. Cho, H.-c. Kim, D.~Seo, and M.~You, ``{Classification of Fermionic
  Topological Orders from Congruence Representations},''
  \href{http://arxiv.org/abs/2210.03681}{{\ttfamily arXiv:2210.03681
  [cond-mat.str-el]}}.

\bibitem{Costello:2020ndc}
K.~Costello, T.~Dimofte, and D.~Gaiotto, ``{Boundary Chiral Algebras and
  Holomorphic Twists},''
  \href{http://dx.doi.org/10.1007/s00220-022-04599-0}{{\em Commun. Math. Phys.}
  {\bfseries 399} no.~2, (2023) 1203--1290},
  \href{http://arxiv.org/abs/2005.00083}{{\ttfamily arXiv:2005.00083
  [hep-th]}}.

\bibitem{atkin1971modular}
A.~O.~L. Atkin and H.~P.~F. Swinnerton-Dyer, ``{Modular forms on noncongruence
  subgroups},'' {\em Combinatorics} {\bfseries 19} (1971) 1--25.

\bibitem{calegari2021unbounded}
C.~Frank, D.~Vesselin, and T.~Yunqing, ``{The Unbounded Denominators
  Conjecture},'' \href{http://arxiv.org/abs/2109.09040}{{\ttfamily
  arXiv:2109.09040}}.

\bibitem{Kaidi:2021ent}
J.~Kaidi, Y.-H. Lin, and J.~Parra-Martinez, ``{Holomorphic modular bootstrap
  revisited},'' \href{http://dx.doi.org/10.1007/JHEP12(2021)151}{{\em JHEP}
  {\bfseries 12} (2021) 151}, \href{http://arxiv.org/abs/2107.13557}{{\ttfamily
  arXiv:2107.13557 [hep-th]}}.

\bibitem{Bae:2021jkc}
J.-B. Bae, Z.~Duan, and S.~Lee, ``{Can the energy bound E\ensuremath{\geq}0
  imply supersymmetry?},''
  \href{http://dx.doi.org/10.1103/PhysRevD.107.045018}{{\em Phys. Rev. D}
  {\bfseries 107} no.~4, (2023) 045018},
  \href{http://arxiv.org/abs/2112.14130}{{\ttfamily arXiv:2112.14130
  [hep-th]}}.

\bibitem{Duan:2022kxr}
Z.~Duan, K.~Lee, S.~Lee, and L.~Li, ``{On classification of fermionic rational
  conformal field theories},''
  \href{http://dx.doi.org/10.1007/JHEP02(2023)079}{{\em JHEP} {\bfseries 02}
  (2023) 079}, \href{http://arxiv.org/abs/2210.06805}{{\ttfamily
  arXiv:2210.06805 [hep-th]}}.

\bibitem{Gadde:2013wq}
A.~Gadde, S.~Gukov, and P.~Putrov, ``{Walls, Lines, and Spectral Dualities in
  3d Gauge Theories},'' \href{http://dx.doi.org/10.1007/JHEP05(2014)047}{{\em
  JHEP} {\bfseries 05} (2014) 047},
  \href{http://arxiv.org/abs/1302.0015}{{\ttfamily arXiv:1302.0015 [hep-th]}}.

\bibitem{Gadde:2013sca}
A.~Gadde, S.~Gukov, and P.~Putrov, ``{Fivebranes and 4-manifolds},''
  \href{http://dx.doi.org/10.1007/978-3-319-43648-7_7}{{\em Prog. Math.}
  {\bfseries 319} (2016) 155--245},
  \href{http://arxiv.org/abs/1306.4320}{{\ttfamily arXiv:1306.4320 [hep-th]}}.

\bibitem{Sugishita:2013jca}
S.~Sugishita and S.~Terashima, ``{Exact Results in Supersymmetric Field
  Theories on Manifolds with Boundaries},''
  \href{http://dx.doi.org/10.1007/JHEP11(2013)021}{{\em JHEP} {\bfseries 11}
  (2013) 021}, \href{http://arxiv.org/abs/1308.1973}{{\ttfamily arXiv:1308.1973
  [hep-th]}}.

\bibitem{Dimofte:2017tpi}
T.~Dimofte, D.~Gaiotto, and N.~M. Paquette, ``{Dual boundary conditions in 3d
  SCFT\textquoteright{}s},''
  \href{http://dx.doi.org/10.1007/JHEP05(2018)060}{{\em JHEP} {\bfseries 05}
  (2018) 060}, \href{http://arxiv.org/abs/1712.07654}{{\ttfamily
  arXiv:1712.07654 [hep-th]}}.

\bibitem{Cheng:2018vpl}
M.~C.~N. Cheng, S.~Chun, F.~Ferrari, S.~Gukov, and S.~M. Harrison, ``{3d
  Modularity},'' \href{http://dx.doi.org/10.1007/JHEP10(2019)010}{{\em JHEP}
  {\bfseries 10} (2019) 010}, \href{http://arxiv.org/abs/1809.10148}{{\ttfamily
  arXiv:1809.10148 [hep-th]}}.

\bibitem{Cheng:2022rqr}
M.~C.~N. Cheng, S.~Chun, B.~Feigin, F.~Ferrari, S.~Gukov, S.~M. Harrison, and
  D.~Passaro, ``{3-Manifolds and VOA Characters},''
  \href{http://arxiv.org/abs/2201.04640}{{\ttfamily arXiv:2201.04640
  [hep-th]}}.

\bibitem{Chung:2023qth}
H.-J. Chung, ``{3d-3d Correspondence and 2d $\mathcal{N}=(0,2)$ Boundary
  Conditions},'' \href{http://arxiv.org/abs/2307.10125}{{\ttfamily
  arXiv:2307.10125 [hep-th]}}.

\bibitem{Kedem:1993ze}
R.~Kedem, T.~R. Klassen, B.~M. McCoy, and E.~Melzer, ``{Fermionic sum
  representations for conformal field theory characters},''
  \href{http://dx.doi.org/10.1016/0370-2693(93)90194-M}{{\em Phys. Lett. B}
  {\bfseries 307} (1993) 68--76},
  \href{http://arxiv.org/abs/hep-th/9301046}{{\ttfamily arXiv:hep-th/9301046}}.

\bibitem{Nahm:1994vas}
W.~Nahm, ``{Conformal field theory, dilogarithms, and three-dimensional
  manifolds},'' {\em Adv. Appl. Clifford Algebras} {\bfseries 4} no.~S1, (1994)
  179--191.

\bibitem{Berkovich:1994es}
A.~Berkovich and B.~M. McCoy, ``{Continued fractions and Fermionic
  representations for characters of M(p,p-prime) minimal models},''
  \href{http://dx.doi.org/10.1007/BF00400138}{{\em Lett. Math. Phys.}
  {\bfseries 37} (1996) 49--66},
  \href{http://arxiv.org/abs/hep-th/9412030}{{\ttfamily arXiv:hep-th/9412030}}.

\bibitem{Nahm:2004ch}
W.~Nahm, \href{http://dx.doi.org/10.1007/978-3-540-30308-4_2}{``{Conformal
  field theory and torsion elements of the Bloch group},''} in {\em {Les
  Houches School of Physics: Frontiers in Number Theory, Physics and
  Geometry}}, pp.~67--132.
\newblock 2007.
\newblock \href{http://arxiv.org/abs/hep-th/0404120}{{\ttfamily
  arXiv:hep-th/0404120}}.

\bibitem{welsh2005fermionic}
T.~A. Welsh, {\em Fermionic expressions for minimal model Virasoro characters}.
\newblock American Mathematical Soc., 2005.

\bibitem{Zagier:2007knq}
D.~Zagier, \href{http://dx.doi.org/10.1007/978-3-540-30308-4_1}{``{The
  Dilogarithm Function},''} in {\em {Les Houches School of Physics: Frontiers
  in Number Theory, Physics and Geometry}}, pp.~3--65.
\newblock 2007.

\bibitem{Mukhi:2019xjy}
S.~Mukhi, ``{Classification of RCFT from Holomorphic Modular Bootstrap: A
  Status Report},'' in {\em {Pollica Summer Workshop 2019}: {Mathematical and
  Geometric Tools for Conformal Field Theories}}.
\newblock 10, 2019.
\newblock \href{http://arxiv.org/abs/1910.02973}{{\ttfamily arXiv:1910.02973
  [hep-th]}}.

\bibitem{Bantay:2007zz}
P.~Bantay and T.~Gannon, ``{Vector-valued modular functions for the modular
  group and the hypergeometric equation},''
  \href{http://dx.doi.org/10.4310/CNTP.2007.v1.n4.a2}{{\em Commun. Num. Theor.
  Phys.} {\bfseries 1} (2007) 651--680}.

\bibitem{Gannon:2013jua}
T.~Gannon, ``{The theory of vector-modular forms for the modular group},''
  \href{http://dx.doi.org/10.1007/978-3-662-43831-2_9}{{\em Contrib. Math.
  Comput. Sci.} {\bfseries 8} (2014) 247--286},
  \href{http://arxiv.org/abs/1310.4458}{{\ttfamily arXiv:1310.4458 [math.NT]}}.

\bibitem{Cheng:2020srs}
M.~C.~N. Cheng, T.~Gannon, and G.~Lockhart, ``{Modular Exercises for Four-Point
  Blocks -- I},'' \href{http://arxiv.org/abs/2002.11125}{{\ttfamily
  arXiv:2002.11125 [hep-th]}}.

\bibitem{Gaiotto:2016wcv}
D.~Gaiotto, ``{Twisted compactifications of 3d $ \mathcal{N} $ = 4 theories and
  conformal blocks},'' \href{http://dx.doi.org/10.1007/JHEP02(2019)061}{{\em
  JHEP} {\bfseries 02} (2019) 061},
  \href{http://arxiv.org/abs/1611.01528}{{\ttfamily arXiv:1611.01528
  [hep-th]}}.

\bibitem{Gaiotto:2017euk}
D.~Gaiotto and M.~Rap\v{c}\'ak, ``{Vertex Algebras at the Corner},''
  \href{http://dx.doi.org/10.1007/JHEP01(2019)160}{{\em JHEP} {\bfseries 01}
  (2019) 160}, \href{http://arxiv.org/abs/1703.00982}{{\ttfamily
  arXiv:1703.00982 [hep-th]}}.

\bibitem{Costello:2018fnz}
K.~Costello and D.~Gaiotto, ``{Vertex Operator Algebras and 3d $ \mathcal{N} $
  = 4 gauge theories},'' \href{http://dx.doi.org/10.1007/JHEP05(2019)018}{{\em
  JHEP} {\bfseries 05} (2019) 018},
  \href{http://arxiv.org/abs/1804.06460}{{\ttfamily arXiv:1804.06460
  [hep-th]}}.

\bibitem{Costello:2018swh}
K.~Costello, T.~Creutzig, and D.~Gaiotto, ``{Higgs and Coulomb branches from
  vertex operator algebras},''
  \href{http://dx.doi.org/10.1007/JHEP03(2019)066}{{\em JHEP} {\bfseries 03}
  (2019) 066}, \href{http://arxiv.org/abs/1811.03958}{{\ttfamily
  arXiv:1811.03958 [hep-th]}}.

\bibitem{Garner:2022vds}
N.~Garner, ``{Twisted Formalism for 3d $\mathcal{N}=4$ Theories},''
  \href{http://arxiv.org/abs/2204.02997}{{\ttfamily arXiv:2204.02997
  [hep-th]}}.

\bibitem{Garner:2023pmt}
N.~Garner and W.~Niu, ``{Line Operators in $U(1|1)$ Chern-Simons Theory},''
  \href{http://arxiv.org/abs/2304.05414}{{\ttfamily arXiv:2304.05414
  [hep-th]}}.

\bibitem{Yoshida:2023wyt}
Y.~Yoshida, ``{Fermionic extensions of $W$-algebras via 3d $\mathcal{N}=4$
  gauge theories with a boundary},''
  \href{http://arxiv.org/abs/2304.03270}{{\ttfamily arXiv:2304.03270
  [hep-th]}}.

\bibitem{Beem:2023dub}
C.~Beem and A.~E.~V. Ferrari, ``{Free field realisation of boundary vertex
  algebras for Abelian gauge theories in three dimensions},''
  \href{http://arxiv.org/abs/2304.11055}{{\ttfamily arXiv:2304.11055
  [hep-th]}}.

\bibitem{Dimofte:2011py}
T.~Dimofte, D.~Gaiotto, and S.~Gukov, ``{3-Manifolds and 3d Indices},''
  \href{http://dx.doi.org/10.4310/ATMP.2013.v17.n5.a3}{{\em Adv. Theor. Math.
  Phys.} {\bfseries 17} no.~5, (2013) 975--1076},
  \href{http://arxiv.org/abs/1112.5179}{{\ttfamily arXiv:1112.5179 [hep-th]}}.

\bibitem{Lee:2013ida}
S.~Lee and M.~Yamazaki, ``{3d Chern-Simons Theory from M5-branes},''
  \href{http://dx.doi.org/10.1007/JHEP12(2013)035}{{\em JHEP} {\bfseries 12}
  (2013) 035}, \href{http://arxiv.org/abs/1305.2429}{{\ttfamily arXiv:1305.2429
  [hep-th]}}.

\bibitem{Yagi:2013fda}
J.~Yagi, ``{3d TQFT from 6d SCFT},''
  \href{http://dx.doi.org/10.1007/JHEP08(2013)017}{{\em JHEP} {\bfseries 08}
  (2013) 017}, \href{http://arxiv.org/abs/1305.0291}{{\ttfamily arXiv:1305.0291
  [hep-th]}}.

\bibitem{Cordova:2013cea}
C.~Cordova and D.~L. Jafferis, ``{Complex Chern-Simons from M5-branes on the
  Squashed Three-Sphere},''
  \href{http://dx.doi.org/10.1007/JHEP11(2017)119}{{\em JHEP} {\bfseries 11}
  (2017) 119}, \href{http://arxiv.org/abs/1305.2891}{{\ttfamily arXiv:1305.2891
  [hep-th]}}.

\bibitem{Dimofte:2014zga}
T.~Dimofte, ``{Complex Chern\textendash{}Simons Theory at Level k via the
  3d\textendash{}3d Correspondence},''
  \href{http://dx.doi.org/10.1007/s00220-015-2401-1}{{\em Commun. Math. Phys.}
  {\bfseries 339} no.~2, (2015) 619--662},
  \href{http://arxiv.org/abs/1409.0857}{{\ttfamily arXiv:1409.0857 [hep-th]}}.

\bibitem{HIKAMI_2001}
K.~HIKAMI, ``{HYPERBOLIC} {STRUCTURE} {ARISING} {FROM} a {KNOT} {INVARIANT},''
  \href{https://doi.org/10.1142%2Fs0217751x0100444x}{{\em International Journal
  of Modern Physics A} {\bfseries 16} no.~19, (Jul, 2001) 3309--3333}.

\bibitem{hikami2007generalized}
K.~Hikami, ``Generalized volume conjecture and the a-polynomials: the
  neumann--zagier potential function as a classical limit of the partition
  function,'' {\em Journal of Geometry and Physics} {\bfseries 57} no.~9,
  (2007) 1895--1940.

\bibitem{Dimofte:2009yn}
T.~Dimofte, S.~Gukov, J.~Lenells, and D.~Zagier, ``{Exact Results for
  Perturbative Chern-Simons Theory with Complex Gauge Group},''
  \href{http://dx.doi.org/10.4310/CNTP.2009.v3.n2.a4}{{\em Commun. Num. Theor.
  Phys.} {\bfseries 3} (2009) 363--443},
  \href{http://arxiv.org/abs/0903.2472}{{\ttfamily arXiv:0903.2472 [hep-th]}}.

\bibitem{EllegaardAndersen:2011vps}
J.~Ellegaard~Andersen and R.~Kashaev, ``{A TQFT from Quantum Teichm\"uller
  Theory},'' \href{http://dx.doi.org/10.1007/s00220-014-2073-2}{{\em Commun.
  Math. Phys.} {\bfseries 330} (2014) 887--934},
  \href{http://arxiv.org/abs/1109.6295}{{\ttfamily arXiv:1109.6295 [math.QA]}}.

\bibitem{Dimofte:2012qj}
T.~D. Dimofte and S.~Garoufalidis, ``{The Quantum content of the gluing
  equations},'' {\em Geom. Topol.} {\bfseries 17} (2013) 1253--1316,
  \href{http://arxiv.org/abs/1202.6268}{{\ttfamily arXiv:1202.6268 [math.GT]}}.

\bibitem{Dimofte:2011gm}
T.~Dimofte, ``{Quantum Riemann Surfaces in Chern-Simons Theory},''
  \href{http://dx.doi.org/10.4310/ATMP.2013.v17.n3.a1}{{\em Adv. Theor. Math.
  Phys.} {\bfseries 17} no.~3, (2013) 479--599},
  \href{http://arxiv.org/abs/1102.4847}{{\ttfamily arXiv:1102.4847 [hep-th]}}.

\bibitem{2004math......6242G}
F.~{Gueritaud} and D.~{Futer}, ``{On canonical triangulations of once-punctured
  torus bundles and two-bridge link complements},''
  \href{http://dx.doi.org/10.48550/arXiv.math/0406242}{{\em arXiv Mathematics
  e-prints} (June, 2004) math/0406242},
  \href{http://arxiv.org/abs/math/0406242}{{\ttfamily arXiv:math/0406242
  [math.GT]}}.

\bibitem{neumann1985volumes}
W.~D. Neumann and D.~Zagier, ``Volumes of hyperbolic three-manifolds,'' {\em
  Topology} {\bfseries 24} no.~3, (1985) 307--332.

\end{thebibliography}\endgroup

\end{document}